%% file: Draft9.tex
\documentclass[a4 paper, 11pt]{article}
\pdfoutput=1
\usepackage{comment}
\usepackage{jheppub}
\usepackage{subcaption}
\usepackage{amsmath}
\usepackage{mathrsfs}
\usepackage{amssymb}
\usepackage{amscd}
\usepackage{dsfont}
\usepackage{enumerate}
\usepackage{amsfonts}
\usepackage{epsfig}
\usepackage{mathtools}
\usepackage{yfonts}
\usepackage{bbold}
\usepackage{breqn}
\usepackage[utf8]{inputenc}
\usepackage[english]{babel}
\usepackage{graphicx}
\allowdisplaybreaks[4]

\input{preamble.tex}


\title{ABJM to BMN in a Double Scale Limit: Indices and Bubbling Geometries}

\author[\dagger,\ddagger]{Zhengyuan Du}

\affiliation[\dagger]{School of Physics, Peking University,
Beijing 100871, China}

\affiliation[\ddagger]{Center for High Energy Physics, Peking University,
Beijing 100871, China}

\emailAdd{duzhengyuan@pku.edu.cn}

\abstract{
We study a double scale limit of $U(N)_k\times U(N)_{-k}$ ABJM theory, with $N,k\to\infty$ and $N/k^2\to\nu\in(0,\infty)$, at fixed positive total monopole charge $q$. In this limit, the Penrose geometry retains a compact null circle of finite radius, and the sector of monopole charge $q$ carries $q$ units of longitudinal momentum. It is therefore naturally associated with the rank-$q$ BMN matrix model, while $\nu$ fixes its dimensionless coupling. We establish this relation from both gravity and the supersymmetric index. On the gravity side, we take the double scale limit of Hopf quotients of the Donos--Sim\'on half-BPS geometries. The disk carrying the growing background flux becomes an infinite conducting plane, while the remaining finite disks reproduce the Lin--Maldacena electrostatic problem. Their quantized fluxes map directly to the partition data labeling BMN vacua. On the field-theory side, starting from the finite-$N$, finite-$k$ ABJM localization formula, we prove that, at fixed positive monopole charge $q$, the ABJM superconformal index factorizes coefficientwise in the transverse fugacities into a universal neutral contribution and the refined Witten index of the rank-$q$ BMN matrix model, together with the longitudinal momentum weight. The neutral contribution is precisely the limiting zero-monopole-charge index. Dividing by this universal factor therefore isolates the BMN index summed over all its supersymmetric vacua.
}

\begin{document}
\maketitle

\section{Introduction}
\label{sec:introduction-draft4}

The Banks--Fischler--Shenker--Susskind (BFSS) proposal
describes M-theory through supersymmetric matrix quantum
mechanics \cite{Banks:1996vh}.  In discrete light-cone
quantization (DLCQ), a finite-rank \(U(q)\) model describes the sector
with \(q\) units of longitudinal momentum
\cite{Susskind:1997cw,Seiberg:1997ad}.\footnote{For a review
of Matrix theory, its membrane interpretation and DLCQ
formulation, see \cite{Taylor:2001vb}.}

The Berenstein--Maldacena--Nastase (BMN) model extends this
matrix description to the maximally supersymmetric
eleven-dimensional pp-wave \cite{Berenstein:2002jq}.  It is
a mass deformation of the BFSS model and has gauge group
\(U(q)\).  Its supersymmetric vacua are fuzzy spheres
labelled by partitions of \(q\).  Their perturbative spectra
were computed in \cite{Dasgupta:2002hx,Kim:2002if}, and the
vacuum states and certain excited multiplets are protected
non-perturbatively \cite{Dasgupta:2002ru}.

AdS/CFT gives a complementary description of M-theory and
string theory in terms of conformal field theory
\cite{Maldacena:1997re,Gubser:1998bc,Witten:1998qj}.  A
Penrose limit on the gravity side should therefore select a
corresponding large-charge sector of the boundary theory.
This suggests a direct way to connect the two frameworks:
one may try to obtain a finite-rank BMN model as a controlled
sector of a holographic CFT.

ABJM theory provides a natural setting for this question.
The \(U(N)_k\times U(N)_{-k}\) theory is dual to M-theory on
\(\mathrm{AdS}_4\times S^7/\mathbb Z_k\)
\cite{Aharony:2008ug}.  Momentum along the Hopf circle is
measured by the monopole charge in ABJM.  Moreover, half-BPS
monopole states, BMN vacua and bubbling geometries share the
same Young-diagram labels \cite{Sheikh-Jabbari:2009vjj}.
Protected fluctuations around dressed monopoles give
further evidence for this relation
\cite{Berenstein:2009sa,Kim:2009ia}.  The question is then
whether a correlated limit of \(N\), \(k\) and the Hopf
momentum can keep the BMN rank finite while retaining the
compact light-cone circle and finite excitation energies.

Earlier work studied several related regimes.  A type IIA
Penrose limit of \(\mathrm{AdS}_4\times\mathbb{CP}^3\)
selects BMN-like operators \cite{Nishioka:2008gz}.  At
\(k\) of order one, large-angular-momentum monopole states
are related to membrane excitations of the BMN model
\cite{Kovacs:2013una}.  The Penrose limit applies to the
full supergravity fields \cite{Gueven:2000ru}.  On the
\(\mathrm{AdS}_4\times S^7\) covering space it gives the
maximally supersymmetric pp-wave \cite{Blau:2002dy}.
A DLCQ interpretation also requires tracking the global
identifications \cite{Shomer:2003tj}.

In this paper we take the double scale limit
\begin{equation}
    N,k\longrightarrow\infty,\qquad
    \frac{N}{k^2}\longrightarrow\nu\in(0,\infty),
    \qquad
    J_{\mathrm H}=kq,\qquad q\in\mathbb Z_{>0}\ \text{fixed},
    \label{eq:intro-double-scaling-draft4}
\end{equation}
Here \(J_{\mathrm H}\) is the Hopf angular momentum, and
\(q\) is the common total Goddard--Nuyts--Olive (GNO)
charge of the two ABJM gauge groups.  We keep \(q\) fixed
while \(N\) and \(k\) diverge.  At fixed Planck length
\(\ell_p\) and pp-wave mass parameter \(\mu\), the
orbifold action becomes a finite translation along the
null circle.  Its radius and the BMN coupling are
\begin{equation}
    R_-=\frac{\mu\ell_p^2}{6}(32\pi^2\nu)^{1/3},
    \qquad
    \frac{g_{\mathrm B}^2}{\mu^3}=\frac{\nu}{27},
    \label{eq:intro-DLCQ-parameters-draft6}
\end{equation}
in the normalization of \eqref{eq:BMN-coupling-nu-draft4}.
For states with bounded light-cone energy and transverse
charges, \(p^+R_-\to q\).  The two ranks therefore have
different roles.  The large ABJM rank \(N\) sets the parent
geometry, whereas the fixed charge \(q\) becomes the BMN
matrix rank.  The ratio \(\nu\) fixes a finite BMN coupling;
it does not impose weak coupling.  A complementary analysis
at strong BMN coupling reproduces the pp-wave supergravity
spectrum, subject to a bound-state assumption
\cite{Komatsu:2024vnb}.

BMN vacua also have a geometric description.  They can be
viewed as configurations of spherical membranes and
transverse fivebranes \cite{Maldacena:2002rb}.  Their early
type IIA duals involve polarized D2- and NS5-brane shells
\cite{Lin:2004kw}.  Lin, Lunin and Maldacena classified the
corresponding half-BPS geometries \cite{Lin:2004nb}, and
Lin and Maldacena (LM) expressed the BMN branch as an
electrostatic problem with conducting disks above a plane
\cite{Lin:2005nh}.

This electrostatic description also emerges from matrix
model localization.  In suitable large-rank regimes, the
localized eigenvalue density obeys the same equation as the
charge density on the supergravity disks
\cite{Asano:2014vba,Asano:2014eca}.  The scalar eigenvalue
distribution also captures spherical transverse M5-branes
\cite{Asano:2017nxw}.  These results relate the BMN model
directly to LM electrostatics.  We ask a different question:
how does the LM boundary problem arise from an
asymptotically \(\mathrm{AdS}_4\times S^7/\mathbb Z_k\)
geometry?

Our first result answers this question.  We start from the
axisymmetric Donos--Sim\'on geometries
\cite{Donos:2010va}.  Their electrostatic data consist of a
semi-infinite line charge and a set of conducting disks.  In
the double scale limit, the disk at the line endpoint grows
into an infinite conducting plane.  The other disks remain
finite.  After an affine subtraction and a rescaling, the
potential becomes the LM potential: its cubic background
comes from the growing disk, and its image charges come from
the response of that disk.  The metric, four-form and circle
period have the corresponding LM limits.  The M2 and M5
fluxes become the D2 and NS5 integers that specify the BMN
vacuum partition.

Our second result concerns the supersymmetric index.  The
ABJM index was first compared with the dual graviton Fock
space in a large-\(N\), large-\(k\) regime
\cite{Bhattacharya:2008bja}.  Its localization formula
includes all monopole sectors \cite{Kim:2009wb}.  The BMN
index is a sum of matrix integrals, one for each
partition-labelled vacuum \cite{Chang:2024lkw}.  A recent
ABJM--BMN comparison keeps \(N\) and \(k\) fixed while the
longitudinal momenta grow with fixed differences
\cite{Chang:2026yeg}.  Our limit instead first keeps the total
monopole charge \(q\) fixed and compares ABJM with the
all-vacuum index of the rank-\(q\) BMN model.  We then sum the
fixed-charge relations to obtain a grand canonical ensemble
of BMN rank.

Let \(\mathcal I^{\mathrm{ABJM}}_{N,k}
(\delta;\boldsymbol\Delta)\) be the full ABJM index written
in the variables adapted to the Penrose limit.  Here
\(\boldsymbol\Delta=(\Delta_1,\Delta_2,\Delta_3)\), and the
longitudinal chemical potential is
\(\sigma=\delta/k\), with \(\operatorname{Re}\delta>0\).
In the double scale limit, the index contains a universal
neutral factor.  This factor is the limiting index in the
zero-total-charge sector.  The result is
\begin{equation}
    \lim_{\substack{N,k\to\infty\\N/k^2\to\nu}}^{\mathrm{coeff}}
    \mathcal I^{\mathrm{ABJM}}_{N,k}
                   (\delta;\boldsymbol\Delta)
    =
    \mathcal I_{\mathrm{neutral}}(\boldsymbol\Delta)
    \sum_{q=0}^{\infty}e^{-\delta q}
    \sum_{\lambda\vdash q}
        \mathcal I^{\mathrm{BMN}}_{q,\lambda}
                   (\boldsymbol\Delta).
    \label{eq:intro-index-relation-draft4}
\end{equation}
Each partition \(\lambda\vdash q\) labels one BMN vacuum.
The outer sum is the grand canonical ensemble of BMN rank,
and \(e^{-\delta q}\) is the longitudinal-momentum weight.
The equality is coefficientwise in the transverse fugacities
\(u_a=e^{-\Delta_a}\) and is understood as a formal series in
\(e^{-\delta}\): each coefficient fixes \(q\) and a monomial
in the \(u_a\).

The proof works directly at finite \(N\) and \(k\).  Fix a
maximum transverse degree \(L\).  When \(k>L\), the
Gauss-law constraints remove sectors with negative fluxes
or unequal flux distributions.  The surviving sector is a
matched partition of \(q\).  When \(N\geq q+L\), the Haar
integrals over the zero-flux blocks stabilize and produce
the universal neutral factor.  A character projection then
reduces the charged blocks to the BMN integral for the same
partition.  This gives an exact identity through degree
\(L\).  The remaining shifts of the chemical potentials are
proportional to \(\delta/k\) and vanish in the double scale
limit.  This proves
\eqref{eq:intro-index-relation-draft4} coefficientwise,
without exchanging the limit with the infinite fugacity
series.

The geometric and index calculations use the same
dictionary.  The charge \(q\) becomes the BMN rank, and a
partition \(\lambda\vdash q\) labels both the surviving
monopole sector and the BMN vacuum.  The geometric result
concerns classical fields.  The index result is an exact
statement about protected coefficients.  Tests of the full
interacting dynamics remain open.

The paper is organized as follows.
Section~\ref{sec:penrose-index-dictionary} follows the
Penrose limit together with the orbifold identification.  It
derives the compact null radius, identifies \(q\) with the
BMN rank, and matches the charges and fugacities of the two
indices.  Section~\ref{sec:half-BPS-geometric-limit-draft4}
first reviews the Donos--Sim\'on covering geometry and its
diagonal \(\mathbb Z_k\) quotient.  It then computes the
M2, M5 and Hopf charges and takes the double scale limit of
the electrostatic problem.  The result is the normalized LM
metric and four-form, together with the compact-circle
period and the BMN vacuum data.
Section~\ref{sec:double-scaling-index-draft4} proves the
index relation from the finite-\(N\), finite-\(k\)
localization formula.  It selects the monopole sectors that
survive at fixed degree, isolates the neutral factor, and
uses a character projection to obtain the BMN matrix
integral for every vacuum partition.
Section~\ref{sec:discussion-draft4} summarizes the scope of
the results and discusses dynamical tests, gravitational
saddles and instanton effects.
Appendix~\ref{app:ABJM-BMN-index-review} reviews the ABJM
and BMN indices.  Appendix~\ref{app:character-identities-draft5}
collects the character identities used in the proof.

\paragraph*{Note added.}
During the preparation of this manuscript, we became aware
of the work of Komatsu, Lee and Patel
\cite{Komatsu:2026ano}.  Their proposal for the double scale
limit and their coefficientwise ABJM--BMN index relation
overlap with results independently obtained here.  We also
derive the double scale limit of the half-BPS bubbling
geometries.  Our index proof controls the joint \(N,k\)
limit directly from the finite-\(N,k\) localization formula,
with explicit bounds at fixed monopole charge and transverse
degree.

\section{The Penrose limit and the ABJM--BMN index dictionary}
\label{sec:penrose-index-dictionary}

The geometric starting point is the duality between
\(U(N)_k\times U(N)_{-k}\) ABJM theory and M-theory on
\(\mathrm{AdS}_4\times S^7/\mathbb Z_k\)
\cite{Aharony:2008ug}.  On the covering space, the Penrose
limit about a null geodesic carrying angular momentum on
\(S^7\) gives the maximally supersymmetric eleven-dimensional
pp-wave \cite{Blau:2002dy}.  Its compactification depends
on the fate of the orbifold identification.  We consider
the double scale limit
\begin{equation}
    N,k\longrightarrow\infty,\qquad
    \frac{N}{k^2}\longrightarrow\nu\in(0,\infty),
    \label{eq:joint-Penrose-scaling-draft4}
\end{equation}
in which the quotient generator becomes a finite translation
along the null coordinate \(x^-\), while its shift of
\(x^+\) and its transverse rotation vanish.  The resulting
pp-wave retains a compact null circle,
\(x^-\sim x^-+2\pi R_-\).

The BMN matrix model is proposed to describe M-theory on
this background in DLCQ \cite{Berenstein:2002jq}.
The orbifold projection \(J_{\mathrm H}=kq\) relates the
Hopf angular momentum to the common total GNO charge
\(q\) of the two ABJM gauge groups.  For states with
bounded light-cone energy and transverse charges, we will
find \(p^+R_-\to q\).  A sector with fixed positive \(q\)
therefore has the momentum appropriate to the rank-\(q\)
BMN model, even as the parent color rank \(N\) diverges.
This identifies the ABJM charge sector
relevant to a finite-rank matrix description.

The supersymmetric index provides a protected observable
with which to test this relation.  Its comparison requires
matching the BPS condition and the commuting charges that
refine the two indices.  We first derive the compact
pp-wave geometry and its circle normalization, then follow
the conserved charges and fugacity weights through the
same limit.  This will identify the transverse BMN
refinements and retain the longitudinal momentum weight,
including the finite-\(k\) corrections.  The equality of
protected multiplicities is a separate question, addressed
by localization in section~\ref{sec:double-scaling-index-draft4}.

\subsection{The Penrose limit and null compactification}
\label{subsec:compact-pp-wave-draft4}

We take \(k>0\) and keep the Planck length \(\ell_p\)
fixed.  The parent metric and four-form are normalized as
\begin{equation}
    ds_{11}^2
    =
    \frac{R^2}{4}\,ds_{\mathrm{AdS}_4}^2
    +R^2ds_{S^7/\mathbb Z_k}^2,
    \qquad
    G_4
    =
    \frac{3R^3}{8}\,\omega_{\mathrm{AdS}_4},
    \qquad
    R^6=32\pi^2Nk\,\ell_p^6.
    \label{eq:ABJM-background-draft4}
\end{equation}
Here \(R\) is the radius of the covering seven-sphere,
and \(\omega_{\mathrm{AdS}_4}\)
is the volume form of the unit-radius AdS metric.  In global
coordinates,
\begin{equation}
    ds_{\mathrm{AdS}_4}^2
    =
    -\cosh^2\rho\,dt^2
    +d\rho^2
    +\sinh^2\rho\,d\Omega_2^2.
    \label{eq:global-AdS4-metric-draft4}
\end{equation}
Parametrize the covering seven-sphere by
\begin{equation}
    Z_4=\cos\vartheta\,e^{i\chi},
    \qquad
    (Z_1,Z_2,Z_3)=\sin\vartheta\,\boldsymbol u,
    \qquad
    \boldsymbol u\in S^5\subset\mathbb C^3,
    \label{eq:S7-adapted-coordinates-draft4}
\end{equation}
so that
\begin{equation}
    ds_{S^7}^2
    =
    d\vartheta^2
    +\cos^2\vartheta\,d\chi^2
    +\sin^2\vartheta\,ds_{S^5}^2.
    \label{eq:S7-adapted-metric-draft4}
\end{equation}
We choose the null geodesic
\begin{equation}
    \rho=0,
    \qquad
    \vartheta=0,
    \qquad
    \chi=\frac{t}{2}.
    \label{eq:Penrose-null-geodesic-draft4}
\end{equation}
The factor \(1/2\) follows from the relative radii in
\eqref{eq:ABJM-background-draft4}: on this trajectory the
\((t,\chi)\) metric is \(R^2(-dt^2/4+d\chi^2)\).
To resolve its neighborhood, introduce
\begin{equation}
    r=\frac{R}{2}\rho,
    \qquad
    z=R\vartheta,
    \label{eq:Penrose-transverse-coordinates-draft4}
\end{equation}
and, for a fixed \(\mu>0\),
\begin{equation}
    x^+
    =
    \frac{3}{2\mu}(t+2\chi),
    \qquad
    x^-
    =
    \frac{\mu R^2}{12}(t-2\chi).
    \label{eq:Penrose-light-cone-coordinates-draft4}
\end{equation}
The inverse transformation is
\begin{equation}
    t
    =
    \frac{\mu}{3}x^+
    +\frac{6}{\mu R^2}x^-,
    \qquad
    \chi
    =
    \frac{\mu}{6}x^+
    -\frac{3}{\mu R^2}x^-.
    \label{eq:Penrose-light-cone-inverse-draft4}
\end{equation}
Taking \(R\to\infty\) at fixed \(x^\pm,r,z\) gives the
maximally supersymmetric eleven-dimensional pp-wave
\cite{Blau:2002dy},
\begin{equation}
    \begin{aligned}
        ds_{11}^2={}&-2dx^+dx^-+d\boldsymbol z_3^2+d\boldsymbol z_6^2-
        \left[
            \left(\frac{\mu}{3}\right)^2\boldsymbol z_3^2
            +
            \left(\frac{\mu}{6}\right)^2\boldsymbol z_6^2
        \right](dx^+)^2,
        \\
        G_4={}&
        \mu\,dx^+\wedge dz^1\wedge dz^2\wedge dz^3,
    \end{aligned}
    \label{eq:maximal-pp-wave-draft4}
\end{equation}
where
\begin{equation}
    d\boldsymbol z_3^2=dr^2+r^2d\Omega_2^2,
    \qquad
    d\boldsymbol z_6^2=dz^2+z^2d\Omega_5^2.
\end{equation}
The orientation of the parent four-form is chosen so that
\(G_{+123}=\mu\).

To determine the global identification, follow the orbifold
action through the same change of coordinates.  Its generator
acts on the covering sphere as
\begin{equation}
    (Z_1,Z_2,Z_3,Z_4)
    \mapsto
    e^{2\pi i/k}(Z_1,Z_2,Z_3,Z_4),
    \label{eq:ABJM-orbifold-action-draft4}
\end{equation}
In \eqref{eq:S7-adapted-coordinates-draft4}, this sends
\(\chi\mapsto\chi+2\pi/k\) and
\(\boldsymbol u\mapsto e^{2\pi i/k}\boldsymbol u\).
Equation \eqref{eq:Penrose-light-cone-coordinates-draft4}
therefore gives
\begin{equation}
    x^+
    \mapsto
    x^++\frac{6\pi}{\mu k},
    \qquad
    x^-
    \mapsto
    x^--\frac{\pi\mu R^2}{3k},
    \label{eq:orbifold-light-cone-shifts-draft4}
\end{equation}
together with a transverse rotation through \(2\pi/k\).
A finite, nonzero translation along \(x^-\) requires
\(R^2/k\) to approach a positive constant.  Thus \(k\)
grows with \(R^2\), and both the \(x^+\) shift and the
transverse rotation vanish.  The surviving identification is
\begin{equation}
    x^-\sim x^-+2\pi R_-,
    \qquad
    R_-=\frac{\mu R^2}{6k},
    \label{eq:null-radius-draft4}
\end{equation}
where we have used the inverse generator to choose the
positive period.  The flux-radius relation in
\eqref{eq:ABJM-background-draft4} gives
\begin{equation}
    R_-
    =
    \frac{\mu\ell_p^2}{6}
    \left(
        \frac{32\pi^2N}{k^2}
    \right)^{1/3}.
    \label{eq:null-radius-scaling-draft4}
\end{equation}
At fixed \(\mu,\ell_p\), retaining a finite null radius
therefore requires the \(N,k\) scaling in
\eqref{eq:joint-Penrose-scaling-draft4}.  The limiting
spacetime is \eqref{eq:maximal-pp-wave-draft4} with the
compact identification \eqref{eq:null-radius-draft4}.

The parent curvature and Hopf-fibre scales behave differently:
\(R/\ell_p\sim k^{1/2}\), whereas
\(R/(k\ell_p)\sim k^{-1/2}\).
Consequently, the smooth limit of the classical fields
does not imply a uniformly valid eleven-dimensional
supergravity approximation on the full parent quotient.
At the level of the classical fields, the double scale limit takes the form
\begin{equation}
    \frac{\mathrm{AdS}_4\times S^7}{\mathbb Z_k}
    \xrightarrow[\;N/k^2\to\nu\;]{\substack{
        N,k\to\infty\\
        \text{local Penrose contraction}
    }}
    \frac{\text{maximally SUSY 11d pp-wave}}
         {x^-\sim x^-+2\pi R_-}.
    \label{eq:ABJM-to-compact-pp-wave-draft4}
\end{equation}

The null radius also fixes the BMN coupling.  In the
normalization of appendix~\ref{app:BMN-index-review}, the
bosonic kinetic term is
\((2g_{\mathrm B}^2)^{-1}\operatorname{Tr}(D_+X^I)^2\).
With membrane tension
\(T_{\mathrm{M2}}=[(2\pi)^2\ell_p^3]^{-1}\), the
matrix-theory relation \cite{Kovacs:2013una} yields
\begin{equation}
    g_{\mathrm B}^2
    =(2\pi T_{\mathrm{M2}})^2R_-^3
    =\frac{R_-^3}{(2\pi)^2\ell_p^6},
    \qquad
    \frac{g_{\mathrm B}^2}{\mu^3}=\frac{\nu}{27}.
    \label{eq:BMN-coupling-nu-draft4}
\end{equation}
The dimensionless coupling is finite and need not be small.

\subsection{Light-cone charges and protected fugacities}
\label{subsec:light-cone-index-dictionary-draft4}

The coordinate transformation also determines the energy
and momentum of the states retained by this compactification.
To express them in ABJM variables, let
\begin{equation}
    E=i\partial_t,
    \qquad
    J_4=-i\partial_\chi
    \label{eq:energy-J4-draft4}
\end{equation}
denote the global AdS energy and the angular momentum along
the geodesic.  The inverse coordinate transformation
\eqref{eq:Penrose-light-cone-inverse-draft4} gives
\begin{equation}
    p^-\equiv i\partial_{x^+}=\frac{\mu}{6}(2E-J_4),
    \qquad
    p^+\equiv i\partial_{x^-}=\frac{3}{\mu R^2}(2E+J_4).
    \label{eq:light-cone-momenta-draft4}
\end{equation}
and hence the dimensionless BMN Hamiltonian
\begin{equation}
    H_{\mathrm{BMN}}
    \equiv
    \frac{6}{\mu}p^-
    =
    2E-J_4.
    \label{eq:BMN-Hamiltonian-draft4}
\end{equation}
For the four complex coordinates, write
\begin{equation}
    Z_A=|Z_A|e^{i\phi_A},
    \qquad
    J_A=-i\partial_{\phi_A},
    \qquad
    J_{\mathrm H}=\sum_{A=1}^{4}J_A.
    \label{eq:geometric-angular-momenta-draft4}
\end{equation}
with \(\phi_4=\chi\).  Invariance under
\eqref{eq:ABJM-orbifold-action-draft4} imposes
\begin{equation}
    J_{\mathrm H}=kq,
    \qquad
    q\in\mathbb Z.
    \label{eq:Hopf-GNO-relation-draft4}
\end{equation}
The ABJM monopole dictionary identifies this integer with
the common total GNO charge,
\begin{equation}
    \boldsymbol n=(n_1,\ldots,n_N),
    \qquad
    \widetilde{\boldsymbol n}
    =(\widetilde n_1,\ldots,\widetilde n_N),
    \qquad
    q
    =
    \sum_{i=1}^{N}n_i
    =
    \sum_{i=1}^{N}\widetilde n_i.
    \label{eq:total-GNO-charge-draft4}
\end{equation}
We take \(q>0\), without restricting the signs of the
individual fluxes.  Writing \(J_\perp=J_1+J_2+J_3\),
the charge constraint and \eqref{eq:BMN-Hamiltonian-draft4}
imply
\begin{equation}
    J_4=kq-J_\perp,
    \qquad
    E
    =
    \frac{kq}{2}
    +\frac12\left(H_{\mathrm{BMN}}-J_\perp\right).
    \label{eq:energy-charge-dictionary-draft4}
\end{equation}
Using \eqref{eq:null-radius-draft4}, the longitudinal
momentum becomes
\begin{equation}
    p^+
    =
    \frac{q}{R_-}
    -\frac{J_\perp}{kR_-}
    +\frac{H_{\mathrm{BMN}}}{2kR_-}.
    \label{eq:pplus-fixed-q-draft4}
\end{equation}
For fixed \(q\) and bounded \(H_{\mathrm{BMN}},J_a\),
\(a=1,2,3\), this gives
\begin{equation}
    p^+R_-\longrightarrow q.
    \label{eq:q-DLCQ-momentum-draft4}
\end{equation}
This gives the momentum assignment required by the
rank-\(q\) BMN description.  To apply it to the index,
we must identify the protected gradings and establish
that fixing them bounds the light-cone energy.

Use the \(SU(4)_R\) scalar basis of
appendix~\ref{app:ABJM-index-review},
\begin{equation}
    Y^A
    =
    (A_1,A_2,B^{\dagger\dot 1},B^{\dagger\dot 2}),
    \qquad
    A=1,\ldots,4.
    \label{eq:ABJM-scalar-basis-draft4}
\end{equation}
and identify the angular charges of \(Z_A\) with those of
\(Y^A\).  The chosen geodesic then lies along
\(Y^4=B^{\dagger\dot2}\).  The ABJM Cartans are
\begin{equation}
    \begin{aligned}
        h_1&=\frac12(J_1-J_2-J_3+J_4),
        &
        h_2&=\frac12(J_1-J_2+J_3-J_4),
        \\
        h_3&=\frac12(-J_1-J_2+J_3+J_4),
        &
        h_4&=\frac12(J_1+J_2+J_3+J_4).
    \end{aligned}
    \label{eq:ABJM-geometric-Cartans-draft4}
\end{equation}
In particular, the total-charge constraint gives
\begin{equation}
    h_4=\frac{J_{\mathrm H}}{2}=\frac{kq}{2}.
    \label{eq:h4-fixed-q-draft4}
\end{equation}
Choose \(\mathcal Q=Q_{34,-}\) and its radial adjoint
\(\mathcal S=S^{34,-}\), as in the ABJM superconformal
index \cite{Bhattacharya:2008zy,Kim:2009wb}.  Their positive
anticommutator is
\begin{equation}
    \Delta_{\mathcal Q}
    \equiv
    \{\mathcal Q,\mathcal S\}
    =
    E-h_3-j_3
    \geq0.
    \label{eq:ABJM-positive-operator-draft4}
\end{equation}
With the transverse BMN Cartans chosen as
\begin{equation}
    M^{12}=j_3,
    \qquad
    M^{45}=-J_1,
    \qquad
    M^{67}=-J_2,
    \qquad
    M^{89}=J_3.
    \label{eq:BMN-Cartans-draft4}
\end{equation}
the corresponding BMN operator satisfies
\begin{equation}
    \begin{aligned}
        \Delta_{\mathrm{BMN}}
        &\equiv
        H_{\mathrm{BMN}}
        -2M^{12}-M^{45}-M^{67}-M^{89}
        \\
        &=
        H_{\mathrm{BMN}}-2j_3+J_1+J_2-J_3
        =
        2\Delta_{\mathcal Q}.
    \end{aligned}
    \label{eq:positive-operator-dictionary-draft4}
\end{equation}
The two BPS conditions therefore agree.  The three
commuting transverse gradings are
\begin{equation}
    \ell_1\equiv j_3-J_1,
    \qquad
    \ell_2\equiv j_3-J_2,
    \qquad
    \ell_3\equiv j_3+J_3,
    \label{eq:transverse-gradings-draft4}
\end{equation}
which coincide with the BMN refinements
\eqref{eq:BMN-refinements-review-draft4}.

At fixed \(\ell_a\), the BPS states also satisfy the
finite-energy conditions used above.  Spatial and
R-symmetry rotations of
\eqref{eq:ABJM-positive-operator-draft4} give
\(E-h_3+j_3\geq0\) and \(E-h_1-j_3\geq0\)
\cite{Kim:2009wb}.  On the BPS locus \(E=h_3+j_3\),
these imply \(2j_3\geq0\) and
\(h_3-h_1=\ell_1+\ell_3-2j_3\geq0\).  Hence
\begin{equation}
    0\leq 2j_3\leq\ell_1+\ell_3,
    \qquad
    H_{\mathrm{BMN}}=\ell_1+\ell_2+\ell_3-j_3.
    \label{eq:finite-degree-BPS-energy-draft4}
\end{equation}
Fixed transverse gradings thus bound \(j_3\), all three
\(J_a\), and \(H_{\mathrm{BMN}}\), ensuring that
\eqref{eq:q-DLCQ-momentum-draft4} applies to the states
whose index coefficients we compare.

Having identified the protected charges and the states
retained by the limit, we can compare their fugacity
weights.  The ABJM index is
\begin{equation}
    \mathcal I^{\mathrm{ABJM}}_{N,k}(x,y_1,y_2,y_3)
    =
    \operatorname{Tr}_{\mathcal H_{\mathrm{ABJM}}}
    \left[
        (-1)^F
        e^{-\beta\Delta_{\mathcal Q}}
        x^{E+j_3}
        y_1^{h_1}y_2^{h_2}y_3^{h_4}
    \right].
    \label{eq:ABJM-index-draft4}
\end{equation}
Using the transverse fugacities \(u_a=e^{-\Delta_a}\)
introduced above and a longitudinal chemical potential
\(\sigma\), set
\begin{equation}
    \begin{aligned}
        x&=e^{-\sigma/2}\sqrt{u_1u_2u_3},
        &
        y_1&=e^{-\sigma/2}\sqrt{\frac{u_2}{u_1u_3}},
        \\
        y_2&=e^{\sigma/2}\sqrt{\frac{u_2u_3}{u_1}},
        &
        y_3&=e^{-\sigma/2}\sqrt{\frac{u_3}{u_1u_2}},
    \end{aligned}
    \label{eq:Penrose-fugacity-map-draft4}
\end{equation}
with \(u_a^{1/2}=e^{-\Delta_a/2}\).
The Cartan map \eqref{eq:ABJM-geometric-Cartans-draft4}
then gives the exact identity
\begin{equation}
    \begin{aligned}
        x^{E+j_3}y_1^{h_1}y_2^{h_2}y_3^{h_4}
        ={}&
        \exp\left[
            -\frac12(\sigma+\Delta_1+\Delta_2+\Delta_3)
            \Delta_{\mathcal Q}
        \right]
        e^{-\sigma(J_4+j_3)}
        \\
        &\times
        e^{-\Delta_1(j_3-J_1)
           -\Delta_2(j_3-J_2)
           -\Delta_3(j_3+J_3)}.
    \end{aligned}
    \label{eq:exact-index-weight-draft4}
\end{equation}
On contributing states, \(\Delta_{\mathcal Q}=0\), so
the first exponential is unity.  Moreover,
\(J_4+j_3=kq+\ell_1+\ell_2-\ell_3\).
Keeping the longitudinal weight finite therefore requires
\begin{equation}
    \sigma=\frac{\delta}{k},
    \qquad
    \operatorname{Re}\delta>0,
    \label{eq:longitudinal-fugacity-scaling-draft4}
\end{equation}
and the BPS insertion becomes
\begin{equation}
    e^{-\delta q}
    e^{-\left(\Delta_1+\frac{\delta}{k}\right)\ell_1}
    e^{-\left(\Delta_2+\frac{\delta}{k}\right)\ell_2}
    e^{-\left(\Delta_3-\frac{\delta}{k}\right)\ell_3}.
    \label{eq:fixed-q-Penrose-weight-draft4}
\end{equation}
The \(O(1/k)\) shifts vanish at each fixed transverse
degree, leaving the BMN refinements and the momentum
weight \(e^{-\delta q}\).

Let \(\mathcal I^{\mathrm{ABJM}}_{N,k;q}
(\delta;\boldsymbol\Delta)\) be the total-charge-\(q\)
component of \eqref{eq:ABJM-index-draft4}, evaluated with
\eqref{eq:Penrose-fugacity-map-draft4} and
\(\sigma=\delta/k\).  For \(q\geq0\), its physical
expansion after gauge projection is
\begin{equation}
    \mathcal I^{\mathrm{ABJM}}_{N,k;q}
    =
    e^{-\delta q}
    \sum_{\boldsymbol L\in\mathbb Z_{\geq0}^3}
    c_{N,k;q}(\boldsymbol L)\,
    e^{-\frac{\delta}{k}(L_1+L_2-L_3)}
    u_1^{L_1}u_2^{L_2}u_3^{L_3},
    \label{eq:physical-transverse-expansion-draft4}
\end{equation}
Here \(\boldsymbol L=(L_1,L_2,L_3)\) gives the values of
the transverse gradings \(\ell_a\), and
\(c_{N,k;q}(\boldsymbol L)\) is the signed multiplicity at
those charges.  The support \(L_a\geq0\) is a property of
the gauge-projected index and follows directly from the
localization expansion, as shown after
\eqref{eq:selection-monomial-degree-decomposition-draft4}.
Individual letters before projection need not carry integral
transverse gradings.
For a family of generating functions \(F_{N,k}\), we define
the double scale limit coefficientwise by
\begin{equation}
    \lim_{\substack{N,k\to\infty\\ N/k^2\to\nu}}
    [u_1^{L_1}u_2^{L_2}u_3^{L_3}]
    F_{N,k}
    =
    [u_1^{L_1}u_2^{L_2}u_3^{L_3}]F,
    \label{eq:coefficient-extraction-limit-draft4}
\end{equation}
for every fixed \(\boldsymbol L\), with
\(\delta,q,\nu\) held fixed.  This prescription makes no
interchange between the limit and the infinite fugacity sum.

Convergence of the indices as functions of the fugacities
is a stronger question.  Suppose that a \(k\)-independent
\(\epsilon>0\) satisfies
\begin{equation}
    \operatorname{Re}\left(\Delta_1+\frac{\delta}{k}\right)\geq\epsilon,
    \qquad
    \operatorname{Re}\left(\Delta_2+\frac{\delta}{k}\right)\geq\epsilon,
    \qquad
    \operatorname{Re}\left(\Delta_3-\frac{\delta}{k}\right)\geq\epsilon
    \label{eq:safe-Penrose-chamber-draft4}
\end{equation}
for all sufficiently large \(k\).  After factoring out
\(e^{-\delta q}\), the fugacity weight multiplying
\(c_{N,k;q}(\boldsymbol L)\) in
\eqref{eq:physical-transverse-expansion-draft4} has absolute
value at most \(e^{-\epsilon(L_1+L_2+L_3)}\).
The full sum also depends on the growth of
\(\lvert c_{N,k;q}(\boldsymbol L)\rvert\) with transverse
degree and on its \(N,k\) dependence.  Controlling this
growth sufficiently to establish a limit of functions is
left to future work; the results below are coefficientwise.

Let \(\mathcal I_q^{\mathrm{BMN}}\) denote the sum over
all rank-\(q\) vacua appearing in
\eqref{eq:intro-index-relation-draft4}, with conventions
reviewed in appendix~\ref{app:BMN-index-review}
\cite{Chang:2024lkw}.  The charge and fugacity dictionary
suggests the following tentative fixed-\(q\) component of
the grand canonical ensemble of BMN rank:
\begin{equation}
    \lim_{\substack{N,k\to\infty\\N/k^2\to\nu}}^{\mathrm{coeff}}
    \mathcal I^{\mathrm{ABJM}}_{N,k;q}(\delta;\boldsymbol\Delta)
    \overset{?}{=}
    e^{-\delta q}\,
    \mathcal I_q^{\mathrm{BMN}}(\boldsymbol\Delta),
    \qquad q>0.
    \label{eq:naive-fixed-q-index-comparison-draft4}
\end{equation}
Here \(e^{-\delta q}\) is the grand canonical weight, with
\(\delta\) conjugate to the longitudinal charge \(q\).
The charge map leaves the protected multiplicities undetermined.
Section~\ref{sec:double-scaling-index-draft4} will derive
the precise relation from localization: an additional
neutral contribution survives the limit and factorizes
from the BMN index, supplying the correction to
\eqref{eq:naive-fixed-q-index-comparison-draft4}.

\section{Half-BPS geometries in the double scale limit}
\label{sec:half-BPS-geometric-limit-draft4}

The double scale limit described in
section~\ref{sec:penrose-index-dictionary}
determines the compact pp-wave background and the protected
charge dictionary.  We now apply it to backreacted
half-BPS solutions in the eleven-dimensional LLM class
\cite{Lin:2004nb}.  The axisymmetric electrostatic description
of Donos and Sim\'on \cite{Donos:2010va} makes it possible
to follow their flux-carrying cycles as the asymptotic AdS
radius grows.  We show how the diagonal Hopf quotients of
these solutions approach the eleven-dimensional uplift of the
Lin--Maldacena backgrounds associated with BMN vacua
\cite{Lin:2005nh}.

The parameters retain the conventions of
section~\ref{subsec:compact-pp-wave-draft4}: we take
\(N,k\to\infty\) with \(N/k^2\to\nu\in(0,\infty)\)
and fixed \(q\).  One conducting disk carries the growing
background flux,
while the other disks retain finite rescaled data.  The
large disk becomes a conducting plane, and its induced
charge supplies the images required by the LM boundary
problem.  We first determine the parent geometry and its
charges, then implement the double scale limit in electrostatic
variables to obtain the normalized fields and quantized vacuum data.

The construction concerns classical solutions; the validity
of the supergravity approximation depends separately on the
chosen fluxes.  The corresponding index relation
\eqref{eq:intro-index-relation-draft4} is established
independently from finite-\(N\), finite-\(k\) localization
in section~\ref{sec:double-scaling-index-draft4}.

\subsection{Parent geometries and their electrostatic data}
\label{subsec:parent-electrostatic-data-draft4}

The Donos--Sim\'on construction \cite{Donos:2010va}
describes the covering half-BPS geometries through an
axisymmetric electrostatic potential sourced by a
semi-infinite line charge and conducting disks.  The disk
charges and separations encode the quantized fluxes, while
regularity is imposed through electrostatic boundary
conditions.  To obtain the parent backgrounds relevant to
ABJM theory, we must also specify the global circle
identification and its effect on the physical charges.

We first present the covering metric and three-form,
together with the boundary conditions, M2 and M5 fluxes,
and asymptotic charges.  We then construct the diagonal
Hopf \(\mathbb Z_k\) quotient, establish its
\(\mathrm{AdS}_4\times S^7/\mathbb Z_k\) asymptotics,
and determine the circle period and charge normalization.
The resulting relations between the electrostatic data,
the ABJM rank \(N\), and the total monopole charge \(q\)
specify the fixed-charge families used in
section~\ref{subsec:conducting-plane-limit-draft4} to
implement the double scale limit in electrostatic variables.

\subsubsection{The Donos--Sim\'on construction on the covering space}
\label{subsubsec:Donos-covering-geometry-draft5}

The axisymmetric LLM solutions of Donos and Sim\'on
\cite{Lin:2004nb,Donos:2010va} have local
\(\mathbb R\times SO(3)\times SO(6)\times U(1)\) symmetry.
Their eleven-dimensional coordinates consist of an
electrostatic time \(\tau\), an angular coordinate
\(\beta\), the auxiliary half-plane \(\rho\geq0\),
\(\eta\in\mathbb R\), and unit two- and five-spheres.
The global AdS time \(t\) used in
section~\ref{subsec:compact-pp-wave-draft4} is related to
these coordinates by \(t=2\tau+\beta\).
The solution is determined by an axisymmetric potential
\(V(\rho,\eta)\) satisfying
\begin{equation}
    \frac1\rho\partial_\rho(\rho\partial_\rho V)
    +\partial_\eta^2V=0,
    \qquad
    \dot V\equiv\rho\partial_\rho V,
    \qquad V'\equiv\partial_\eta V,
    \label{eq:parent-Laplace-conventions-draft4}
\end{equation}
away from the sources.  Repeated dots and primes denote
repeated application of these derivatives.  The coordinates
\(\rho,\eta\) and the potential \(V\) have dimensions of
length cubed; \(\tau,\beta\) and the sphere angles
are dimensionless.

Define
\begin{equation}
    \mathcal D_V=(\dot V')^2-\ddot V V'',
    \qquad
    \Delta_V=(2\dot V-\ddot V)V''+(\dot V')^2,
    \qquad
    v=v_\beta d\beta,
    \quad v_\beta=1-\frac{\dot V'}{\mathcal D_V}.
    \label{eq:parent-derivative-combinations-draft5}
\end{equation}
The covering-space metric is
\begin{equation}
    \begin{aligned}
        ds_{11,\mathrm{cov}}^2
        ={}&\left(-\frac{\dot V\Delta_V}{2V''}\right)^{1/3}
        \Bigg[
            4d\Omega_5^2-
            \frac{2V''\dot V}{\Delta_V}d\Omega_2^2
        \\
        &\quad-\frac{2V''}{\dot V}
            \left(d\rho^2+d\eta^2+
                  \frac{\rho^2}{\mathcal D_V}d\beta^2\right)
            -\frac{4\mathcal D_V}{\Delta_V}
                  (d\tau+v)^2
        \Bigg].
    \end{aligned}
    \label{eq:parent-covering-metric-draft5}
\end{equation}
Let \(R_2\) and \(R_5\) denote the proper radii of the
two- and five-sphere orbits.  Their coefficients in
\eqref{eq:parent-covering-metric-draft5} give the LLM
coordinate \cite{Lin:2004nb,Donos:2010va}
\begin{equation}
    y_{\mathrm{LLM}}=\rho\partial_\rho V=\dot V
                   =\frac14 R_2R_5^2.
    \label{eq:parent-LLM-radius-product-draft6}
\end{equation}
Here \(y_{\mathrm{LLM}}\) is the noncompact LLM coordinate.
In a smooth geometry, the sphere radii and therefore
\(y_{\mathrm{LLM}}\) must be continuous.
In the interior, we require \(\dot V>0\), \(V''<0\),
\(\mathcal D_V>0\) and \(\Delta_V>0\).
Where a sphere or circle shrinks, we impose the smoothness
conditions described below.

We normalize the two-sphere volume form by
\(\int_{S^2}\omega_2=4\pi\).
With the four-form normalization of
\eqref{eq:ABJM-background-draft4}, a local three-form
potential in the coordinates \((\tau,\beta)\) is
\begin{equation}
    C_3^{(11)}
    =\left[
        \frac{4\dot V^{\,2}V''}{\Delta_V}(d\tau+v)
        +2\left(\frac{\dot V\dot V'}{\mathcal D_V}-\eta\right)
          d\beta
      \right]\wedge\omega_2,
    \qquad G_4=dC_3^{(11)}.
    \label{eq:LM-parent-three-form-normalization-draft4}
\end{equation}
The covering-space identification in this conformal frame is
\begin{equation}
    (\tau,\beta)
    \sim(\tau-2\pi,\beta+4\pi).
    \label{eq:parent-covering-identifications-draft4}
\end{equation}
It keeps the global time \(t=2\tau+\beta\) fixed.
In particular, \(\beta\) is not independently identified
at fixed \(\tau\).

For \(\mathrm{AdS}_4\times S^7\) asymptotics, the
electrostatic background is a semi-infinite line charge,
\begin{equation}
    \left.\dot V\right|_{\rho=0}
    =2\eta\,\Theta(\eta),
    \qquad
    V_b(\rho,\eta)
    =\sqrt{\rho^2+\eta^2}
     +\eta\log\frac{\sqrt{\rho^2+\eta^2}-\eta}{\ell_p^3}.
    \label{eq:parent-line-background-draft4}
\end{equation}
The reference length in the logarithm fixes the affine
freedom \(V\mapsto V+c_0+c_1\eta\), which leaves the local
fields unchanged.  Number the conducting disks from the one
closest to the line endpoint:
\begin{equation}
    \mathcal D_i:\quad
    \eta=-d_i,\quad 0\leq\rho\leq a_i,
    \qquad
    0\leq d_0<d_1<\cdots<d_{n_{\mathrm d}}.
    \label{eq:parent-disk-configuration-draft4}
\end{equation}
The line endpoint fixes the origin of \(\eta\).  To see
why regularity requires \(d_i\geq0\), suppose a disk intersected
the charged axis at \(\eta=\eta_*>0\).  Approaching the
intersection along the disk, the equipotential condition
would give \(\dot V=0\); approaching along the charged
axis, \eqref{eq:parent-line-background-draft4} would instead
give \(\dot V\to2\eta_*>0\).  By
\eqref{eq:parent-LLM-radius-product-draft6}, these would be
different limits of the same geometric quantity
\(R_2R_5^2/4\).  This product would have no unique
continuous limit at the putative intersection, excluding
a smooth completion of the metric.
Consequently,
\begin{equation}
    \eta_i=-d_i\leq0,\qquad d_i\geq0.
    \label{eq:parent-disk-position-regularity-draft6}
\end{equation}
At the line endpoint \(\eta=0\), both limits vanish, so
contact there is compatible with this condition.
Such contact is not required:
\(d_0>0\) leaves a gap between the line endpoint and the
first disk and is allowed by regularity \cite{Donos:2010va}.
The signed disk densities and electrostatic charges are
\begin{equation}
    \sigma_i(\rho)
    =\partial_\eta V(\rho,-d_i+0)
     -\partial_\eta V(\rho,-d_i-0),
    \qquad
    Q_i=2\pi\int_0^{a_i}\rho\,\sigma_i(\rho)\,d\rho.
    \label{eq:parent-density-charge-draft4}
\end{equation}
Writing \(V=V_b+\sum_{i=0}^{n_{\mathrm d}}V_{d_i}\),
the disk potentials are
\begin{equation}
    V_{d_i}(\rho,\eta)
    =-\frac1{4\pi}\int_0^{a_i}\rho'\,d\rho'
      \int_0^{2\pi}d\varphi'\,
      \frac{\sigma_i(\rho')}{
      \sqrt{\rho^2+\rho^{\prime 2}-2\rho\rho'\cos\varphi'
                  +(\eta+d_i)^2}}.
    \label{eq:parent-disk-potential-draft4}
\end{equation}
Each conductor is an equipotential, and smoothness removes
the edge singularity:
\begin{equation}
    V(\rho,-d_i)=V_{0i}\quad(0\leq\rho\leq a_i),
    \qquad \sigma_i(a_i)=0.
    \label{eq:parent-conductor-boundary-conditions-draft4}
\end{equation}
These conditions apply to the total potential.  The disk
densities and potentials are determined jointly, and the
radii, charges and potentials are not independent data.
We retain the regular solutions with Lorentzian signature
described in \cite{Donos:2010va}.

The configuration with a single conducting disk at the
line-charge endpoint \(\eta=0\) gives global
\(\mathrm{AdS}_4\times S^7\) \cite{Donos:2010va}.

For a finite disk distribution, the monopole and dipole
coefficients are \cite{Donos:2010va}
\begin{equation}
    F=\frac1{4\pi}\sum_{i=0}^{n_{\mathrm d}}Q_i,
    \qquad
    P=-\frac1{4\pi}\sum_{i=0}^{n_{\mathrm d}}d_iQ_i.
    \label{eq:parent-dipole-charge-draft6}
\end{equation}
The far-field potential then has the expansion
\begin{equation}
    \begin{gathered}
        \mathcal R=\sqrt{\rho^2+\eta^2},\qquad
        V=V_b-\frac{F}{\mathcal R}
             -\frac{P\eta}{\mathcal R^3}
             +O(\mathcal R^{-3}),
        \\
        F=-\frac{R^6}{512}.
    \end{gathered}
    \label{eq:parent-far-field-potential-draft5}
\end{equation}
The last relation fixes the asymptotic radius.  In the
asymptotic global coordinates of \cite{Donos:2010va},
\(\varrho=\sinh\rho_{\mathrm{AdS}}\) is the dimensionless
AdS radial coordinate, \(\theta\) is the seven-sphere
polar angle, and \(\phi_4=-\tau\).
Their radius \(L_4\) equals \(R/2\) in our conventions.
The first dipole correction gives
\begin{equation}
    \begin{aligned}
        ds_{11,\mathrm{cov}}^2\simeq{}&
        \frac{R^2}{4}\Bigg[
            \left(1-\frac{2^{14}P(1+2\cos2\theta)}{R^9\varrho}\right)
            \varrho^2(-dt^2+d\Omega_2^2)
        \\
        &\qquad+
            \left(1+\frac{2^{13}P(1+2\cos2\theta)}{R^9\varrho}\right)
            \frac{d\varrho^2}{\varrho^2}
        \Bigg]
        \\
        &+R^2\left(1+\frac{2^{13}P(1+2\cos2\theta)}{R^9\varrho}\right)
        \Bigg[d\theta^2+\sin^2\theta\,d\Omega_5^2
        \\
        &\hspace{3.5cm}
            +\cos^2\theta\left(
                d\phi_4+\frac{3\cdot2^{12}P}{R^9\varrho}\,dt
            \right)^2\Bigg].
    \end{aligned}
    \label{eq:parent-far-field-metric-data-draft5}
\end{equation}
Here the displayed coefficients are retained through
\(1/\varrho\); omitted diagonal corrections are of relative
order \(\varrho^{-2}\), and the omitted terms in the angular
connection are \(O(\varrho^{-2})\).  The leading metric
is therefore \(\mathrm{AdS}_4\times S^7\), with radii
\(R/2\) and \(R\), and the four-form approaches the value
in \eqref{eq:ABJM-background-draft4}.  Higher multipoles
contribute only at subleading orders.

The fluxes and asymptotic charges can now be read from these
electrostatic data \cite{Donos:2010va}.  Orient the cycles
so that negative \(Q_i\) correspond to positive M2 numbers.
The disk fluxes and total M2 flux on the covering space are
\begin{equation}
    \begin{gathered}
        N_{2,\mathrm{cov}}^{(i)}
        =-\frac{4Q_i}{\pi^3\ell_p^6}\in\mathbb Z_{>0},
        \\
        N_{\mathrm{cov}}
        =\sum_{i=0}^{n_{\mathrm d}}N_{2,\mathrm{cov}}^{(i)}
        =-\frac4{\pi^3\ell_p^6}\sum_iQ_i
        =\frac{R^6}{32\pi^2\ell_p^6}.
    \end{gathered}
    \label{eq:parent-covering-M2-flux-draft6}
\end{equation}
The M5 periods measure the gap from the line endpoint to
the first disk and the subsequent separations:
\begin{equation}
    \begin{aligned}
        N_{5,\mathrm{cov}}^{(0)}
        &=\frac{4d_0}{\pi\ell_p^3}\in\mathbb Z_{\geq0},
        \\
        N_{5,\mathrm{cov}}^{(i)}
        &=\frac{4(d_i-d_{i-1})}{\pi\ell_p^3}
          \in\mathbb Z_{>0},
        \qquad i=1,\ldots,n_{\mathrm d}.
    \end{aligned}
    \label{eq:parent-covering-M5-periods-draft4}
\end{equation}
When \(d_0=0\), the first cycle collapses and its period
vanishes.  A nonzero gap \(d_0>0\) instead carries the
M5 flux \(N_{5,\mathrm{cov}}^{(0)}\).

In the analysis of Donos and Sim\'on \cite{Donos:2010va},
the R-charge is determined from the asymptotic mass.
In our conventions, its nonnegative magnitude
\(J_{\mathrm{cov}}\) satisfies
\begin{equation}
    J_{\mathrm{cov}}=\frac{64P}{\pi^3\ell_p^9}
    =\sum_{i=0}^{n_{\mathrm d}}N_{2,\mathrm{cov}}^{(i)}
      \sum_{j=0}^{i}N_{5,\mathrm{cov}}^{(j)},
    \qquad
    E_{\mathrm{cov}}=\frac12J_{\mathrm{cov}}.
    \label{eq:parent-covering-R-charge-draft6}
\end{equation}
Here \(E_{\mathrm{cov}}\) is the dimensionless global AdS
energy above the vacuum.  The last equality is the half-BPS
relation.  The \(SO(6)\)-invariant solution carries no
angular momentum on the five-sphere, so its nonzero angular
charge lies entirely in the remaining \(U(1)\).

\subsubsection{The Hopf \texorpdfstring{\(\mathbb Z_k\)}{Zk} quotient}
\label{subsubsec:parent-Hopf-quotient-draft5}

To implement the diagonal Hopf quotient of the covering
solution, write the five-sphere as
\begin{equation}
    d\Omega_5^2
    =ds_{\mathbb{CP}^2}^2+(d\psi+\mathcal A)^2,
    \qquad \psi\sim\psi+2\pi,
    \qquad d\mathcal A=2J_{\mathbb{CP}^2}.
    \label{eq:parent-S5-Hopf-fibration-draft4}
\end{equation}
On the asymptotic seven-sphere, choose
\begin{equation}
    (Z_1,Z_2,Z_3)=\sin\theta\,\boldsymbol u,
    \qquad Z_4=\cos\theta\,e^{i\phi_4},
    \qquad \boldsymbol u\in S^5\subset\mathbb C^3,
    \label{eq:parent-asymptotic-complex-coordinates-draft5}
\end{equation}
where \(\psi\) is the common phase of \(\boldsymbol u\).
The diagonal cyclic quotient
\eqref{eq:ABJM-orbifold-action-draft4} is generated by shifting both
\(\phi_4\) and \(\psi\) by \(2\pi/k\) at fixed \(t\).
Using \(\phi_4=-\tau\) and \(t=2\tau+\beta\), its
generator on the covering geometry is
\begin{equation}
    \xi_{\mathrm H}
    =2\partial_\beta-\partial_\tau+\partial_\psi.
    \label{eq:parent-Hopf-generator-draft4}
\end{equation}
The local fields are invariant under this Killing vector.
Its flow through \(2\pi\) is a covering-space identification,
so the quotient is generated by the order-\(k\) action
\begin{equation}
    (\tau,\beta,\psi)
    \mapsto
    \left(\tau-\frac{2\pi}{k},
          \beta+\frac{4\pi}{k},
          \psi+\frac{2\pi}{k}\right).
    \label{eq:parent-discrete-Hopf-action-draft5}
\end{equation}
In particular, the shift of \(\tau\) does not identify
global AdS time: \(t=2\tau+\beta\) is unchanged.

Introduce coordinates adapted to the compact action,
\begin{equation}
    t=2\tau+\beta,
    \qquad \vartheta_{\mathrm H}=\frac\beta2,
    \qquad \varphi=\psi-\frac\beta2.
    \label{eq:parent-Hopf-adapted-coordinates-draft4}
\end{equation}
On the cover, \(t\) is noncompact and both
\(\vartheta_{\mathrm H}\) and \(\varphi\) have period
\(2\pi\).  The generator is
\(\xi_{\mathrm H}=\partial_{\vartheta_{\mathrm H}}|_{t,\varphi}\).
On the quotient we use
\begin{equation}
    \vartheta_{\mathrm H}\sim\vartheta_{\mathrm H}+\frac{2\pi}{k},
    \qquad y=k\vartheta_{\mathrm H},
    \qquad y\sim y+2\pi.
    \label{eq:parent-quotient-circle-draft4}
\end{equation}
In these coordinates, the quotient metric is
\begin{equation}
    \begin{aligned}
        ds_{11,k}^2
        ={}&\left(-\frac{\dot V\Delta_V}{2V''}\right)^{1/3}
        \Bigg[
            4\left\{ds_{\mathbb{CP}^2}^2+
              \left(d\varphi+\mathcal A+\frac{dy}{k}\right)^2\right\}
            -\frac{2V''\dot V}{\Delta_V}d\Omega_2^2
        \\
        &\quad-\frac{2V''}{\dot V}
              \left(d\rho^2+d\eta^2+
                    \frac{4\rho^2}{k^2\mathcal D_V}dy^2\right)
            -\frac{4\mathcal D_V}{\Delta_V}
              \left(\frac12dt+\frac{2v_\beta-1}{k}dy\right)^2
        \Bigg].
    \end{aligned}
    \label{eq:parent-quotient-metric-draft5}
\end{equation}
Using \(d\tau=\tfrac12dt-\tfrac1kdy\),
\(d\beta=\tfrac2kdy\), and
\(\Delta_V=\mathcal D_V+2\dot V V''\), the potential
\eqref{eq:LM-parent-three-form-normalization-draft4} becomes
\begin{equation}
    C_3^{(11)}=\left[
        \frac{2\dot V^{\,2}V''}{\Delta_V}\,dt
        +\frac4k\left(
            \frac{\dot V^{\,2}V''+\dot V\dot V'}{\Delta_V}
            -\eta\right)dy
        \right]\wedge\omega_2.
    \label{eq:parent-quotient-three-form-draft6}
\end{equation}
Both fields are determined by the same potential
\(V(\rho,\eta)\) and electrostatic boundary conditions
as on the cover.

Asymptotically,
\begin{equation}
    \phi_4=\frac yk-\frac t2,\qquad
    \psi=\varphi+\frac yk.
    \label{eq:parent-quotient-asymptotic-angles-draft5}
\end{equation}
The period of \(y\) acts on
\eqref{eq:parent-asymptotic-complex-coordinates-draft5} as
\(Z_A\mapsto e^{2\pi i/k}Z_A\), while leaving the AdS
coordinates fixed.  Applying this identification to the
leading metric in \eqref{eq:parent-far-field-metric-data-draft5},
with the flux normalization \eqref{eq:ABJM-background-draft4}, gives
\begin{equation}
    ds_{11,k}^2\sim
    \frac{R^2}{4}\,ds_{\mathrm{AdS}_4}^2
    +R^2ds_{S^7/\mathbb Z_k}^2,
    \qquad
    G_4\sim\frac{3R^3}{8}\,\omega_{\mathrm{AdS}_4}.
    \label{eq:parent-quotient-AdS-asymptotics-draft5}
\end{equation}
The omitted terms vanish relative to the leading fields in
an asymptotic orthonormal frame.  The identification is the
diagonal Hopf quotient of the asymptotic seven-sphere, on
which it acts freely.
At finite \(k\), the continuous global symmetries inherited
from the cover commute with the discrete Hopf action.

Away from the line-charge endpoint, freeness follows from
the degeneration structure of the regular covering solution
\cite{Donos:2010va}.  On the charged axis
\(\rho=0,\eta>0\), \(v_\beta\to\tfrac12\), so the
circle generated at fixed global time by
\(2\partial_\beta-\partial_\tau\) collapses, while
\(S^5\) remains finite.  The generator
\eqref{eq:parent-Hopf-generator-draft4} then reduces to
the free Hopf action \(\partial_\psi\).
On the uncharged axis \(\rho=0,\eta<0\), including
the regular disk centers, \(S^5\) collapses while the
complementary circle survives.  On the disk faces and
rims at \(\rho>0\), the five-sphere remains finite.
The surviving circle actions have primitive period
\(2\pi\), by \eqref{eq:parent-covering-identifications-draft4}
and \eqref{eq:parent-S5-Hopf-fibration-draft4}; hence no
rotation through \(2\pi j/k\), \(1\leq j<k\), fixes a
point on these strata.  We treat the line-charge endpoint
below, after imposing the fixed-\(q\) condition.

The quotient geometry determines the normalization of its
charges.  On an invariant seven-cycle the covering map has
degree \(k\), so the M2 flux carried by a disk is
\begin{equation}
    N_2^{(i)}=\frac{N_{2,\mathrm{cov}}^{(i)}}{k}
             =-\frac{4Q_i}{k\pi^3\ell_p^6}
             \in\mathbb Z_{>0}.
    \label{eq:parent-M2-flux-quotient-draft4}
\end{equation}
The total quotient flux is the ABJM color rank:
\begin{equation}
    \begin{gathered}
        N=\frac{N_{\mathrm{cov}}}{k}
         =\sum_{i=0}^{n_{\mathrm d}}N_2^{(i)},
        \\
        \sum_{i=0}^{n_{\mathrm d}}Q_i
        =-\frac{k\pi^3\ell_p^6}{4}N
        =-\frac{\pi R^6}{128}.
    \end{gathered}
    \label{eq:parent-total-charge-draft4}
\end{equation}
Thus \(R^6=32\pi^2Nk\ell_p^6\), as in
\eqref{eq:ABJM-background-draft4}.

At fixed \(y\), the reference orbit has
\(\phi_4=-t/2\), opposite to the angular orientation in
\eqref{eq:Penrose-null-geodesic-draft4}.  We identify the
ABJM complex coordinates with the complex conjugates of
the \(Z_A\) used here.  This reverses the Hopf generator
without changing its cyclic quotient, and fixes the
positive signs of \(J_{\mathrm H}\) and \(q\) in
\eqref{eq:parent-Hopf-charge-draft6}.
The quotient reduces each asymptotic charge integral by
\(1/k\).  In addition, the normalized circle coordinate
\(y=k\vartheta_{\mathrm H}\) gives the momentum
\(q=J_{\mathrm H}/k\).  In the positive-momentum orientation,
\begin{equation}
    \begin{gathered}
        J_{\mathrm H}=\frac{J_{\mathrm{cov}}}{k},
        \qquad E=\frac{E_{\mathrm{cov}}}{k}
               =\frac{J_{\mathrm H}}2=\frac{kq}{2},
        \\
        q=\frac{J_{\mathrm H}}{k}
         =\frac{64P}{k^2\pi^3\ell_p^9}
         =-\frac{16}{k^2\pi^4\ell_p^9}
           \sum_{i=0}^{n_{\mathrm d}}d_iQ_i.
    \end{gathered}
    \label{eq:parent-Hopf-charge-draft6}
\end{equation}
Every disk with \(d_i>0\), including the first disk when
\(d_0>0\), contributes to \(q\).  A disk at \(d_i=0\)
contributes only to \(N\).  With \(Q_i<0\) and
\(d_i\geq0\), the momentum \(q\) is nonnegative.

To relate the geometric charges to fixed-monopole-charge
sectors, define the scaled cumulative heights
\begin{equation}
    s_i=\frac{4d_i}{k\pi\ell_p^3}
       =\frac1k\sum_{j=0}^{i}N_{5,\mathrm{cov}}^{(j)},
    \qquad i=0,\ldots,n_{\mathrm d}.
    \label{eq:parent-scaled-height-labels-draft6}
\end{equation}
For \(i>0\), integrating the field strength of
\eqref{eq:parent-quotient-three-form-draft6} over the
four-cycle formed by \(S^2\), a path between adjacent
disks and one winding around the quotient circle gives the
M5 charge \(n_5^{(i)}\).
We keep the following integers fixed:
\begin{equation}
    \begin{gathered}
        n_5^{(i)}=\frac{N_{5,\mathrm{cov}}^{(i)}}{k},
        \qquad i=0,\ldots,n_{\mathrm d},
        \\
        n_5^{(0)}=s_0\in\mathbb Z_{\geq0},
        \qquad
        n_5^{(i)}=s_i-s_{i-1}\in\mathbb Z_{>0}
        \quad(i>0).
    \end{gathered}
    \label{eq:plane-limit-M5-scaling-draft4}
\end{equation}
The quotient momentum can then be written as
\begin{equation}
    q=\sum_{i=0}^{n_{\mathrm d}}s_iN_2^{(i)}.
    \label{eq:parent-rank-momentum-draft6}
\end{equation}
The partition \(\lambda\vdash q\) has \(N_2^{(i)}\) parts of
size \(s_i\) for each \(s_i>0\), in the convention of the
matched GNO sectors \eqref{eq:selection-surviving-fluxes-draft4}.
The flux calculation in
section~\ref{subsec:LM-fields-vacua-draft4} will identify
these same data with the BMN vacuum labels.

We now specialize to the double scale limit
\eqref{eq:intro-double-scaling-draft4} at fixed \(q\).
Since the integer
heights obey \(s_i\geq s_0\geq0\),
\eqref{eq:parent-total-charge-draft4} and
\eqref{eq:parent-rank-momentum-draft6} imply
\begin{equation}
    q=\sum_{i=0}^{n_{\mathrm d}}s_iN_2^{(i)}
      \geq s_0N,
    \qquad
    N>q\ \Longrightarrow\ s_0=0\ \Longrightarrow\ d_0=0.
    \label{eq:parent-fixed-charge-endpoint-draft6}
\end{equation}
The first disk therefore lies at the line-charge endpoint
when \(N>q\).  This follows from the fixed-\(q\) condition
and the integer heights; regularity alone only requires
\(d_0\geq0\).
Since \(s_i\geq1\) for \(i>0\), the endpoint disk carries
\(N_2^{(0)}\geq N-q>0\) and has nonzero radius.

The charged and uncharged axes end at the centers of
opposite faces of this disk.  These are different
spacetime points, although both have electrostatic
coordinates \((\rho,\eta)=(0,0)\), as in the single-disk
\(\mathrm{AdS}_4\times S^7\) solution \cite{Donos:2010va}.
At the charged-axis endpoint, the circle generated by
\(2\partial_\beta-\partial_\tau\) shrinks and \(S^5\)
remains finite.  At the other endpoint, \(S^5\) shrinks
and the complementary circle remains finite.
The other disks contribute smooth fields near these
points and preserve this pattern in the regular covering
geometry.  Hence the two circle factors in
\(\xi_{\mathrm H}\) never shrink at the same spacetime
point.  Each surviving circle has minimum period
\(2\pi\), so no nontrivial element of \(\mathbb Z_k\)
fixes a point.  The quotient action is therefore free
for the fixed-\(q\) families with \(N>q\).

For the families of interest, \(n_{\mathrm d}\) and
\(N_2^{(i)}\), \(i>0\), remain fixed, whereas the endpoint
disk carries the growing flux
\(N_2^{(0)}=N-\sum_{i>0}N_2^{(i)}\).
We select the hierarchy
\begin{equation}
    \frac{a_i}{a_0}\longrightarrow0,
    \qquad \frac{d_i}{a_0}\longrightarrow0,
    \qquad i>0,
    \qquad \frac{N}{k^2}\longrightarrow\nu.
    \label{eq:parent-large-disk-hierarchy-draft4}
\end{equation}
The families studied below obey this hierarchy.
At finite \(N,k\), the endpoint radius and density are
determined by the full conductor problem, including the
response to the other disks.  This response must therefore
be retained when taking the limit.

\subsection{The double scale limit in electrostatic variables}
\label{subsec:conducting-plane-limit-draft4}

We study how the electrostatic potential of the
Donos--Sim\'on \(\mathbb Z_k\) quotient changes in the
double scale limit \eqref{eq:intro-double-scaling-draft4}.
The monopole charge \(q\) and the Planck length \(\ell_p\)
remain fixed.  For the disk hierarchy
\eqref{eq:parent-large-disk-hierarchy-draft4}, the endpoint
disk becomes a conducting plane, while the other disks remain
as finite sources in the LM potential.  To keep their
positions and sizes finite, introduce the rescaled coordinates
\begin{equation}
    \Lambda=\frac{k\ell_p^3}{2},\qquad
    \rho=\Lambda r,\qquad \eta=-\Lambda z,\qquad
    b_i=\frac{a_i}{\Lambda},\qquad z_i=\frac{d_i}{\Lambda}.
    \label{eq:plane-limit-coordinates-draft4}
\end{equation}
The endpoint disk lies at \(z=0\), and the other disks
lie in \(z>0\).  The quantized heights and separations in
\eqref{eq:parent-scaled-height-labels-draft6} and
\eqref{eq:plane-limit-M5-scaling-draft4} become
\begin{equation}
    z_i=\frac\pi2s_i,\qquad
    z_i-z_{i-1}=\frac\pi2n_5^{(i)},\qquad z_0=0.
    \label{eq:plane-limit-quantized-heights-draft4}
\end{equation}
The finite disks therefore stay at fixed heights as \(k\)
grows.  We take \(b_0\to\infty\), with \(b_i\), \(i>0\),
approaching finite positive limits and their M2 integers
\(N_2^{(i)}\) fixed.  We rescale their surface charge
densities as
\begin{equation}
    \Sigma_i(r)=4b_0^2\sigma_i(\Lambda r),\qquad i>0.
    \label{eq:plane-limit-rescaled-densities-draft4}
\end{equation}
Before expanding the potential, we determine the leading
size and potential of the endpoint disk.  We compare it
with the single-disk vacuum of the same radius \(a_0\),
whose disk potential and charge are \(\pi a_0/2\) and
\(-2\pi a_0^2\), respectively \cite{Donos:2010va}.
We denote this reference potential, including the line
charge, by \(V_{\mathrm{ref}}\).

Each finite disk has charge
\(Q_i=-k\pi^3\ell_p^6N_2^{(i)}/4\), which grows
linearly with \(k\).  The charge it induces on the large
disk has the same order, since the source radius and height
are small compared with \(a_0\).  The total charge instead
grows as \(kN\).  The relative change in the endpoint
charge is therefore of order \(1/N\), or \(1/k^2\).
Its leading relation \(Q_0\sim-2\pi a_0^2\) then gives
\(a_0\propto k^{3/2}\) and \(b_0^2\propto k\).

Smoothness at the rim fixes the disk potential.  The
regular disk solution \cite{Donos:2010va} bounds the shift
from its reference value by a constant times
\(\Lambda\sum_{i>0}|Q_i|/a_0^2\).  Since \(\Lambda\)
and the finite-disk charges grow linearly with \(k\), this
shift is of order \(1/k\).  For the actual endpoint disk,
the two estimates give
\begin{equation}
    V_{00}=\frac{\pi a_0}{2}
           +O\!\left(\frac1k\right),
    \qquad
    Q_0=-2\pi a_0^2
        \left[1+O\!\left(\frac1{k^2}\right)\right].
    \label{eq:plane-limit-response-bounds-draft4}
\end{equation}
The charge estimate has a relative error of order \(1/k^2\).
Combining it with the total flux
\eqref{eq:parent-total-charge-draft4} fixes the radius:
\begin{equation}
    b_0^2=\frac{\pi^2N}{2k}
          \left[1+O\!\left(\frac1{k^2}\right)\right],
    \qquad a_0\sim\frac{R^3}{16}.
    \label{eq:plane-limit-background-radius-draft4}
\end{equation}
With this scale determined, the reference potential has the
following expansion at fixed \((r,z)\):
\begin{equation}
    V_{\mathrm{ref}}(\Lambda r,-\Lambda z)
    =\frac{\pi a_0}{2}-\Lambda z\log\frac{2a_0}{\ell_p^3}
    +\frac{\Lambda}{4b_0^2}\left(r^2z-\frac23z^3\right)
    +O\!\left(\frac1k\right).
    \label{eq:plane-limit-reference-expansion-draft4}
\end{equation}
The expansion also holds for derivatives on fixed compact
sets away from the rim.  The constant and linear terms leave
the local supergravity fields unchanged.  We subtract these
terms from the full potential \(V\) and normalize the cubic
term to define
\begin{equation}
    W(r,z)=\frac{4b_0^2}{\Lambda}
    \left[
        V(\Lambda r,-\Lambda z)-\frac{\pi a_0}{2}
        +\Lambda z\log\frac{2a_0}{\ell_p^3}
    \right].
    \label{eq:plane-limit-potential-draft4}
\end{equation}
We use \(W\) for both the rescaled potential and its limit.
The reference solution contributes the cubic background
\(W_{\mathrm{bg}}=r^2z-\tfrac23z^3\), with corrections
of order \(1/k\).  On the actual endpoint disk,
\eqref{eq:plane-limit-response-bounds-draft4} gives
\(W(r,0)=4b_0^2(V_{00}-\pi a_0/2)/\Lambda=O(1/k)\).
Since \(b_0\to\infty\), every fixed \(r\) eventually lies
on the disk.  Its limit is therefore a conducting plane
with \(W(r,0)=0\).

The same radius relation fixes the normalization of the
potential and the finite-disk charges.  Using
\(N/k^2\to\nu\), we obtain
\begin{equation}
    \begin{aligned}
        \frac{\Lambda}{4b_0^2}
        &\longrightarrow\frac{\ell_p^3}{4\pi^2\nu},
        \\
        2\pi\int_0^{b_i}r\,\Sigma_i(r)\,dr
        &=\frac{4b_0^2}{\Lambda^2}Q_i
         =-4\pi^3\frac{b_0^2}{k}N_2^{(i)}
         \longrightarrow\mathcal Q_i
         =-2\pi^5\nu N_2^{(i)}.
    \end{aligned}
    \label{eq:plane-limit-finite-charges-draft4}
\end{equation}
These limits fix the normalization of the fields and fluxes
derived in section~\ref{subsec:LM-fields-vacua-draft4}.

The induced charge on the endpoint disk also contributes to
\(W\).  A finite source and its induced charge are described
by the Dirichlet Green function for the conducting disk.
As the disk grows, this Green function approaches that of
the half-space.  Each source at height \(z_i\) then has
an opposite image at \(-z_i\), which enforces the
zero-potential boundary condition.

To control this limit, subtract the two Green functions.
Their source singularities cancel, so the difference is
harmonic.  On the part of \(z=0\) outside the disk, its
boundary values are \(O(1/\sqrt{k})\).  The maximum
principle and interior estimates give the same bound for
the difference and its derivatives on fixed compact sets
away from the sources and boundary.  The remaining disk
potential is \(O(1/k)\) and vanishes as well.  Combining
the cubic background with the finite sources and their
images gives
\begin{equation}
    \begin{aligned}
        W(r,z)&=r^2z-\frac23z^3
        -\frac1{4\pi}\sum_{i>0}
          \int_0^{b_i}r'\Sigma_i(r')\,dr'
          \int_0^{2\pi}d\varphi'
        \\
        &\quad\times\left[
          \frac1{\sqrt{r^2+r^{\prime 2}-2rr'\cos\varphi'+(z-z_i)^2}}
          -\frac1{\sqrt{r^2+r^{\prime 2}-2rr'\cos\varphi'+(z+z_i)^2}}
        \right].
    \end{aligned}
    \label{eq:plane-limit-LM-potential-draft4}
\end{equation}
The second kernel is the image contribution from the
endpoint disk.  The conductor boundary conditions become
\begin{equation}
    W(r,0)=0,\qquad
    \partial_rW(r,z_i)=0\quad(0<r<b_i),\qquad
    \Sigma_i(b_i)=0.
    \label{eq:plane-limit-boundary-problem-draft4}
\end{equation}
The first condition fixes the plane potential; the other two
make each finite disk an equipotential with a smooth rim.
Here \(\Sigma_i\) is the jump of \(\partial_zW\) in the
direction of increasing \(z\), consistently with
\(\eta=-\Lambda z\) and
\eqref{eq:plane-limit-rescaled-densities-draft4}.
Together with the fixed charges
\eqref{eq:plane-limit-finite-charges-draft4}, these conditions
determine the finite radii and densities.  They reproduce the
BMN branch of the Lin--Maldacena electrostatic problem
\cite{Lin:2005nh}.  We next use this potential and its
normalization to derive the limiting fields, circle period
and vacuum fluxes.

\subsection{The Lin--Maldacena fields and vacuum data}
\label{subsec:LM-fields-vacua-draft4}

We now study how the Donos--Sim\'on \(\mathbb Z_k\)
quotient geometry changes in the double scale limit.
Using the potential scaling derived in
section~\ref{subsec:conducting-plane-limit-draft4}, we obtain
the limiting metric, three-form and circle period.
Reduction along the circle then gives the LM fields, whose
quantized fluxes determine the BMN vacuum data.

For the compact direction, introduce at finite \(k\) the
rescaled circle coordinate
\(\zeta=(8b_0^2/k)y\).  The radius estimate
\eqref{eq:plane-limit-background-radius-draft4} gives
\(b_0^2/k\to\pi^2\nu/2\), so in the double scale limit
\begin{equation}
    \zeta=4\pi^2\nu\,y,
    \qquad
    \zeta\sim\zeta+8\pi^3\nu.
    \label{eq:LM-circle-scale-draft4}
\end{equation}
We use \(\zeta\) for both the finite-\(k\) coordinate and
its limit.  The potential normalization is fixed by
\eqref{eq:plane-limit-finite-charges-draft4}.

With dots denoting \(r\partial_r\) and primes \(\partial_z\), define
\begin{equation}
    \mathcal D_W=(\dot W')^2-\ddot W W'',
    \qquad
    \Delta_W=(\ddot W-2\dot W)W''-(\dot W')^2.
    \label{eq:LM-derivative-combinations-draft4}
\end{equation}
Here \(\Delta_W\) uses the LM sign convention,
which is opposite to the corresponding combination in the
parent metric \cite{Donos:2010va,Lin:2005nh}.

At finite \(k\), the potential definition
\eqref{eq:plane-limit-potential-draft4} and the parent
combinations \eqref{eq:parent-derivative-combinations-draft5}
give the exact identities
\begin{equation}
    \begin{gathered}
        \dot V=\frac{\Lambda}{4b_0^2}\dot W,\qquad
        \ddot V=\frac{\Lambda}{4b_0^2}\ddot W,
        \\
        V''=\frac{1}{4b_0^2\Lambda}W'',\qquad
        \dot V'=-\frac{1}{4b_0^2}\dot W',
        \\
        \mathcal D_V=\frac{\mathcal D_W}{16b_0^4},
        \qquad
        \Delta_V=-\frac{\Delta_W}{16b_0^4}.
    \end{gathered}
    \label{eq:LM-scaled-derivatives-draft4}
\end{equation}
The minus sign in the mixed derivative follows from
\(\eta=-\Lambda z\).
The quotient coordinates
\eqref{eq:parent-Hopf-adapted-coordinates-draft4} and
\eqref{eq:parent-quotient-circle-draft4} give
\begin{equation}
    \begin{aligned}
        d\psi+\mathcal A
        &=d\varphi+\mathcal A+\frac{1}{8b_0^2}\,d\zeta,
        \\
        d\tau+v
        &=\frac12dt+
          \left(\frac{\dot W'}{\mathcal D_W}
                         +\frac{1}{8b_0^2}\right)d\zeta.
    \end{aligned}
    \label{eq:LM-vanishing-angular-twists-draft4}
\end{equation}
Substituting into the parent metric
\eqref{eq:parent-covering-metric-draft5} and taking the
double scale limit gives
\begin{equation}
    \begin{aligned}
        ds_{11,\mathrm{lim}}^2
        ={}&\frac{\ell_p^2}{(4\pi^2\nu)^{2/3}}
        \left(\frac{\dot W\Delta_W}{2W''}\right)^{1/3}
        \Bigg[
            4d\Omega_5^2+
            \frac{2W''\dot W}{\Delta_W}d\Omega_2^2
        \\
        &\quad-\frac{2W''}{\dot W}
        \left(dr^2+dz^2+
                     \frac{r^2}{\mathcal D_W}d\zeta^2\right)
        +\frac{4\mathcal D_W}{\Delta_W}
        \left(\frac12dt+\frac{\dot W'}{\mathcal D_W}d\zeta\right)^2
        \Bigg].
    \end{aligned}
    \label{eq:LM-limiting-eleven-metric-draft4}
\end{equation}
Here \(d\Omega_5^2=ds_{\mathbb{CP}^2}^2+
(d\varphi+\mathcal A)^2\) has its standard angular periods.
The vanishing twist in
\eqref{eq:LM-vanishing-angular-twists-draft4} restores
the round five-sphere without an additional quotient on it.

The same limit of the quotient potential
\eqref{eq:parent-quotient-three-form-draft6} gives
\begin{equation}
    C_{3,\mathrm{lim}}^{(11)}
    =\frac{\ell_p^3}{4\pi^2\nu}
    \left[
        -\frac{2\dot W^{\,2}W''}{\Delta_W}\,dt
        +2\left(z+\frac{\dot W\dot W'}{\Delta_W}\right)d\zeta
    \right]\wedge\omega_2,
    \qquad
    G_{4,\mathrm{lim}}=dC_{3,\mathrm{lim}}^{(11)}.
    \label{eq:LM-limiting-three-form-draft4}
\end{equation}
The metric and three-form above hold on each nonsingular
coordinate patch.  The conductor boundary conditions
complete the fields smoothly at the sphere-degeneration
loci.

The identification with LM is most transparent after
completing the square in the time--circle sector.
Using \(\ddot W+r^2W''=0\), one finds
\begin{equation}
    \begin{aligned}
        &-\frac{2W''}{\dot W}
          \frac{r^2}{\mathcal D_W}d\zeta^2
        +\frac{4\mathcal D_W}{\Delta_W}
          \left(\frac12dt+\frac{\dot W'}{\mathcal D_W}d\zeta\right)^2
        \\
        &\qquad=
        -\frac{\ddot W}{\ddot W-2\dot W}dt^2
        -\frac{2(\ddot W-2\dot W)}{\dot W\Delta_W}
          \left(d\zeta-\frac{\dot W\dot W'}{\ddot W-2\dot W}dt\right)^2 .
    \end{aligned}
    \label{eq:LM-KK-completed-square-draft4}
\end{equation}
We reduce along the \(2\pi\)-periodic coordinate
\(y=\zeta/(4\pi^2\nu)\).
Writing the ten-dimensional metric and potentials in
string units, our reduction convention is
\begin{equation}
    \begin{aligned}
        \frac{ds_{11}^2}{\ell_p^2}
        &=e^{-2\Phi/3}\,d\widehat s_{10,\mathrm{str}}^2
          +e^{4\Phi/3}(dy+\widehat C_1)^2,
        \\
        \frac{C_3^{(11)}}{\ell_p^3}
        &=\widehat C_3+\widehat B_2\wedge dy .
    \end{aligned}
    \label{eq:LM-KK-units-draft4}
\end{equation}
Here \(d\widehat s_{10,\mathrm{str}}^2=ds_{10,\mathrm{str}}^2/
\alpha'\), \(\widehat B_2=B_2/\alpha'\), and
\(\widehat C_p=C_p/(\alpha')^{p/2}\).
Equations \eqref{eq:LM-limiting-eleven-metric-draft4}--%
\eqref{eq:LM-KK-units-draft4} then give the fully normalized
string-frame fields
\begin{equation}
    \begin{aligned}
        d\widehat s_{10,\mathrm{str}}^2
        ={}&\left(\frac{\ddot W-2\dot W}{-W''}\right)^{1/2}
        \Bigg[
            -\frac{\ddot W}{\ddot W-2\dot W}dt^2
            -\frac{2W''}{\dot W}(dr^2+dz^2)
        \\
        &\hspace{3.1cm}
            +4d\Omega_5^2+
            \frac{2W''\dot W}{\Delta_W}d\Omega_2^2
        \Bigg],
        \\
        e^{4\Phi}
        ={}&(4\pi^2\nu)^4
            \frac{4(\ddot W-2\dot W)^3}{(-W'')\dot W^{\,2}\Delta_W^2},
        \qquad
        \widehat C_1
        =-\frac{\dot W\dot W'}{4\pi^2\nu(\ddot W-2\dot W)}\,dt,
        \\
        \widehat C_3
        ={}&-\frac{\dot W^{\,2}W''}{2\pi^2\nu\Delta_W}
                   dt\wedge\omega_2,
        \qquad
        \widehat B_2
        =2\left(z+\frac{\dot W\dot W'}{\Delta_W}\right)\omega_2 .
    \end{aligned}
    \label{eq:LM-normalized-IIA-fields-draft4}
\end{equation}
These are the LM fields, written with global-time
normalization \(t\), and with electrostatic potential
\begin{equation}
    \mathcal V=\frac{W}{4\pi^2\nu},
    \qquad
    \mathcal V_{\mathrm{bg}}
    =\frac1{4\pi^2\nu}\left(r^2z-\frac23z^3\right).
    \label{eq:LM-potential-normalization-draft4}
\end{equation}
Indeed, multiplying the electrostatic potential by a
constant leaves the string-frame metric and \(B_2\)
unchanged, while rescaling the RR potentials and the
dilaton as in \eqref{eq:LM-normalized-IIA-fields-draft4}
\cite{Lin:2005nh}.  The factor \(1/(4\pi^2\nu)\) in
\eqref{eq:LM-potential-normalization-draft4} is thus fixed
by the inherited circle period.

The quantized fluxes of these LM fields determine the
BMN vacuum data.
Let \(\mathcal C_3^{(i)}\) be the three-cycle
obtained by fibering \(S^2\) over a path between consecutive
conductors.  At its endpoints
\(\widehat B_2=2z_i\omega_2\), up to the gauge shifts
needed to use regular potentials on the shrinking spheres.
Consequently,
\begin{equation}
    n_{\mathrm{NS5}}^{(i)}
    =\frac1{4\pi^2}\int_{\mathcal C_3^{(i)}}d\widehat B_2
    =\frac{2(z_i-z_{i-1})}{\pi}
    =\frac{N_{5,\mathrm{cov}}^{(i)}}{k}
    =n_5^{(i)}.
    \label{eq:LM-NS5-flux-draft4}
\end{equation}
This is the M5 charge \(n_5^{(i)}\) in
\eqref{eq:plane-limit-M5-scaling-draft4}, expressed as an
NS5 charge after reduction along the limiting \(y\)-circle.

The LM electrostatic charge is conventionally defined as
the coefficient of the positive Coulomb kernel.  In terms
of our signed surface charge it is
\begin{equation}
    Q_i^{\mathrm{LM}}
    =-\frac{\mathcal Q_i}{16\pi^3\nu}
    =\frac{\pi^2}{8}N_2^{(i)},
    \qquad
    n_{\mathrm{D2}}^{(i)}
    =\frac{8Q_i^{\mathrm{LM}}}{\pi^2}=N_2^{(i)}.
    \label{eq:LM-D2-charge-draft4}
\end{equation}
The last equality uses the LM flux quantization
\cite{Lin:2005nh}.  In particular, the normalization
in \eqref{eq:LM-potential-normalization-draft4} is also
required to retain the parent finite M2 integers as the
D2 vacuum labels.

The height of a disk above the plane, rather than its
separation from the preceding disk, determines the
dimension of the associated irreducible \(SU(2)\)
representation.  In the notation of
appendix~\ref{app:BMN-index-review},
\begin{equation}
    s_i=\frac{2z_i}{\pi}
       =\sum_{j=1}^i n_{\mathrm{NS5}}^{(j)}
       =\sum_{j=1}^i n_5^{(j)}.
    \label{eq:LM-BMN-vacuum-dictionary-draft4}
\end{equation}
Together with \eqref{eq:LM-D2-charge-draft4}, this identifies
\(N_2^{(i)}\) as the multiplicity of the
\(s_i\)-dimensional representation.
The partition of \(q\) in
\eqref{eq:parent-rank-momentum-draft6} therefore specifies
the BMN vacuum decomposition
\eqref{eq:BMN-vacuum-decomposition-review-draft4}.
With the plane fixing the origin of height and the
corresponding \(B_2\) gauge, \(q\) is the total D0 charge,
equivalently the longitudinal momentum in units of
\(1/R_-\) \cite{Lin:2005nh}.
The parent Hopf momentum, the limiting D0 charge and the
BMN rank are therefore the same integer \(q\).
The growing parent rank \(N\) sets the background plane,
while the finite disk data determine the BMN vacuum.
The following section derives the corresponding partition
labels and the full protected-index relation directly from
the ABJM localization formula.

\section{The double scale limit of the ABJM index}
\label{sec:double-scaling-index-draft4}

We determine the correction to
\eqref{eq:naive-fixed-q-index-comparison-draft4} by controlling
the transverse expansion of the ABJM localization formula at finite
\(N\) and \(k\) \cite{Kim:2009wb}.  Fix the total monopole charge
\(q>0\), the longitudinal chemical potential \(\delta\), and a monomial
\(u_1^{L_1}u_2^{L_2}u_3^{L_3}\), with
\(\boldsymbol L\in\mathbb Z_{\geq0}^3\) and finite total degree
\(L=L_1+L_2+L_3\).  We first derive identities for its coefficient,
valid once \(N\) and \(k\) satisfy bounds depending on \(q\) and \(L\).
The same bounds will control every coefficient of total degree at most
\(L\).  Only after establishing these finite-parameter statements do
we take \(N,k\to\infty\) with \(N/k^2\to\nu\in(0,\infty)\).

To keep the finite-\(k\) dependence explicit, we absorb the shifts in
\eqref{eq:fixed-q-Penrose-weight-draft4} into
\begin{equation}
    \widehat u_1=e^{-\sigma}u_1,
    \qquad
    \widehat u_2=e^{-\sigma}u_2,
    \qquad
    \widehat u_3=e^{\sigma}u_3,
    \qquad
    \sigma=\frac{\delta}{k},
    \label{eq:localization-shifted-fugacities-draft4}
\end{equation}
and define
\begin{equation}
    \widehat{\mathfrak u}
    \equiv\widehat u_1\widehat u_2\widehat u_3
    =x^2=e^{-\sigma}\mathfrak u,
    \qquad
    \mathfrak u=u_1u_2u_3.
    \label{eq:localization-x-product-draft4}
\end{equation}
The fixed-charge insertion is then
\(e^{-\delta q}\widehat u_1^{\ell_1}
\widehat u_2^{\ell_2}\widehat u_3^{\ell_3}\).
We assign degree one to each \(\widehat u_a\), so that
\(\widehat{\mathfrak u}\) has total degree three.  Converting a
coefficient of degree \(\boldsymbol L\) from \(\widehat u_a\) to
\(u_a\) multiplies it by \(e^{-\sigma(L_1+L_2-L_3)}\), which tends
to unity in the double scale limit at fixed \(\delta\) and
\(\boldsymbol L\).  It therefore suffices to control
the expansion in the shifted fugacities.

The Chern--Simons term assigns electric gauge charge to a monopole,
which must be canceled by charged matter to obtain a gauge-invariant
state.  We use the scalar \(Y^4=B^{\dagger\dot 2}\) selected by the
geodesic as a reference for this matter dressing.  Its weight in the
\(y_3\)-independent function \(f^+\) of
\eqref{eq:ABJM-matter-letters-review-draft4} is
\begin{equation}
    A\equiv\sqrt{\frac{xy_1}{y_2}}
    =\left(\frac{\widehat u_1\widehat u_2}{\widehat u_3}\right)^{1/4},
    \qquad
    y_3^{1/2}A=e^{-\sigma}.
    \label{eq:localization-background-letter-draft4}
\end{equation}
All fractional powers are fixed by the chemical potentials in
\eqref{eq:Penrose-fugacity-map-draft4}.  The reference weight for
\(kq\) units of this gauge charge combines with the topological
fugacity as
\begin{equation}
    y_3^{kq/2}A^{kq}=e^{-\delta q}.
    \label{eq:localization-background-weight-draft4}
\end{equation}
The holonomy constraints will fix the net matter charge to \(kq\),
as shown below.  Separating the factor \(A\) from the two matter
functions gives
\begin{equation}
    \begin{aligned}
        g^+(\widehat{\mathbf u})\equiv A^{-1}f^+
        &=
        \frac{1+\widehat u_3-\widehat u_1\widehat u_3
              -\widehat u_2\widehat u_3}
             {1-\widehat{\mathfrak u}},
        \\
        g^-(\widehat{\mathbf u})\equiv Af^-
        &=
        \frac{\widehat u_1+\widehat u_2-\widehat u_1\widehat u_2
              -\widehat{\mathfrak u}}
             {1-\widehat{\mathfrak u}}.
    \end{aligned}
    \label{eq:selection-relative-letter-functions-draft4}
\end{equation}
Here \(\widehat{\mathbf u}=(\widehat u_1,\widehat u_2,\widehat u_3)\).
The labels \(+\) and \(-\) distinguish opposite bifundamental gauge
charges.  The constant term of \(g^+\) represents the reference scalar,
while \(g^-\) starts at degree one.  The common denominator sums
BPS-preserving derivatives, each of weight \(\widehat{\mathfrak u}\).
A matter mode joining eigenvalues with fluxes \(n_i\) and
\(\widetilde n_j\) has the additional weight
\(x^{|n_i-\widetilde n_j|}\), determined by its lowest angular
momentum in the monopole background.  Thus its contribution is
\begin{equation}
    f^\pm_{ij}
    =\widehat{\mathfrak u}^{\,|n_i-\widetilde n_j|/2}
      A^{\pm1}g^\pm(\widehat{\mathbf u}).
    \label{eq:localization-fluxed-matter-letters-draft4}
\end{equation}
The vector contributions have the corresponding flux-difference weights,
\begin{equation}
    \begin{aligned}
        f^{\mathrm{adj}}_{ij}
        &=-\left(1-\delta_{n_i n_j}\right)
          \widehat{\mathfrak u}^{\,|n_i-n_j|/2},
        \\
        \widetilde f^{\mathrm{adj}}_{ij}
        &=-\left(1-\delta_{\widetilde n_i\widetilde n_j}\right)
          \widehat{\mathfrak u}^{\,|\widetilde n_i-\widetilde n_j|/2}.
    \end{aligned}
    \label{eq:localization-vector-letters-draft4}
\end{equation}
Vector factors joining eigenvalues with equal flux are already included
in the normalized residual Haar measures
\eqref{eq:ABJM-holonomy-measure-review-draft4}.  With the
zero-point exponent \(E_0\) defined in
\eqref{eq:ABJM-zero-point-energy-review-draft4}, the localization
formula \eqref{eq:ABJM-localization-formula-review-draft4} at fixed
total charge becomes
\begin{equation}
    \begin{aligned}
        \mathcal I^{\mathrm{ABJM}}_{N,k;q}
         (\delta;\boldsymbol\Delta)
        ={}&e^{-\delta q}A^{-kq}
        \sum_{\substack{
            \boldsymbol n,\widetilde{\boldsymbol n}\in\mathbb Z^N/S_N\\
            \sum_i n_i=\sum_i\widetilde n_i=q}}
        \widehat{\mathfrak u}^{\,E_0/2}
        \int[d\boldsymbol\alpha]_{\boldsymbol n}
             [d\widetilde{\boldsymbol\alpha}]_{\widetilde{\boldsymbol n}}
        \\
        &\quad\times
        \exp\left[
            ik\sum_{i=1}^N
            \left(n_i\alpha_i-\widetilde n_i\widetilde\alpha_i\right)
        \right]
        \exp\left[
            \sum_{m=1}^{\infty}\frac{\mathcal F_m}{m}
        \right],
    \end{aligned}
    \label{eq:localization-fixed-charge-formula-draft4}
\end{equation}
The exponential \(\exp[\sum_{m\geq1}\mathcal F_m/m]\) is the
plethystic exponential, which assembles products of the single-letter
modes with their bosonic or fermionic statistics.  Writing
\(g_m^\pm=g^\pm(\widehat u_1^m,\widehat u_2^m,\widehat u_3^m)\),
its kernel is
\begin{equation}
    \begin{aligned}
        \mathcal F_m={}&
        \sum_{i,j=1}^N
        \widehat{\mathfrak u}^{\,m|n_i-\widetilde n_j|/2}
        \left[
            A^m g_m^+e^{im(\widetilde\alpha_j-\alpha_i)}
            +A^{-m}g_m^-e^{im(\alpha_i-\widetilde\alpha_j)}
        \right]
        \\
        &-\sum_{\substack{i,j=1\\n_i\neq n_j}}^N
        \widehat{\mathfrak u}^{\,m|n_i-n_j|/2}
        e^{-im(\alpha_i-\alpha_j)}
        -\sum_{\substack{i,j=1\\\widetilde n_i\neq\widetilde n_j}}^N
        \widehat{\mathfrak u}^{\,m|\widetilde n_i-\widetilde n_j|/2}
        e^{-im(\widetilde\alpha_i-\widetilde\alpha_j)}.
    \end{aligned}
    \label{eq:localization-plethystic-kernel-draft4}
\end{equation}
In the \(m\)-th term of the plethystic exponential, the fugacities and
holonomy eigenvalues are raised to their \(m\)-th powers, while the GNO
charges remain unchanged.  The flux sum in
\eqref{eq:localization-fixed-charge-formula-draft4} includes all pairs
of integer GNO vectors with total charge \(q\); positivity and equality
of the two flux distributions have not been assumed.

The explicit powers in \(\mathcal F_m\) allow us to determine which
sectors can contribute through degree \(L\).  In
section~\ref{subsec:monopole-selection-draft4}, we show that a negative
flux or unequal nonnegative flux distributions force every contributing
term to have degree at least \(k\).  Such sectors are therefore absent
through degree \(L\) whenever \(k>L\).  For the remaining sectors,
section~\ref{subsec:neutral-factor-stabilization-draft4} integrates over
the holonomies associated with zero flux, and
section~\ref{subsec:CS-projection-BMN-draft4} identifies the remaining
integral with a BMN vacuum contribution.  The resulting finite-degree
identities imply the reduced-index relation in the double scale limit.

\subsection{Selection of the surviving monopole sectors}
\label{subsec:monopole-selection-draft4}

We work at finite \(N,k\) and fixed total monopole charge \(q\geq0\).
To control the index through a fixed transverse degree \(L\), we
Taylor-expand the plethystic exponential in
\eqref{eq:localization-fixed-charge-formula-draft4}, keeping the
functions \(g_m^\pm\) unexpanded.  For each \(m\geq1\) and pair
\(i,j\), let \(a^+_{m;ij},a^-_{m;ij}\in\mathbb Z_{\geq0}\)
be the Taylor powers of the two matter terms in \(\mathcal F_m\).
Similarly, \(v_{m;ij},\widetilde v_{m;ij}\in\mathbb Z_{\geq0}\)
are the Taylor powers of the two vector terms.  We set
\(v_{m;ij}=0\) when \(n_i=n_j\) and
\(\widetilde v_{m;ij}=0\) when \(\widetilde n_i=\widetilde n_j\).
These integers label the Taylor expansion of the mode-\(m\)
bilinears, rather than occupations of individual bosonic or
fermionic states.

We also expand the Haar factors in Fourier monomials.  In the first
gauge group, their finite product gives
\begin{equation}
    \begin{aligned}
        \prod_{\substack{i<j\\n_i=n_j}}
        \left(2\sin\frac{\alpha_i-\alpha_j}{2}\right)^2
        &=\prod_{\substack{i\neq j\\n_i=n_j}}
          \left(1-e^{i(\alpha_j-\alpha_i)}\right)
        \\
        &=\sum_{\{h\}}(-1)^{\sum_{i,j}h_{ij}}
          \prod_{\substack{i\neq j\\n_i=n_j}}
          e^{ih_{ij}(\alpha_j-\alpha_i)}.
    \end{aligned}
    \label{eq:selection-Haar-expansion-draft5}
\end{equation}
Here \(h_{ij}\in\{0,1\}\), since each Haar factor is linear
in its Fourier monomial.  The second group has the analogous
indices \(\widetilde h_{ij}\).  We set these indices to zero
outside their respective equal-flux, off-diagonal pairs.
Let \(\mathbf a\) denote the collection of all six families
\(a^+,a^-,v,\widetilde v,h,\widetilde h\); each Taylor term
has finite support in \(m\).

With the residual Weyl factors written explicitly, the index becomes
\begin{equation}
    \begin{aligned}
        \mathcal I^{\mathrm{ABJM}}_{N,k;q}
        ={}&e^{-\delta q}A^{-kq}
        \sum_{\substack{
            \boldsymbol n,\widetilde{\boldsymbol n}\in\mathbb Z^N/S_N\\
            \sum_i n_i=\sum_i\widetilde n_i=q}}
        \frac{\widehat{\mathfrak u}^{\,E_0/2}}
             {|W_{\boldsymbol n}|\,|W_{\widetilde{\boldsymbol n}}|}
        \\
        &\times\sum_{\mathbf a}C(\mathbf a)
        \int\prod_{i=1}^{N}
        \frac{d\alpha_i\,d\widetilde\alpha_i}{(2\pi)^2}
        \exp\left[ik\sum_{i=1}^{N}
        (n_i\alpha_i-\widetilde n_i\widetilde\alpha_i)\right]
        \\
        &\times\prod_{m,i,j}
        \left[\widehat{\mathfrak u}^{\,m|n_i-\widetilde n_j|/2}
        A^m g_m^+e^{im(\widetilde\alpha_j-\alpha_i)}
        \right]^{a^+_{m;ij}}
        \\
        &\times\prod_{m,i,j}
        \left[\widehat{\mathfrak u}^{\,m|n_i-\widetilde n_j|/2}
        A^{-m}g_m^-e^{im(\alpha_i-\widetilde\alpha_j)}
        \right]^{a^-_{m;ij}}
        \\
        &\times\prod_{\substack{m,i,j\\n_i\neq n_j}}
        \left[\widehat{\mathfrak u}^{\,m|n_i-n_j|/2}
        e^{im(\alpha_j-\alpha_i)}\right]^{v_{m;ij}}
        \\
        &\times\prod_{\substack{m,i,j\\\widetilde n_i\neq\widetilde n_j}}
        \left[\widehat{\mathfrak u}^{\,m|\widetilde n_i-\widetilde n_j|/2}
        e^{im(\widetilde\alpha_j-\widetilde\alpha_i)}
        \right]^{\widetilde v_{m;ij}}
        \\
        &\times\prod_{i,j}
        e^{ih_{ij}(\alpha_j-\alpha_i)
           +i\widetilde h_{ij}(\widetilde\alpha_j-\widetilde\alpha_i)}.
    \end{aligned}
    \label{eq:selection-occupation-expansion-draft5}
\end{equation}
Unless a restriction is displayed, \(m\geq1\) and
\(i,j=1,\ldots,N\).  The numerical coefficient is
\begin{equation}
    \begin{aligned}
        C(\mathbf a)
        ={}&(-1)^{\sum_{i,j}(h_{ij}+\widetilde h_{ij})}
        \prod_{m,i,j}
        \frac{1}{m^{a^+_{m;ij}+a^-_{m;ij}}
                  a^+_{m;ij}!\,a^-_{m;ij}!}
        \\
        &\times\prod_{m,i,j}
        \frac{(-1)^{v_{m;ij}+\widetilde v_{m;ij}}}
             {m^{v_{m;ij}+\widetilde v_{m;ij}}
              v_{m;ij}!\,\widetilde v_{m;ij}!}.
    \end{aligned}
    \label{eq:selection-Taylor-coefficient-draft5}
\end{equation}
The powers of \(m^{-1}\) and the factorials come from the
exponential, while the displayed signs come from the vector and
Haar factors.  The fermionic matter signs remain inside
\(g_m^\pm\).  We use
\eqref{eq:selection-occupation-expansion-draft5} as a formal
expansion and impose the holonomy constraints before extracting
transverse coefficients.

The holonomy integrals retain a term only if every Fourier
exponent vanishes.  The resulting constraints are
\begin{equation}
    \begin{aligned}
        k n_i
        &=\sum_{m,j}m
          \left(a^+_{m;ij}-a^-_{m;ij}
                +v_{m;ij}-v_{m;ji}\right)
          +\sum_j(h_{ij}-h_{ji}),
        \\
        k\widetilde n_j
        &=\sum_{m,i}m
          \left(a^+_{m;ij}-a^-_{m;ij}
                +\widetilde v_{m;ij}-\widetilde v_{m;ji}\right)
          +\sum_i(\widetilde h_{ij}-\widetilde h_{ji}).
    \end{aligned}
    \label{eq:selection-occupation-constraints-draft5}
\end{equation}
In particular, a matter factor in mode \(m\) contributes \(m\)
units of Fourier charge.  Define the total matter charge units by
\begin{equation}
    P=\sum_{m,i,j}m a^+_{m;ij},
    \qquad
    M=\sum_{m,i,j}m a^-_{m;ij}.
    \label{eq:selection-matter-occupations-draft5}
\end{equation}
Summing either family of constraints cancels the vector and Haar
terms and yields
\begin{equation}
    P-M=kq.
    \label{eq:selection-total-matter-charge-draft4}
\end{equation}
Every term surviving the holonomy projection therefore supplies
\(A^{P-M}=A^{kq}\), canceling the prefactor \(A^{-kq}\).

The total exponent of the flux-difference factors is
\begin{equation}
    \begin{aligned}
        \mathscr H={}&
        \sum_{m,i,j}m(a^+_{m;ij}+a^-_{m;ij})|n_i-\widetilde n_j|
        \\
        &+\sum_{m,i,j}m v_{m;ij}|n_i-n_j|
         +\sum_{m,i,j}m\widetilde v_{m;ij}
                            |\widetilde n_i-\widetilde n_j|.
    \end{aligned}
    \label{eq:selection-occupation-harmonic-exponent-draft5}
\end{equation}
Haar factors contribute no transverse degree.  Up to the numerical
coefficient \(C(\mathbf a)\) and Weyl normalization, the projected
Taylor term is \(e^{-\delta q}\mathcal W\), with
\begin{equation}
    \mathcal W=
    \widehat{\mathfrak u}^{\,(E_0+\mathscr H)/2}
    \prod_{m,i,j}(g_m^+)^{a^+_{m;ij}}(g_m^-)^{a^-_{m;ij}}.
    \label{eq:selection-monomial-degree-decomposition-draft4}
\end{equation}
Here \(\mathcal W\) is a fugacity series; the factors
\(g_m^\pm\) have not been expanded into individual letters.

Equation \eqref{eq:selection-monomial-degree-decomposition-draft4}
also proves the nonnegative transverse support used in
\eqref{eq:physical-transverse-expansion-draft4}.  This does
not require the individual fluxes to be nonnegative or the
two flux distributions to agree.  For every term surviving
the holonomy projection,
\eqref{eq:selection-total-matter-charge-draft4} gives
\(P-M=kq\).  Hence the factor \(A^{P-M}\) from the matter
insertions cancels the prefactor \(A^{-kq}\).  All remaining
fugacity dependence is contained in \(\mathcal W\).

The zero-point exponent \(E_0\) is nonnegative
\cite{Kim:2009wb}, and \(\mathscr H\geq0\) follows directly
from \eqref{eq:selection-occupation-harmonic-exponent-draft5}.
Moreover, every monomial in \(g_m^\pm\) has a nonnegative
power of each \(\widehat u_a\), as is clear from
\eqref{eq:selection-relative-letter-functions-draft4}.
Thus no projected contribution contains a negative power of
any transverse fugacity.

The powers are also integral.  Since the two GNO vectors
have the same total charge, \(|a-b|\equiv a+b\pmod2\)
in \eqref{eq:ABJM-zero-point-energy-review-draft4} gives
\(E_0\equiv0\pmod2\).  Multiply the two Fourier constraints
in \eqref{eq:selection-occupation-constraints-draft5} by
\(n_i\) and \(\widetilde n_j\), respectively, and subtract
their sums.  The Haar terms cancel because they join equal
fluxes.  Reducing the remaining terms modulo two gives
\[
    \mathscr H\equiv
    k\left(\sum_i n_i^2-\sum_j\widetilde n_j^2\right)
    \equiv k(q-q)\equiv0\pmod2.
\]
Hence \((E_0+\mathscr H)/2\) is a nonnegative integer.
The conversion from \(\widehat u_a\) to \(u_a\) leaves
these exponents unchanged, so the gauge-projected index
has support on \(\mathbb Z_{\geq0}^3\), as stated in
\eqref{eq:physical-transverse-expansion-draft4}.

For the total-degree estimate below, every monomial in
\(g_m^+\) has nonnegative total degree, while every monomial
in \(g_m^-\) has total degree at least \(m\).

Since \(\widehat{\mathfrak u}=\widehat u_1\widehat u_2\widehat u_3\)
has total degree three, every monomial in \(\mathcal W\) has total
degree at least \(M+\tfrac32(E_0+\mathscr H)\).  A Taylor term
can therefore contribute through degree \(L\) only if
\begin{equation}
    M+\frac32(E_0+\mathscr H)\leq L.
    \label{eq:selection-total-degree-bound-draft4}
\end{equation}
In particular,
\begin{equation}
    M\leq L,
    \qquad P=kq+M\leq kq+L.
    \label{eq:selection-finite-matter-bound-draft5}
\end{equation}
These restrictions hold for each Taylor term at finite \(N,k\),
before cancellations among the coefficients \(C(\mathbf a)\).

\subsubsection{Exclusion of negative monopole fluxes}
\label{subsubsec:negative-flux-exclusion-draft4}

Consider a term satisfying the Fourier constraints
\eqref{eq:selection-occupation-constraints-draft5} and the degree
bound \eqref{eq:selection-total-degree-bound-draft4}.  Let
\(S_-=\{i:n_i<0\}\) and
\(\widetilde S_-=\{j:\widetilde n_j<0\}\), and define
\begin{equation}
    r_-=-\sum_{i\in S_-}n_i,
    \qquad
    \widetilde r_-=-\sum_{j\in\widetilde S_-}\widetilde n_j.
    \label{eq:selection-negative-flux-charge-draft4}
\end{equation}
Summing the first Fourier constraint over \(i\in S_-\) eliminates
the vector terms with both indices in \(S_-\).  The Haar terms
also cancel: \(h_{ij}\) can be nonzero only when \(n_i=n_j\),
so a nonzero Haar term with \(i\in S_-\) necessarily has
\(j\in S_-\).  Multiplying the resulting equation by \(-1\)
gives the exact identity
\begin{equation}
    kr_-=
    \sum_{\substack{m,\,i\in S_-\\j=1,\ldots,N}}
    m\left(a^-_{m;ij}-a^+_{m;ij}\right)
    +\sum_{\substack{m,\,i\in S_-\\j\notin S_-}}
    m\left(v_{m;ji}-v_{m;ij}\right).
    \label{eq:selection-negative-cut-balance-draft4}
\end{equation}
All Taylor indices are nonnegative.  Moreover,
\(|n_j-n_i|\geq1\) for \(i\in S_-\) and \(j\notin S_-\).
The terms \(m v_{m;ji}\) with these indices are therefore
bounded by their contributions
\(m v_{m;ji}|n_j-n_i|\) to \(\mathscr H\).
Discarding the negative summands in
\eqref{eq:selection-negative-cut-balance-draft4} yields
\begin{equation}
    \begin{aligned}
        kr_-
        &\leq
        \sum_{\substack{m,\,i\in S_-\\j=1,\ldots,N}}
        m a^-_{m;ij}
        +\sum_{\substack{m,\,i\in S_-\\j\notin S_-}}
        m v_{m;ji}
        \\
        &\leq M+\mathscr H\leq L.
    \end{aligned}
    \label{eq:selection-negative-cut-estimate-draft4}
\end{equation}
The last inequality follows from
\eqref{eq:selection-total-degree-bound-draft4} and
\(E_0,\mathscr H\geq0\).

Applying the second Fourier constraint to
\(j\in\widetilde S_-\) gives, by the same calculation,
\begin{equation}
    \begin{aligned}
        k\widetilde r_-
        &\leq
        \sum_{\substack{m,\,j\in\widetilde S_-\\i=1,\ldots,N}}
        m a^-_{m;ij}
        +\sum_{\substack{m,\,j\in\widetilde S_-\\i\notin\widetilde S_-}}
        m\widetilde v_{m;ji}
        \\
        &\leq M+\mathscr H\leq L.
    \end{aligned}
    \label{eq:selection-tilded-negative-estimate-draft5}
\end{equation}
Thus every term contributing through degree \(L\) obeys
\begin{equation}
    L\geq k\max\{r_-,\widetilde r_-\}.
    \label{eq:selection-negative-degree-bound-draft4}
\end{equation}
Since the fluxes are integers, any negative entry makes
\(\max\{r_-,\widetilde r_-\}\geq1\).  All sectors containing
negative fluxes therefore have zero coefficients through degree
\(L\) whenever
\begin{equation}
    k>L.
    \label{eq:selection-rank-independent-cutoff-draft4}
\end{equation}
The threshold depends only on the chosen degree \(L\), independently
of \(N\) and of the magnitudes of the fluxes.

At fixed \(N,k\), the bounds above ensure that only finitely many
flux sectors and Taylor-index configurations contribute through
degree \(L\).  At \(q=0\) and \(k>L\), the remaining nonnegative
fluxes have zero total charge, so both flux vectors vanish.

\subsubsection{Exclusion of unequal flux distributions}
\label{subsubsec:unequal-flux-exclusion-draft4}

Assume \(k>L\), so that
section~\ref{subsubsec:negative-flux-exclusion-draft4} restricts
both flux vectors to nonnegative entries.  At fixed total charge
\(q>0\), define the multiplicities of each positive flux value by
\begin{equation}
    r_s=\#\{i:n_i=s\},
    \qquad
    \widetilde r_s=\#\{j:\widetilde n_j=s\},
    \qquad s\geq1.
    \label{eq:selection-positive-flux-multiplicities-draft5}
\end{equation}
They satisfy \(\sum_{s\geq1}sr_s
=\sum_{s\geq1}s\widetilde r_s=q\).
We show that the Fourier constraints and the degree bound force
\(r_s=\widetilde r_s\) for every \(s\).

Fix a positive integer \(s\).  Multiply the first constraint in
\eqref{eq:selection-occupation-constraints-draft5} by
\(\delta_{n_i,s}\), multiply the second by
\(\delta_{\widetilde n_j,s}\), and subtract their sums over
\(i\) and \(j\), respectively.  Here \(\delta_{a,s}\) is the
Kronecker delta.  The Haar terms vanish because \(h_{ij}\)
is supported on \(n_i=n_j\), where
\(\delta_{n_i,s}-\delta_{n_j,s}=0\); the tilded terms vanish
in the same way.  Relabeling indices in the vector sums gives
\begin{equation}
    \begin{aligned}
        ks(r_s-\widetilde r_s)
        ={}&\sum_{m,i,j}m
        (a^+_{m;ij}-a^-_{m;ij})
        (\delta_{n_i,s}-\delta_{\widetilde n_j,s})
        \\
        &+\sum_{m,i,j}m v_{m;ij}
        (\delta_{n_i,s}-\delta_{n_j,s})
        \\
        &+\sum_{m,i,j}m\widetilde v_{m;ij}
        (\delta_{\widetilde n_i,s}-\delta_{\widetilde n_j,s}).
    \end{aligned}
    \label{eq:selection-fixed-flux-balance-draft5}
\end{equation}
For integer \(a,b\),
\begin{equation}
    |\delta_{a,s}-\delta_{b,s}|\leq |a-b|.
    \label{eq:selection-Kronecker-bound-draft5}
\end{equation}
Indeed, a nonzero difference on the left requires \(a\neq b\),
and hence \(|a-b|\geq1\).  All Taylor indices are nonnegative,
so \(|a^+_{m;ij}-a^-_{m;ij}|
\leq a^+_{m;ij}+a^-_{m;ij}\).  Taking the absolute value of
\eqref{eq:selection-fixed-flux-balance-draft5}, applying the
triangle inequality and then
\eqref{eq:selection-Kronecker-bound-draft5}, we obtain
\begin{equation}
    \begin{aligned}
        ks|r_s-\widetilde r_s|
        \leq{}&\sum_{m,i,j}m
        (a^+_{m;ij}+a^-_{m;ij})|n_i-\widetilde n_j|
        \\
        &+\sum_{m,i,j}m v_{m;ij}|n_i-n_j|
        +\sum_{m,i,j}m\widetilde v_{m;ij}
                         |\widetilde n_i-\widetilde n_j|
        \\
        ={}&\mathscr H\leq\frac{2L}{3}.
    \end{aligned}
    \label{eq:selection-unequal-partition-bound-draft4}
\end{equation}
The last inequality follows from
\eqref{eq:selection-total-degree-bound-draft4} and
\(M,E_0\geq0\).  For \(k>L\) and \(s\geq1\), it implies
\(|r_s-\widetilde r_s|\leq2L/(3ks)<1\).
Since the multiplicities are integers, they must agree for every
positive flux value.  The numbers of zero entries also agree,
because both vectors have length \(N\).  Thus unequal nonnegative
flux distributions cannot contribute through degree \(L\).

Combining this result with the exclusion of negative fluxes, the
only possible sectors through degree \(L\) have
\begin{equation}
    \boldsymbol n=\widetilde{\boldsymbol n}
    =(0^{N_0},1^{r_1},2^{r_2},\ldots),
    \qquad
    \sum_{s\geq1}sr_s=q,
    \qquad
    N_0=N-\sum_{s\geq1}r_s,
    \label{eq:selection-surviving-fluxes-draft4}
\end{equation}
up to independent Weyl permutations.  These sectors are labelled
by partitions \(\lambda=(1^{r_1}2^{r_2}\cdots)\vdash q\).
Denote their contributions by
\(\mathcal I^{\mathrm{ABJM}}_{N,k;\lambda}\).
For \(N\geq q\), every partition of \(q\) fits within the rank.
Writing \(\widehat{\mathbf u}^{\,\boldsymbol\gamma}
=\prod_a\widehat u_a^{\gamma_a}\) and
\(|\boldsymbol\gamma|=\sum_a\gamma_a\), we therefore obtain
the exact coefficient identity
\begin{equation}
    \begin{gathered}
        [\widehat{\mathbf u}^{\,\boldsymbol\gamma}]
        \mathcal I^{\mathrm{ABJM}}_{N,k;q}
        =\sum_{\lambda\vdash q}
        [\widehat{\mathbf u}^{\,\boldsymbol\gamma}]
        \mathcal I^{\mathrm{ABJM}}_{N,k;\lambda},
        \\
        \boldsymbol\gamma\in\mathbb Z_{\geq0}^3,
        \qquad |\boldsymbol\gamma|\leq L,
        \qquad N\geq q,\quad k>L.
    \end{gathered}
    \label{eq:selection-fixed-coefficient-sum-draft4}
\end{equation}
Both exclusions apply to each Taylor term before cancellations.
Consequently, \eqref{eq:selection-fixed-coefficient-sum-draft4}
holds simultaneously for all coefficients through any fixed finite
degree \(L\), at finite \(N,k\).  Its sufficient conditions are
eventually satisfied along the double scale limit at fixed \(q,L\).
Together with the all-zero-sector result at \(q=0\), this reduces
the remaining calculation to the matched sectors
\eqref{eq:selection-surviving-fluxes-draft4} and the neutral denominator.

\subsection{Finite-rank stabilization and the neutral factor}
\label{subsec:neutral-factor-stabilization-draft4}

In a matched sector \(\lambda\) of
\eqref{eq:selection-surviving-fluxes-draft4}, \(E_0=0\).
We integrate over the two zero-flux holonomies through degree
\(L\) at finite \(N,k\).  The required Haar moments will be
shown to stabilize once \(N_0\geq L\).

Denote the zero-flux holonomies by \(U_0,\widetilde U_0\) and
the charged holonomies by \(V_s,\widetilde V_s\in U(r_s)\).
Their power sums are
\begin{equation}
    \begin{aligned}
        v_{s,\pm m}&=\operatorname{Tr}V_s^{\pm m},
        &\widetilde v_{s,\pm m}
        &=\operatorname{Tr}\widetilde V_s^{\pm m},
        \qquad s\geq1,
        \\
        v_{0,\pm m}&=\rho_{\pm m}
        =\operatorname{Tr}U_0^{\pm m},
        &\widetilde v_{0,\pm m}&=\widetilde\rho_{\pm m}
        =\operatorname{Tr}\widetilde U_0^{\pm m}.
    \end{aligned}
    \label{eq:stabilization-block-power-sums-draft4}
\end{equation}
The labels \(s,t\) denote flux values, and traces of unoccupied
blocks vanish.  Grouping
\eqref{eq:localization-plethystic-kernel-draft4} by these values gives
\begin{equation}
    \begin{aligned}
        \mathcal S_m={}&
        \sum_{s,t\geq0}\widehat{\mathfrak u}^{\,m|s-t|/2}
        \left(
            A^m g_m^+v_{s,-m}\widetilde v_{t,m}
            +A^{-m}g_m^-\widetilde v_{s,-m}v_{t,m}
        \right)
        \\
        &-\sum_{\substack{s,t\geq0\\s\neq t}}
        \widehat{\mathfrak u}^{\,m|s-t|/2}
        \left(v_{s,-m}v_{t,m}
             +\widetilde v_{s,-m}\widetilde v_{t,m}\right).
    \end{aligned}
    \label{eq:stabilization-exact-block-action-draft4}
\end{equation}
With the equal-flux factors included in the normalized Haar
measures, the sector contribution is
\begin{equation}
    \mathcal I^{\mathrm{ABJM}}_{N,k;\lambda}
    =e^{-\delta q}A^{-kq}
    \int[dU_0][d\widetilde U_0]
    \prod_{s:r_s>0}[dV_s][d\widetilde V_s]
    \left(\frac{\det V_s}{\det\widetilde V_s}\right)^{ks}
    \exp\left[\sum_{m=1}^{\infty}\frac{\mathcal S_m}{m}\right].
    \label{eq:stabilization-exact-sector-integral-draft4}
\end{equation}
We retain \(A\) as a formal charge parameter.
Equation~\eqref{eq:selection-total-matter-charge-draft4} fixes its
net matter power to \(A^{kq}\), which cancels \(A^{-kq}\) in
\eqref{eq:stabilization-exact-sector-integral-draft4}.
Transverse degrees are counted after this cancellation.

Summing the Fourier constraints
\eqref{eq:selection-occupation-constraints-draft5} over each
zero-flux block cancels the Haar contributions and requires equal
total positive and negative trace powers in that block.
Denote their common weighted sums by
\(d\) for \(U_0\) and \(\widetilde d\) for \(\widetilde U_0\),
counting \(m\) for each occurrence of a trace with power \(\pm m\).
The Taylor expansion
\eqref{eq:selection-occupation-expansion-draft5} gives
\begin{equation}
    \begin{aligned}
        d
        &=\sum_{\substack{m,i,j\\n_i=0}}m a^-_{m;ij}
          +\sum_{\substack{m,i,j\\n_i=0<n_j}}m v_{m;ji}
         =\sum_{\substack{m,i,j\\n_i=0}}m a^+_{m;ij}
          +\sum_{\substack{m,i,j\\n_i=0<n_j}}m v_{m;ij},
        \\[2pt]
        \widetilde d
        &=\sum_{\substack{m,i,j\\\widetilde n_j=0}}m a^-_{m;ij}
          +\sum_{\substack{m,i,j\\\widetilde n_j=0<\widetilde n_i}}
           m\widetilde v_{m;ji}
         =\sum_{\substack{m,i,j\\\widetilde n_j=0}}m a^+_{m;ij}
          +\sum_{\substack{m,i,j\\\widetilde n_j=0<\widetilde n_i}}
           m\widetilde v_{m;ij}.
    \end{aligned}
    \label{eq:stabilization-zero-block-Taylor-powers-draft5}
\end{equation}
For \(d\), the first and second expressions count positive and
negative powers, respectively; for \(\widetilde d\), the order
is reversed.
Haar factors remain in the measures and do not enter these trace
degrees.  In the expressions involving \(a^-_{m;ij}\), the matter
sums are bounded by \(M\); each vector term
has a nonzero integer flux difference, so its Fourier power is
bounded by its contribution to \(\mathscr H\).  Consequently,
\eqref{eq:selection-total-degree-bound-draft4} implies
\begin{equation}
    \max\{d,\widetilde d\}\leq M+\mathscr H\leq L.
    \label{eq:stabilization-zero-block-degree-bound-draft5}
\end{equation}
The zero-flux trace degrees are therefore bounded independently
of the growing reference charge \(P=kq+M\).

For nonnegative integer sequences \(c_m,e_m\) of finite support,
the normalized Haar moment satisfies
\begin{equation}
    \begin{gathered}
        \int_{U(N_0)}[dU]\,
        \prod_{m\geq1}
        (\operatorname{Tr}U^m)^{c_m}
        (\operatorname{Tr}U^{-m})^{e_m}
        =\prod_{m\geq1}\delta_{c_m e_m}\,m^{c_m}c_m!,
        \\
        N_0\geq\max\left\{\sum_m mc_m,\sum_m me_m\right\}.
    \end{gathered}
    \label{eq:stabilization-Haar-moment-draft4}
\end{equation}
Unequal total powers give zero by central \(U(1)\) invariance.
For equal sums \(d\), associate the partitions
\(\eta=(1^{c_1}2^{c_2}\cdots)\) and
\(\zeta=(1^{e_1}2^{e_2}\cdots)\).  The Frobenius expansion
\eqref{eq:characters-Frobenius-appB-draft5} and unitary-character
orthogonality \eqref{eq:characters-unitary-orthogonality-appB-draft5}
express the integral as
\[
    \sum_{\substack{\tau\vdash d\\\ell(\tau)\leq N_0}}
       \chi^\tau(\eta)\chi^\tau(\zeta),
\]
where \(\chi^\tau(\eta)\) is the \(S_d\) character labelled
by \(\tau\), evaluated on cycle type \(\eta\), and
\(\ell(\tau)\) is the number of parts of \(\tau\).
These symmetric-group characters are defined in
appendix~\ref{app:character-identities-draft5}.
When \(N_0\geq d\), the restriction removes no partitions,
and \eqref{eq:characters-symmetric-orthogonality-appB-draft5} gives
\eqref{eq:stabilization-Haar-moment-draft4}.

By \eqref{eq:stabilization-zero-block-degree-bound-draft5},
\(N_0\geq L\) places every relevant moment in this stable range.
Since \(N_0\geq N-q\), \(N\geq q+L\) is sufficient uniformly
over the partitions of \(q\).  The right-hand side of
\eqref{eq:stabilization-Haar-moment-draft4} is precisely the Wick
moment of independent complex Gaussian modes with covariance
\begin{equation}
    \langle\rho_m\rho_{-n}\rangle_{\mathrm G}
    =\langle\widetilde\rho_m\widetilde\rho_{-n}\rangle_{\mathrm G}
    =m\delta_{mn}.
    \label{eq:stabilization-Gaussian-covariance-draft4}
\end{equation}
The two families are independent, and pairings of two positive
or two negative modes vanish.  The Gaussian replacement is made
inside the fixed-charge coefficient: both functionals enforce
the zero-block central projections and leave the charged Fourier
weights unchanged.  Terms excluded by the charged projections
remain absent, while the bound above places all remaining moments
in the stable range.  Their generating function may therefore
be evaluated with the Gaussian moment functional.

Define the charged source polynomials and the zero-flux kernel by
\begin{equation}
    \begin{gathered}
        \boldsymbol\rho_m=
        \begin{pmatrix}\rho_m\\\widetilde\rho_m\end{pmatrix},
        \quad
        \overline{\boldsymbol\rho}_m=
        \begin{pmatrix}\rho_{-m}&\widetilde\rho_{-m}\end{pmatrix},
        \quad
        \boldsymbol L_m=
        \begin{pmatrix}L_m\\\widetilde L_m\end{pmatrix},
        \quad
        \overline{\boldsymbol L}_m=
        \begin{pmatrix}L_{-m}&\widetilde L_{-m}\end{pmatrix},
        \\
        K_m=
        \begin{pmatrix}1&-A^m g_m^+\\-A^{-m}g_m^-&1\end{pmatrix},
        \\
        L_{\pm m}=\sum_{s\geq1}
           \widehat{\mathfrak u}^{\,ms/2}v_{s,\pm m},
        \qquad
        \widetilde L_{\pm m}=\sum_{s\geq1}
           \widehat{\mathfrak u}^{\,ms/2}\widetilde v_{s,\pm m}.
    \end{gathered}
    \label{eq:stabilization-zero-block-kernel-draft4}
\end{equation}
The fugacities are not conjugated in the barred vectors.
Including the Gaussian weight, the
zero-flux terms complete to
\begin{equation}
    \begin{aligned}
        &-\overline{\boldsymbol\rho}_mK_m\boldsymbol\rho_m
         -\overline{\boldsymbol\rho}_mK_m\boldsymbol L_m
         -\overline{\boldsymbol L}_mK_m\boldsymbol\rho_m
        \\
        &\qquad=
        -(\overline{\boldsymbol\rho}_m+\overline{\boldsymbol L}_m)
         K_m(\boldsymbol\rho_m+\boldsymbol L_m)
         +\overline{\boldsymbol L}_mK_m\boldsymbol L_m,
    \end{aligned}
    \label{eq:stabilization-completing-square-draft4}
\end{equation}
with a common factor \(1/m\) in the exponent.  Within the
fixed-charge coefficient, the zero-flux Haar integral therefore
factorizes through degree \(L\) for \(N_0\geq L\):
\begin{equation}
    \begin{gathered}
        \int[dU_0][d\widetilde U_0]\,
        \exp\!\left[\sum_{m=1}^{\infty}\frac{
            \overline{\boldsymbol\rho}_m(\mathbf1-K_m)\boldsymbol\rho_m
            -\overline{\boldsymbol\rho}_mK_m\boldsymbol L_m
            -\overline{\boldsymbol L}_mK_m\boldsymbol\rho_m
        }{m}\right]
        =\prod_{m=1}^{\infty}\mathcal Z_m^{(0)},
        \\[4pt]
    \begin{aligned}
        \mathcal Z_m^{(0)}
        &=\int_{\mathbb C^2}
          \frac{d^2\rho_m\,d^2\widetilde\rho_m}{(\pi m)^2}
          \exp\!\left[-\frac{
              \overline{\boldsymbol\rho}_mK_m\boldsymbol\rho_m
              +\overline{\boldsymbol\rho}_mK_m\boldsymbol L_m
              +\overline{\boldsymbol L}_mK_m\boldsymbol\rho_m
          }{m}\right]
        \\
        &=\frac{1}{\det K_m}
          \exp\left[\frac1m
              \overline{\boldsymbol L}_mK_m\boldsymbol L_m\right],
        \qquad \det K_m=1-g_m^+g_m^-.
    \end{aligned}
    \end{gathered}
    \label{eq:stabilization-zero-block-integral-draft4}
\end{equation}
For Hermitian positive-definite \(K_m\),
the Gaussian integral in
\eqref{eq:stabilization-zero-block-integral-draft4} follows by
completing the square.  Its coefficientwise extension is well
defined since \(g_m^+g_m^-\) and all source bilinears have
positive transverse degree.

The determinant is independent of the charged holonomies and
of \(A\).  Substituting
\eqref{eq:selection-relative-letter-functions-draft4} and
multiplying over \(m\) gives
\begin{equation}
    \begin{aligned}
        \mathcal I_{\mathrm{neutral}}^{(k)}
        &\equiv\prod_{m=1}^{\infty}\frac{1}{1-g_m^+g_m^-}
        \\
        &=\prod_{m=1}^{\infty}
        \frac{(1-\widehat{\mathfrak u}^{\,m})^2}{
            (1-\widehat u_1^m)(1-\widehat u_2^m)
            (1-(\widehat u_1\widehat u_3)^m)
            (1-(\widehat u_2\widehat u_3)^m)}.
    \end{aligned}
    \label{eq:stabilization-neutral-product-draft4}
\end{equation}
Only modes \(m\leq L\) affect this product through degree \(L\),
and its constant term is one.  It is the ABJM zero-flux
determinant \cite{Kim:2009wb}, now obtained in the finite-rank
stable range.  Its \(k\)-dependence enters through
\(\widehat u_a\); its coefficientwise limit
\(\mathcal I_{\mathrm{neutral}}\) is the same product at \(u_a\).

For \(q=0\), the flux bound
\eqref{eq:selection-negative-degree-bound-draft4} leaves only the
all-zero flux sector through degree \(L\) when \(k>L\).
Then \(N_0=N\) and the sources vanish, yielding
\begin{equation}
    \begin{gathered}
        [\widehat{\mathbf u}^{\,\boldsymbol\gamma}]
        \mathcal I^{\mathrm{ABJM}}_{N,k;0}
        =
        [\widehat{\mathbf u}^{\,\boldsymbol\gamma}]
        \mathcal I_{\mathrm{neutral}}^{(k)},
        \\
        |\boldsymbol\gamma|\leq L,
        \qquad N\geq L,\quad k>L.
    \end{gathered}
    \label{eq:stabilization-neutral-sector-identity-draft4}
\end{equation}
This identifies the neutral denominator at the same cutoff.
Its unit constant term also ensures equality of the formal
inverses through degree \(L\).

For \(q>0\), expand the source term
\(\overline{\boldsymbol L}_mK_m\boldsymbol L_m\) and add it to
the charged part of \(\mathcal S_m\).  Its coefficients carry
\(\widehat{\mathfrak u}^{\,m(s+t)/2}\), with negative signs
for matter bilinears and positive signs for same-group bilinears.
Combining them with the direct weight
\(\widehat{\mathfrak u}^{\,m|s-t|/2}\) gives
\begin{equation}
    H_{st}^{(m)}
    =\widehat{\mathfrak u}^{\,m|s-t|/2}
     -\widehat{\mathfrak u}^{\,m(s+t)/2}
    \label{eq:stabilization-induced-subtraction-draft4}
\end{equation}
and the effective action
\begin{equation}
    \begin{aligned}
        \mathcal S_m^{\mathrm{eff}}={}&
        \sum_{s\geq1}
        \left(v_{s,-m}v_{s,m}
             +\widetilde v_{s,-m}\widetilde v_{s,m}\right)
        \\
        &+\sum_{s,t\geq1}H_{st}^{(m)}
        \Big[
            A^m g_m^+v_{s,-m}\widetilde v_{t,m}
            +A^{-m}g_m^-\widetilde v_{s,-m}v_{t,m}
            \\
        &\hspace{23mm}
            -v_{s,-m}v_{t,m}
            -\widetilde v_{s,-m}\widetilde v_{t,m}
        \Big].
    \end{aligned}
    \label{eq:stabilization-effective-charged-action-draft4}
\end{equation}
Thus the sector factorizes coefficientwise as
\begin{equation}
    \begin{aligned}
        [\widehat{\mathbf u}^{\,\boldsymbol\gamma}]
        \mathcal I^{\mathrm{ABJM}}_{N,k;\lambda}
        ={}&e^{-\delta q}
        [\widehat{\mathbf u}^{\,\boldsymbol\gamma}]
        \Bigg\{
        A^{-kq}\mathcal I_{\mathrm{neutral}}^{(k)}
        \int\prod_{s:r_s>0}
        \left(
            [dV_s][d\widetilde V_s]\,
            \frac{(\det V_s)^{ks}}
                 {(\det\widetilde V_s)^{ks}}
        \right)
        \\
        &\hspace{16mm}\times
        \exp\left[\sum_{m=1}^{\infty}
                   \frac{\mathcal S_m^{\mathrm{eff}}}{m}\right]
        \Bigg\},
        \qquad |\boldsymbol\gamma|\leq L,\quad N\geq L+q.
    \end{aligned}
    \label{eq:stabilization-charged-sector-integral-draft4}
\end{equation}
The remaining charged Haar integrals retain their finite ranks
\(r_s\) and determinant charges \(ks\).  Their projection is
evaluated in the next subsection.

\subsection{Chern--Simons projection and the BMN matrix integral}
\label{subsec:CS-projection-BMN-draft4}

We now evaluate the charged-block integrals in
\eqref{eq:stabilization-charged-sector-integral-draft4} for a
matched sector \(\lambda\) of
\eqref{eq:selection-surviving-fluxes-draft4}.  We keep the
transverse degree cutoff \(L\) fixed and assume
\(N\geq L+q\) and \(k>L\), so that both the sector selection
and the zero-flux Haar reduction apply.  The remaining ranks
\(r_s\) are finite.  We show that their determinant insertions
reduce the two holonomies in each occupied block to one Haar
integral, and then identify its kernel with the BMN vacuum
kernel at the shifted fugacities.

\subsubsection{Determinant projection at finite degree}
\label{subsubsec:finite-degree-CS-projection-draft5}

Each occupied flux value \(s\) in
\eqref{eq:stabilization-charged-sector-integral-draft4}
still carries two \(U(r_s)\) holonomies.  We show that,
through degree \(L\), the determinant insertions reduce
their integrals exactly to one diagonal Haar integral.
We first separate the reference scalar, which supplies the
Chern--Simons charge at zero transverse degree, from the
remaining excitations.  The latter have bounded holonomy
powers at fixed degree, allowing an exact character projection.
The factors \(e^{-\delta q}\mathcal I_{\mathrm{neutral}}^{(k)}\)
will be restored at the end.

Keeping \(A\) independent of the transverse fugacities,
rescale the second holonomy in each occupied block:
\begin{equation}
    W_s=A\widetilde V_s,
    \qquad w_{s,\pm m}=\operatorname{Tr}W_s^{\pm m},
    \qquad \widetilde v_{s,\pm m}=A^{\mp m}w_{s,\pm m}.
    \label{eq:projection-holonomy-rescaling-draft5}
\end{equation}
The determinant factors obey
\begin{equation}
    \frac{(\det V_s)^{ks}}{(\det\widetilde V_s)^{ks}}
    =A^{ksr_s}\frac{(\det V_s)^{ks}}{(\det W_s)^{ks}},
    \qquad
    A^{-kq}\prod_{s:r_s>0}A^{ksr_s}=1,
    \label{eq:projection-rescaled-determinants-draft5}
\end{equation}
where we used \(\sum_s sr_s=q\).
This rescaling is algebraic at the level of the constant-term
projection: the logarithmic eigenvalue differentials and the
Vandermonde ratios are unchanged, so the normalized Haar
projection functional is preserved.  With the pushed-forward
normalized Haar measure, character homogeneity cancels the
factors of \(A\) in
\(\chi_\alpha(W_s)\chi_\beta(W_s^{-1})\) after projection,
preserving \eqref{eq:characters-unitary-orthogonality-appB-draft5}.
The charged factor in
\eqref{eq:stabilization-charged-sector-integral-draft4}
therefore becomes
\begin{equation}
    \begin{aligned}
        &A^{-kq}\int\prod_{s:r_s>0}
          \left([dV_s][d\widetilde V_s]\,
            \frac{(\det V_s)^{ks}}{(\det\widetilde V_s)^{ks}}\right)
          \exp\left[\sum_{m=1}^{\infty}
            \frac{\mathcal S_m^{\mathrm{eff}}(V,\widetilde V)}{m}\right]
        \\
        &\qquad=\int\prod_{s:r_s>0}
          \left([dV_s][dW_s]\,
            \frac{(\det V_s)^{ks}}{(\det W_s)^{ks}}\right)
          \exp\left[\sum_{m=1}^{\infty}
            \frac{\mathcal S_m^{\mathrm{eff}}(V,A^{-1}W)}{m}\right].
    \end{aligned}
    \label{eq:projection-rescaled-charged-integral-draft5}
\end{equation}
Here each matrix argument denotes the collection of occupied
blocks.  All subsequent expressions for
\(\mathcal S_m^{\mathrm{eff}}\) use the rescaled holonomies
\(V,W\), in which the explicit dependence on \(A\) has canceled.

Combining the same-group terms in
\eqref{eq:stabilization-effective-charged-action-draft4}
leaves a unique degree-zero contribution,
\(\sum_s v_{s,-m}w_{s,m}\), supplied by the constant term
of \(g_m^+\) at equal flux.  Separate this reference
contribution by writing
\begin{equation}
    \mathcal S_m^{\mathrm{eff}}
    =\sum_{s\geq1}v_{s,-m}w_{s,m}+\mathcal R_m(V,W).
    \label{eq:projection-remainder-draft4}
\end{equation}
It follows that
\begin{equation}
    \exp\left[\sum_{m=1}^{\infty}
        \frac{\mathcal S_m^{\mathrm{eff}}}{m}\right]
    =\left[\prod_{s:r_s>0}\mathcal C(V_s,W_s)\right]
     \exp\left[\sum_{m=1}^{\infty}
        \frac{\mathcal R_m(V,W)}{m}\right],
    \label{eq:projection-reference-factorization-draft5}
\end{equation}
where the Cauchy kernel collects arbitrary insertions of the
reference scalar in each block:
\begin{equation}
    \begin{aligned}
        \mathcal C(V_s,W_s)
        &\equiv\exp\left[\sum_{m=1}^{\infty}\frac1m
            \operatorname{Tr}V_s^{-m}\operatorname{Tr}W_s^m\right]
        \\
        &=\sum_{\eta:\,\ell(\eta)\leq r_s}
          s_\eta(V_s^{-1})s_\eta(W_s).
    \end{aligned}
    \label{eq:projection-Cauchy-kernel-draft4}
\end{equation}
Here \(\eta\) is a partition and \(s_\eta\) is the Schur
polynomial \eqref{eq:characters-Schur-definition-appB-draft5},
equivalently a polynomial \(U(r_s)\) character.
We use the Cauchy identity
\eqref{eq:characters-Cauchy-appB-draft5} as a formal character
expansion \cite{bump2006averages}; the Haar projections below
are evaluated term by term.

The remainder is
\begin{equation}
    \begin{aligned}
        \mathcal R_m(V,W)
        =\sum_{s,t\geq1}\Big[{}
            &(H_{st}^{(m)}g_m^+-\delta_{st})v_{s,-m}w_{t,m}
             +H_{st}^{(m)}g_m^-w_{s,-m}v_{t,m}
             \\
            &+(\delta_{st}-H_{st}^{(m)})
              (v_{s,-m}v_{t,m}+w_{s,-m}w_{t,m})
        \Big].
    \end{aligned}
    \label{eq:projection-A-independent-remainder-draft5}
\end{equation}
Every monomial multiplying a mode-\(m\) bilinear in
\eqref{eq:projection-A-independent-remainder-draft5} has
transverse degree at least \(m\).  For \(s=t\), the reference
subtraction removes the constant term of \(H_{ss}^{(m)}g_m^+\):
both \(g_m^+-1\) and \(g_m^-\) start at degree \(m\), while
\(1-H_{ss}^{(m)}=\widehat{\mathfrak u}^{\,ms}\) has higher
degree.  For \(s\neq t\), the factor
\(\widehat{\mathfrak u}^{\,m|s-t|/2}\) already has degree
at least \(3m/2\), and all remaining exponents are nonnegative.

Thus a Taylor term containing \(b_m\) bilinears of mode
\(m\) can contribute through degree \(L\) only if
\begin{equation}
    \sum_{m\geq1}m b_m\leq L.
    \label{eq:projection-mode-degree-bound-draft5}
\end{equation}
For a fixed fugacity coefficient through degree \(L\), define
\begin{equation}
    \begin{gathered}
        \Phi_{\boldsymbol\gamma}(V,W)
        =[\widehat{\mathbf u}^{\,\boldsymbol\gamma}]
          \exp\left[\sum_{m=1}^{\infty}
            \frac{\mathcal R_m(V,W)}{m}\right],
        \\
        \gamma_a\in\tfrac12\mathbb Z_{\geq0},
        \qquad |\boldsymbol\gamma|\leq L.
    \end{gathered}
    \label{eq:projection-finite-Laurent-insertion-draft5}
\end{equation}
This is a finite Laurent polynomial invariant under conjugation
of each block.  A mode-\(m\) bilinear changes the power of
any individual \(W_s\) eigenvalue by at most \(m\), so
\eqref{eq:projection-mode-degree-bound-draft5} places every
such power in \(\Phi_{\boldsymbol\gamma}\) in \([-L,L]\).
These powers are integers; the possible half-integer values
of \(\gamma_a\) concern only the fugacities.

Fix \(\boldsymbol\gamma\) and an occupied block \(s\),
and suppress the other block arguments of
\(\Phi_{\boldsymbol\gamma}(V_s,W_s)\).
The eigenvalue bound makes
\((\det W_s)^{-L}\Phi_{\boldsymbol\gamma}(V_s,W_s)\)
an ordinary symmetric polynomial in the eigenvalues of
\(W_s^{-1}\).  Its Schur expansion and the determinant
shift \eqref{eq:characters-Laurent-expansion-appB-draft5}
therefore give, in terms of the Laurent characters
\eqref{eq:characters-Laurent-definition-appB-draft5}
\cite{bump2006averages},
\begin{equation}
    \Phi_{\boldsymbol\gamma}(V_s,W_s)
    =\sum_\beta c_\beta(V_s)\chi_\beta(W_s^{-1}),
    \qquad c_\beta(V_s)\neq0\ \Longrightarrow\ \beta_{r_s}\geq-L.
    \label{eq:projection-character-insertion-draft5}
\end{equation}
Indeed, a Schur label \(\eta\) becomes
\(\beta=\eta-L\boldsymbol1_{r_s}\), so
\(\beta_{r_s}=\eta_{r_s}-L\geq-L\).
The sum is finite, and the coefficients may depend on the
spectator holonomies.

With \(V_s\) and all spectator variables held fixed, the
projection identity for these insertions is
\begin{equation}
    \int[dW_s]\,
      \frac{(\det V_s)^{ks}}{(\det W_s)^{ks}}
      \mathcal C(V_s,W_s)\Phi_{\boldsymbol\gamma}(V_s,W_s)
    =\Phi_{\boldsymbol\gamma}(V_s,V_s),\qquad ks\geq L.
    \label{eq:projection-determinant-identity-draft4}
\end{equation}
To prove it, apply the determinant shift
\eqref{eq:characters-determinant-shift-appB-draft5} to
\eqref{eq:projection-Cauchy-kernel-draft4}.
Shifting
\(\alpha=\eta-ks\boldsymbol1_{r_s}\) gives
\begin{equation}
    \frac{(\det V_s)^{ks}}{(\det W_s)^{ks}}
    \mathcal C(V_s,W_s)
    =\sum_{\alpha:\,\alpha_{r_s}\geq-ks}
      \chi_\alpha(V_s^{-1})\chi_\alpha(W_s).
    \label{eq:projection-shifted-Cauchy-draft5}
\end{equation}
Normalized Haar orthogonality
\eqref{eq:characters-unitary-orthogonality-appB-draft5}
therefore reduces the left-hand side of
\eqref{eq:projection-determinant-identity-draft4} to
\begin{equation}
    \sum_{\beta:\,\beta_{r_s}\geq-ks}
      c_\beta(V_s)\chi_\beta(V_s^{-1})
    =\sum_\beta c_\beta(V_s)\chi_\beta(V_s^{-1})
    =\Phi_{\boldsymbol\gamma}(V_s,V_s),\qquad ks\geq L.
    \label{eq:projection-character-evaluation-draft5}
\end{equation}
The sums are finite, and the restriction is inactive because
every label in the insertion satisfies
\(\beta_{r_s}\geq-L\geq-ks\).
For the constant coefficient, \(\Phi_{\boldsymbol0}=1\),
so the projection has unit normalization.

Since \(s\geq1\) and \(k>L\),
\eqref{eq:projection-determinant-identity-draft4} applies to
every occupied block.  Each \(W_s\) integration evaluates
that argument at \(V_s\) without changing the exponent
bounds in the other blocks, so interblock couplings impose
no additional condition.  Restoring the longitudinal and
neutral factors gives
\begin{equation}
    \begin{aligned}
        [\widehat{\mathbf u}^{\,\boldsymbol\gamma}]
        \mathcal I^{\mathrm{ABJM}}_{N,k;\lambda}
        ={}&e^{-\delta q}
        [\widehat{\mathbf u}^{\,\boldsymbol\gamma}]
        \Bigg\{
            \mathcal I_{\mathrm{neutral}}^{(k)}
            \int\prod_{s:r_s>0}[dV_s]\,
            \exp\left[\sum_{m=1}^{\infty}
                \frac{\mathcal R_m(V,V)}{m}\right]
        \Bigg\},
        \\
        &|\boldsymbol\gamma|\leq L,\qquad
         N\geq L+q,\qquad k>L.
    \end{aligned}
    \label{eq:projection-diagonal-sector-integral-draft4}
\end{equation}
The neutral factor has nonnegative fugacity exponents, so its
inclusion requires only the coefficients already controlled
through degree \(L\).  The remaining task is to identify the
diagonal kernel \(\mathcal R_m(V,V)\), which we do in
section~\ref{subsubsec:BMN-kernel-identification-draft5}.

\subsubsection{Identification of the BMN kernel}
\label{subsubsec:BMN-kernel-identification-draft5}

Evaluating \(W_s=V_s\) in
\eqref{eq:projection-A-independent-remainder-draft5} gives
\begin{equation}
    \mathcal R_m(V,V)
    =\sum_{s,t\geq1}
      \left[\delta_{st}
            +H_{st}^{(m)}(g_m^++g_m^--2)\right]
      \operatorname{Tr}V_s^{-m}\operatorname{Tr}V_t^m.
    \label{eq:projection-diagonal-action-draft4}
\end{equation}
The reference subtraction removes one of the two diagonal
terms in \eqref{eq:stabilization-effective-charged-action-draft4},
leaving a single \(\delta_{st}\).  The shifted letter functions
\eqref{eq:selection-relative-letter-functions-draft4} satisfy
\begin{equation}
    g_m^++g_m^--2
    =-\frac{\prod_{a=1}^3(1-\widehat u_a^m)}
            {1-\widehat{\mathfrak u}^{\,m}}.
    \label{eq:projection-matter-combination-draft4}
\end{equation}
The subtraction in \(H_{st}^{(m)}\), generated by the
zero-flux integral, truncates the derivative denominator to
a finite sum:
\begin{equation}
    \frac{H_{st}^{(m)}}{1-\widehat{\mathfrak u}^{\,m}}
    =\sum_{n=0}^{\min(s,t)-1}
       \widehat{\mathfrak u}^{\,m(|s-t|/2+n)}
    =\sum_{j=\frac12|s-t|}^{\frac12(s+t)-1}
       \widehat{\mathfrak u}^{\,jm}.
    \label{eq:projection-finite-harmonic-tower-draft4}
\end{equation}
The final sum runs in integer steps and coincides with the
matrix-harmonic range
\eqref{eq:BMN-fuzzy-harmonic-range-review-draft4} between
irreducible blocks of dimensions \(s\) and \(t\).

To use the spin-refinement convention of
\eqref{eq:BMN-block-kernel-review-draft4}, make the
Haar-invariant change of variables \(V_s\mapsto(-1)^sV_s\).
Each trace product acquires \((-1)^{m(s+t)}\).  Since
\(2j\equiv s+t\pmod2\) throughout
\eqref{eq:projection-finite-harmonic-tower-draft4}, the
kernel becomes
\begin{equation}
    \widehat\iota_{st}^{(m)}
    =\delta_{st}
     +\sum_{j=\frac12|s-t|}^{\frac12(s+t)-1}
       (-1)^{2jm+1}\widehat{\mathfrak u}^{\,jm}
       \prod_{a=1}^3(1-\widehat u_a^m).
    \label{eq:projection-BMN-kernel-draft4}
\end{equation}
This is the BMN kernel at \(\widehat u_a\), including the
\(m\)-dependent sign for half-integer spins
\cite{Chang:2024lkw}.  In particular, the constant part of
the \(s=t\), \(j=0\) term cancels \(\delta_{st}\), as in
the normalized BMN matrix integral
\eqref{eq:BMN-vacuum-matrix-integral-review-draft4}.
The remaining \(V_s\) integral in
\eqref{eq:projection-diagonal-sector-integral-draft4} is
therefore \(\mathcal I^{\mathrm{BMN}}_{q,\lambda}\)
evaluated at
\((\Delta_1+\sigma,\Delta_2+\sigma,\Delta_3-\sigma)\).
The same multiplicities \(r_s\) specify the matched flux
sector and the BMN vacuum.

The bounds in
\eqref{eq:projection-diagonal-sector-integral-draft4}
are uniform over the finitely many partitions of fixed
\(q\).  Summing their contributions using
\eqref{eq:selection-fixed-coefficient-sum-draft4} gives
the factorization of the unreduced fixed-charge index.
The \(q=0\) case follows from
\eqref{eq:stabilization-neutral-sector-identity-draft4},
with the empty-vacuum convention
\(\mathcal I^{\mathrm{BMN}}_0=1\).
Writing the BMN index as a function of its fugacities,
we obtain
\begin{equation}
    \begin{gathered}
        [\widehat{\mathbf u}^{\,\boldsymbol\gamma}]
        \mathcal I^{\mathrm{ABJM}}_{N,k;q}
        =e^{-\delta q}
         [\widehat{\mathbf u}^{\,\boldsymbol\gamma}]
         \left\{\mathcal I_{\mathrm{neutral}}^{(k)}
                \mathcal I_q^{\mathrm{BMN}}(\widehat{\mathbf u})\right\},
        \\
        \boldsymbol\gamma\in\mathbb Z_{\geq0}^3,
        \qquad |\boldsymbol\gamma|\leq L,\qquad
        N\geq L+q,\qquad k>L.
    \end{gathered}
    \label{eq:projection-finite-degree-index-relation-draft5}
\end{equation}

For every fixed \(q\geq0\) and \(L\), the displayed
inequalities hold eventually along the double scale limit
\(N,k\to\infty\), \(N/k^2\to\nu\in(0,\infty)\).
At fixed \(\delta\), \(\sigma=\delta/k\to0\), so
\(\widehat u_a\to u_a\) and the coefficient conversion
following \eqref{eq:localization-shifted-fugacities-draft4}
tends to unity.  The neutral factor tends coefficientwise
to \(\mathcal I_{\mathrm{neutral}}\), and each coefficient
of its product with the BMN index involves only a finite
convolution.  Thus
\eqref{eq:projection-finite-degree-index-relation-draft5}
implies
\begin{equation}
    \lim_{\substack{N,k\to\infty\\N/k^2\to\nu}}^{\mathrm{coeff}}
    \mathcal I^{\mathrm{ABJM}}_{N,k;q}
        (\delta;\boldsymbol\Delta)
    =e^{-\delta q}\,
     \mathcal I_{\mathrm{neutral}}(\boldsymbol\Delta)\,
     \mathcal I_q^{\mathrm{BMN}}(\boldsymbol\Delta),
    \qquad q\geq0.
    \label{eq:projection-fixed-charge-limit-draft5}
\end{equation}
The same reasoning applies before summing over \(\lambda\),
so the factorization also holds for each matched vacuum
in \eqref{eq:projection-diagonal-sector-integral-draft4}.

We now assemble the grand canonical ensemble of BMN rank.
Negative total charge is absent through degree \(L\) when
\(k>L\): \eqref{eq:selection-total-matter-charge-draft4}
gives \(M=P-kq\geq k|q|\) for \(q<0\), which violates
\eqref{eq:selection-total-degree-bound-draft4}.
Let \(\mathcal I^{\mathrm{ABJM}}_{N,k}
(\delta;\boldsymbol\Delta)\) denote the full index
\eqref{eq:ABJM-index-draft4} evaluated with
\eqref{eq:Penrose-fugacity-map-draft4} and
\(\sigma=\delta/k\).  Collecting the fixed-charge results
\eqref{eq:projection-fixed-charge-limit-draft5} gives
\begin{equation}
    \lim_{\substack{N,k\to\infty\\N/k^2\to\nu}}^{\mathrm{coeff}}
    \mathcal I^{\mathrm{ABJM}}_{N,k}
        (\delta;\boldsymbol\Delta)
    =\mathcal I_{\mathrm{neutral}}(\boldsymbol\Delta)
     \sum_{q=0}^{\infty}e^{-\delta q}
        \mathcal I_q^{\mathrm{BMN}}(\boldsymbol\Delta).
    \label{eq:projection-final-index-relation-draft4}
\end{equation}
The sum on the right-hand side is the grand canonical
ensemble of BMN rank, with chemical potential \(\delta\)
conjugate to the matrix rank and longitudinal momentum
\(q\).  The equality is coefficientwise in the transverse
fugacities and is understood as a formal series in
\(e^{-\delta}\); the
estimates apply to every finite truncation in charge and
transverse degree.
This is precisely the relation stated in
\eqref{eq:intro-index-relation-draft4}, with the BMN index
expanded as a sum over its vacuum partitions.

\paragraph{Interpretation of the neutral factor.}
At fixed \(q\), the index factorization in
\eqref{eq:projection-fixed-charge-limit-draft5} suggests the
schematic BPS-state relation
\[
    \lim_{\mathrm{double\ scale}}
    \mathcal H^{\mathrm{BPS}}_{\mathrm{ABJM},q}
    \stackrel{?}{\simeq}
    \mathcal H^{\mathrm{BPS}}_{\mathrm{BMN},q}
    \otimes\mathcal H_{\mathrm{neutral}}.
\]
Equation \eqref{eq:projection-fixed-charge-limit-draft5}
is an equality of protected generating functions.  It
suggests a physical picture in which a charged BMN cluster
is accompanied by protected neutral spectators.  The
following harmonic calculation illustrates this picture.

An analogous distinction appears in the original BMN
correspondence between \(\mathcal N=4\) super Yang--Mills
theory and strings on a pp-wave \cite{Sadri:2003pr}.
A double-trace BPS operator has the schematic form
\begin{equation}
    \mathcal T^{J,r}
    \propto
    \operatorname{Tr}Z^{J_1}\operatorname{Tr}Z^{J-J_1},
    \qquad
    r=\frac{J_1}{J},
    \qquad
    (p_1^+,p_2^+)=(r,1-r)p^+.
    \label{eq:neutral-BMN-multitrace-draft8}
\end{equation}
Both traces describe finite-momentum strings when
\(0<r<1\) and the total \(p^+\) are held fixed as
\(J\to\infty\).  If \(J_1\) instead remains
finite, then \(r\to0\).  The first trace remains a valid BPS
operator in the parent theory, but it carries a vanishing
fraction of the longitudinal momentum and does not define a
separate finite-\(p^+\) state in the local pp-wave sector.

We now make the localization near the null geodesic more
explicit.  At fixed positive \(q\) and fixed transverse
charges,
\eqref{eq:energy-charge-dictionary-draft4} and
\eqref{eq:null-radius-draft4} give
\begin{equation}
    J_4=kq\left[1+O\!\left(\frac1k\right)\right],
    \qquad
    \frac{J_4}{R^2}
    \longrightarrow
    \frac{\mu q}{6R_-}>0.
    \label{eq:neutral-large-J-scaling-draft8}
\end{equation}
To illustrate angular localization near the geodesic,
consider scalar harmonics on the covering seven-sphere
\cite{Meremianin2009}.  For \(J\in\mathbb Z_{\geq0}\),
define
\begin{equation}
    \begin{gathered}
        \Psi_{J,\ell,n}
        =\mathcal N_{J,\ell,n}\,
         e^{iJ\chi}Y_\ell(u)
         (\cos\vartheta)^J(\sin\vartheta)^\ell
         P_n^{(\ell+2,J)}(\cos2\vartheta),
        \\
        1=\int_{S^7}d\Omega_7\,|\Psi_{J,\ell,n}|^2,
        \qquad
        \mathcal N_{J,\ell,n}^{-2}
        =2\pi\int_0^{\pi/2}
          W_{J,\ell,n}(\vartheta)d\vartheta,
        \\
        W_{J,\ell,n}(\vartheta)
        =\cos^{2J+1}\!\vartheta\,
         \sin^{2\ell+5}\!\vartheta\,
         \left[P_n^{(\ell+2,J)}(\cos2\vartheta)\right]^2.
    \end{gathered}
    \label{eq:neutral-s7-harmonics-draft8}
\end{equation}
Here \(Y_\ell(u)\) is a unit-normalized scalar spherical
harmonic on \(S^5\), satisfying
\(-\nabla_{S^5}^2Y_\ell=\ell(\ell+4)Y_\ell\), and
\(P_n^{(\ell+2,J)}\) is a Jacobi polynomial.  Moreover,
\(J=-i\partial_\chi\), while \(\ell\) and \(n\) are the
\(S^5\) angular momentum and radial excitation number.
For \(J\to\infty\) with \(\ell,n\) fixed, set
\(x=\sqrt J\,\vartheta\).  The radial weight then satisfies
\begin{equation}
    J^{\ell+3}
    W_{J,\ell,n}\!\left(\frac{x}{\sqrt J}\right)
    \frac{dx}{\sqrt J}
    \longrightarrow
    x^{2\ell+5}e^{-x^2}
    \left[L_n^{(\ell+2)}(x^2)\right]^2dx.
    \label{eq:neutral-large-J-harmonic-limit-draft8}
\end{equation}
Here \(L_n^{(\ell+2)}\) is a generalized Laguerre
polynomial.
The angular width scales as \(J^{-1/2}\), in agreement with the
large-angular-momentum width estimate of
\cite{Kovacs:2013una}.  In the charged sector we set
\(J=J_4\).  Equation
\eqref{eq:neutral-large-J-scaling-draft8} then gives
\(\vartheta=O(1/\sqrt{k})\).  Since \(R\) grows as
\(\sqrt{k}\), \(z=R\vartheta=O(1)\), so the state
remains at finite Penrose distance.

This statement can be expressed directly in terms of the
local norm.  Let
\(\mathcal N_\gamma(C)=\{0\leq\vartheta\leq C/R\}\)
be a fixed neighborhood in the Penrose coordinate
\(z=R\vartheta\).  With
\(\kappa_q=\mu q/(6R_-)\), the exact finite-\(J_4\) norm
and its double scale limit are
\begin{equation}
    \begin{aligned}
        \int_{\mathcal N_\gamma(C)}d\Omega_7\,
            |\Psi_{J_4,\ell,n}|^2
        &=\frac{\displaystyle\int_0^{C/R}
            W_{J_4,\ell,n}(\vartheta)d\vartheta}
           {\displaystyle\int_0^{\pi/2}
            W_{J_4,\ell,n}(\vartheta)d\vartheta},
        \\
        \lim_{\substack{N,k\to\infty\\N/k^2\to\nu}}
        \int_{\mathcal N_\gamma(C)}d\Omega_7\,
            |\Psi_{J_4,\ell,n}|^2
        &=\frac{
        \displaystyle\int_0^{C\sqrt{\kappa_q}}
            x^{2\ell+5}e^{-x^2}
            \left[L_n^{(\ell+2)}(x^2)\right]^2dx}{
        \displaystyle\int_0^\infty
            x^{2\ell+5}e^{-x^2}
            \left[L_n^{(\ell+2)}(x^2)\right]^2dx}.
    \end{aligned}
    \label{eq:neutral-charged-local-norm-draft8}
\end{equation}
For every finite \(C>0\), the right-hand side is a finite
number between zero and one.  It approaches one as
\(C\to\infty\).

The neutral sector behaves differently.  The spectator
itself has zero charge, while the full state remains in the
fixed positive-\(q\) sector.  Its angular momentum
\(J_0\equiv|J_4|=O(1)\), so there is no
large-angular-momentum confinement.  Use the same normalized
harmonic in
\eqref{eq:neutral-s7-harmonics-draft8}, with \(J=J_0\)
fixed.  To express its norm in the same Penrose coordinate,
set \(z=R\vartheta\) in both radial integrals.  Then
\begin{equation}
    \int_{\mathcal N_\gamma(C)}d\Omega_7\,
        |\Psi_{J_0,\ell,n}|^2
    =\frac{\displaystyle\int_0^C
            W_{J_0,\ell,n}(z/R)dz}
           {\displaystyle\int_0^{\pi R/2}
            W_{J_0,\ell,n}(z/R)dz}
    =O\!\left(\frac1{k^{\ell+3}}\right)
    \longrightarrow0.
    \label{eq:neutral-local-norm-vanishes-draft8}
\end{equation}
Indeed, the numerator samples only fixed \(z\), where
\(W_{J_0,\ell,n}(z/R)=O(1/k^{\ell+5/2})\), using
\(R\propto\sqrt{k}\).  The denominator is \(R\) times
the fixed normalization integral in
\eqref{eq:neutral-s7-harmonics-draft8}, giving the stated
power of \(1/k\).
Thus the neutral harmonic has vanishing norm in every fixed
region \(z\leq C\).  Its norm lies outside such regions
as \(R\to\infty\).

The factorization can therefore be pictured as a charged
cluster accompanied by a neutral spectator,
\begin{equation}
    |\Psi_q\rangle
    \sim
    |\psi_q\rangle_{\mathrm{BMN}}
    \otimes
    |\chi\rangle_{\mathrm{neutral}},
    \qquad q_{\mathrm{tot}}=q+0.
    \label{eq:neutral-cluster-state-draft8}
\end{equation}
Fixing the total monopole charge allows such neutral
spectators.  The angular localization calculation supports
interpreting \(\mathcal I_{\mathrm{neutral}}\) as their
protected contribution, which is absent from the local
Penrose sector.  Dividing by this factor isolates the BMN
index.  The index identity does not establish a tensor-product
decomposition of the interacting Hilbert space.  A quantum
derivation of this cluster picture, including operator mixing
and interactions, is left for future work.

\section{Discussion}
\label{sec:discussion-draft4}

The double scale limit connects the protected indices of
ABJM theory and the finite-rank BMN model, and relates their
half-BPS vacuum data through bubbling geometries.  The integer
\(q\) is the total ABJM monopole charge, the number of
longitudinal momentum units \(p^+R_-\) in the limit, and the
BMN matrix rank.  A partition \(\lambda\vdash q\) labels
the surviving ABJM flux sector, the finite conducting disks
and the BMN vacuum.  The geometry identifies these vacuum
data, while the index calculation compares protected
excitations summed over all BMN vacua.

On the gravity side, the starting point is the
Donos--Sim\'on covering geometry together with its diagonal
\(\mathbb Z_k\) quotient.  At fixed \(q\), the endpoint
disk carries the growing M2 flux and approaches an infinite
conducting plane, while the remaining disks retain finite
rescaled positions and charges.  The plane produces the LM
cubic background, and its response to the finite disks
produces the required image charges.  Taking the same limit
of the metric, three-form and circle period gives the
normalized eleven-dimensional LM uplift.  The inherited M2
and M5 fluxes become the D2 and NS5 integers that determine
the BMN vacuum partition.  This derives the LM boundary
problem and its normalization from an asymptotically
\(\mathrm{AdS}_4\times S^7/\mathbb Z_k\) parent.

The index calculation starts instead from the exact ABJM
localization formula at finite \(N\) and \(k\).  Fix a total
charge \(q\) and a maximal transverse degree \(L\).  The
holonomy constraints first show that the projected index has
nonnegative transverse support.  For \(k>L\), sectors with
negative fluxes or unequal flux distributions do not
contribute through degree \(L\).  For \(N\geq q+L\), the
zero-flux Haar moments stabilize and produce the universal
neutral factor.  The remaining Chern--Simons projection then
reduces each occupied flux block to the BMN matrix integral
for the same partition.  After the shifts
\(\widehat u_a\to u_a\) vanish, this proves the fixed-charge
factorization \eqref{eq:projection-fixed-charge-limit-draft5}
coefficientwise in the transverse fugacities.  No
supergravity approximation enters this argument.  The
covering-space harmonic calculation supports interpreting
the neutral factor as a spectator contribution: fixed-level
neutral harmonics have vanishing norm in every fixed region
\(z=R\vartheta\leq C\), whereas modes with \(J_4\sim R^2\)
retain a nonzero limiting norm.  Dividing by the limiting \(q=0\)
component isolates the BMN index; a quantum derivation of
the spectator picture remains open.

\paragraph{Scope and limitations.}
The geometric and index results have different domains of
validity.  The geometry analysis concerns classical fields,
and a reliable supergravity description depends on the flux
regime.  At finite \(N,k\),
\eqref{eq:projection-finite-degree-index-relation-draft5}
is exact through degree \(L\) in the shifted fugacities
\(\widehat u_a\) when \(N\geq q+L\) and \(k>L\).
In the physical fugacities \(u_a\), each coefficient retains
the factor \(e^{-\delta(L_1+L_2-L_3)/k}\); the limiting
identity follows as this factor tends to one.
The parameter \(\nu=N/k^2\) fixes the null-circle
radius and the dimensionless BMN coupling through
\eqref{eq:intro-DLCQ-parameters-draft6}, but it drops out of
the protected index.  Coupling-dependent observables are
therefore required to test the interacting dynamics.
Convergence as a function of the fugacities is a stronger
question.  The chemical-potential condition
\eqref{eq:safe-Penrose-chamber-draft4} suppresses every fixed
monomial, but it does not bound the growth of the signed
coefficients uniformly in \(N\), \(k\) and transverse
degree.  A uniform summable bound would justify passing the
limit through the full fugacity sum.  The same
issue arises if the BMN rank or the transverse degree grows
with the parent parameters.  These limits are not addressed
in the present paper.

\paragraph{Open directions.}
The present results suggest three complementary extensions:
\begin{itemize}
    \item \emph{Gravitational index.}
    A gravitational derivation of the protected index
    relation would require supersymmetric saddles with the
    same charge and chemical potentials.  Their compact
    circle identification and one-loop determinants must
    also be fixed.  The neutral factor should be associated
    with protected modes that are not confined to the local
    Penrose region.  Making this statement precise requires
    a finite-radius treatment of boundary conditions, zero
    modes and renormalized actions; the limiting classical
    fields alone do not determine these data.

    \item \emph{Instanton effects.}
    BMN instantons connecting matrix vacua are known
    \cite{Yee:2003ge}.  Their fermionic zero modes rule out
    direct mixing of supersymmetric ground states but allow
    higher-order processes with suitable insertions.  The
    corresponding ABJM problem is to construct configurations
    on \(S^2\times\mathbb R\) that connect dressed monopole
    backgrounds with the same total charge and different
    partitions \cite{Kovacs:2013una}.  On the gravity side,
    these processes are represented by Euclidean branes
    wrapping cycles of the bubbling geometry, with their
    actions related to the electrostatic data
    \cite{Lin:2006tr}.  Comparing their actions, zero modes
    and selection rules would extend the vacuum dictionary
    to dynamical processes.

    \item \emph{Matrix-model amplitudes.}
    Amplitudes contain information that an index cannot
    determine.  In flat-space Matrix theory, the finite-rank
    three-graviton calculation of Herderschee and Maldacena
    provides an example of such a test
    \cite{Herderschee:2023pza}.  Finding analogous BMN
    observables and deriving their double scale limit from
    ABJM would determine whether the agreement established
    here for vacua and protected coefficients extends to the
    full finite-rank dynamics.
\end{itemize}

\section*{Acknowledgements}

We thank Kimyeong Lee for useful discussions. This work is supported by the Peking University startup Grant No. 7101303985.

\appendix

\section{Review of the ABJM and BMN indices}
\label{app:ABJM-BMN-index-review}

\subsection{ABJM theory and its superconformal index}
\label{app:ABJM-index-review}

We collect the field and charge conventions needed for the finite-rank
localization formula.  ABJM theory is an
\(\mathcal N=6\) Chern--Simons--matter theory with gauge group and levels
\cite{Aharony:2008ug}
\begin{equation}
    U(N)_k\times U(N)_{-k}.
    \label{eq:ABJM-gauge-group-review-draft4}
\end{equation}
In \(\mathcal N=2\) language, it contains two vector multiplets and four
chiral multiplets
\begin{equation}
    \mathscr A_a\in(\boldsymbol N,\overline{\boldsymbol N}),
    \qquad
    \mathscr B_{\dot a}\in(\overline{\boldsymbol N},\boldsymbol N),
    \qquad
    a,\dot a=1,2,
    \label{eq:ABJM-chiral-multiplets-review-draft4}
\end{equation}
coupled by the quartic superpotential fixed by \(\mathcal N=6\)
supersymmetry.  The scalar fields may be assembled into the fundamental of
\(SU(4)_R\),
\begin{equation}
    Y^A=(A_1,A_2,B^{\dagger\dot 1},B^{\dagger\dot 2}),
    \qquad
    Y^A\mapsto UY^A\widetilde U^\dagger,
    \qquad
    A=1,\ldots,4.
    \label{eq:ABJM-Y-basis-review-draft4}
\end{equation}
We use three Cartans \(h_1,h_2,h_3\) of \(SU(4)_R\), together with a
baryon-like charge \(h_4\) commuting with the \(OSp(6|4)\) algebra.  Their
action on the four scalars is
\begin{equation}
\begin{array}{c|rrrr}
     & h_1 & h_2 & h_3 & h_4 \\ \hline
    Y^1 & \frac12 & \frac12 & -\frac12 & \frac12 \\
    Y^2 & -\frac12 & -\frac12 & -\frac12 & \frac12 \\
    Y^3 & -\frac12 & \frac12 & \frac12 & \frac12 \\
    Y^4 & \frac12 & -\frac12 & \frac12 & \frac12
\end{array}
    \label{eq:ABJM-scalar-charge-table-draft4}
\end{equation}
This basis is the one used in
\eqref{eq:ABJM-geometric-Cartans-draft4}.  In particular, the choice of
geodesic made in section~\ref{sec:penrose-index-dictionary} singles out
\(Y^4\).

The index is defined in radial quantization on \(S^2\times\mathbb R\).  Let
\(E\) be the radial Hamiltonian and \(j_3\) the Cartan generator of spatial
rotations.  We choose
\begin{equation}
    \mathcal Q=Q_{34,-},
    \qquad
    \mathcal S=\mathcal Q^{\dagger_{\mathrm{rad}}}=S^{34,-},
    \qquad
    \{\mathcal Q,\mathcal S\}=E-h_3-j_3
    \equiv\Delta_{\mathcal Q}.
    \label{eq:ABJM-supercharge-review-draft4}
\end{equation}
In the order \((E,j_3;h_1,h_2,h_3,h_4)\), the supercharge has charges
\((\tfrac12,-\tfrac12;0,0,1,0)\).  Thus \(E+j_3,h_1,h_2,h_4\) commute
with both \(\mathcal Q\) and \(\mathcal S\).  The superconformal index is
\begin{equation}
    \mathcal I^{\mathrm{ABJM}}_{N,k}(x,y_1,y_2,y_3)
    =
    \operatorname{Tr}_{\mathcal H_{\mathrm{ABJM}}}
    \left[
        (-1)^F e^{-\beta\Delta_{\mathcal Q}}
        x^{E+j_3}y_1^{h_1}y_2^{h_2}y_3^{h_4}
    \right].
    \label{eq:ABJM-index-review-draft4}
\end{equation}
The trace is over the gauge-invariant radial Hilbert space on \(S^2\),
including all monopole sectors.  Positivity of
\(\Delta_{\mathcal Q}=\{\mathcal Q,\mathcal S\}\) and supersymmetric pairing
restrict its contributions to \(\Delta_{\mathcal Q}=0\), making the result
independent of \(\beta\) \cite{Bhattacharya:2008zy,Kim:2009wb}.

Localization reduces \eqref{eq:ABJM-index-review-draft4} to a sum over GNO
fluxes and an integral over holonomies.  We normalize them by
\begin{equation}
    \frac{1}{2\pi}\int_{S^2}F
    =\operatorname{diag}(n_1,\ldots,n_N),
    \qquad
    \frac{1}{2\pi}\int_{S^2}\widetilde F
    =\operatorname{diag}(\widetilde n_1,\ldots,\widetilde n_N),
    \label{eq:ABJM-GNO-fluxes-review-draft4}
\end{equation}
where \(n_i,\widetilde n_i\in\mathbb Z\), and denote the corresponding
holonomies by \(\alpha_i,\widetilde\alpha_i\in[0,2\pi)\).  Integration over
the diagonal gauge holonomy imposes
\begin{equation}
    \sum_{i=1}^N n_i
    =
    \sum_{i=1}^N\widetilde n_i
    \equiv q.
    \label{eq:ABJM-total-flux-review-draft4}
\end{equation}
The Chern--Simons Gauss law then relates the baryon-like charge to the total
flux,
\begin{equation}
    h_4=\frac{J_{\mathrm H}}{2}=\frac{kq}{2}.
    \label{eq:ABJM-h4-flux-review-draft4}
\end{equation}
The full index sums over all \(q\in\mathbb Z\); the index
comparison in the double scale limit uses its \(q>0\)
components and the neutral component.  Even at fixed positive
\(q\), the individual GNO charges may have either sign.

For later use, define the zero-point contribution
\begin{equation}
    E_0(\boldsymbol n,\widetilde{\boldsymbol n})
    =
    \sum_{i,j=1}^N|n_i-\widetilde n_j|
    -\sum_{i<j}|n_i-n_j|
    -\sum_{i<j}|\widetilde n_i-\widetilde n_j|.
    \label{eq:ABJM-zero-point-energy-review-draft4}
\end{equation}
Using \eqref{eq:ABJM-h4-flux-review-draft4}, we extract all \(y_3\)
dependence as the sector weight \(y_3^{kq/2}\).  The remaining
bifundamental single-letter functions are
\begin{equation}
\begin{aligned}
    f^+(x,y_1,y_2)
    &=\frac{
        x^{1/2}\left(\sqrt{y_1/y_2}+\sqrt{y_2/y_1}\right)
        -x^{3/2}\left(\sqrt{y_1y_2}+1/\sqrt{y_1y_2}\right)
       }{1-x^2},
    \\
    f^-(x,y_1,y_2)
    &=\frac{
        x^{1/2}\left(\sqrt{y_1y_2}+1/\sqrt{y_1y_2}\right)
        -x^{3/2}\left(\sqrt{y_1/y_2}+\sqrt{y_2/y_1}\right)
       }{1-x^2}.
\end{aligned}
    \label{eq:ABJM-matter-letters-review-draft4}
\end{equation}
The positive terms count \(B^{\dagger\dot a}\) in \(f^+\) and
\(A^{\dagger a}\) in \(f^-\), while the negative terms count BPS fermions.
The denominator \(1-x^2\) sums the derivatives preserving the chosen BPS
condition.
In a general flux sector they become
\begin{equation}
    f^\pm_{ij}(x,y_1,y_2)
    =x^{|n_i-\widetilde n_j|}f^\pm(x,y_1,y_2).
    \label{eq:ABJM-fluxed-matter-letters-review-draft4}
\end{equation}
The vector multiplets contribute
\begin{equation}
    f^{\mathrm{adj}}_{ij}(x)
    =-\left(1-\delta_{n_i n_j}\right)x^{|n_i-n_j|},
    \qquad
    \widetilde f^{\mathrm{adj}}_{ij}(x)
    =-\left(1-\delta_{\widetilde n_i\widetilde n_j}\right)
      x^{|\widetilde n_i-\widetilde n_j|}.
    \label{eq:ABJM-vector-letters-review-draft4}
\end{equation}
The equal-flux adjoint factors are included in the residual Haar measure
below and are excluded from these letter functions to avoid double counting.
The factors \(x^{|n_i-\widetilde n_j|}\) and their adjoint analogues encode
the lowest monopole-harmonic angular momentum.  They will be essential in
selecting the flux sectors that survive at fixed transverse degree.

Let \(W_{\boldsymbol n}\subset S_N\) be the stabilizer of the flux vector
\(\boldsymbol n\); its order is the product of the factorials of the
multiplicities of each flux value.  We include the residual Weyl factor and
the Vandermonde determinant of the unbroken gauge group in the measure
\begin{equation}
    [d\boldsymbol\alpha]_{\boldsymbol n}
    \equiv
    \frac{1}{|W_{\boldsymbol n}|}
    \prod_{i=1}^N\frac{d\alpha_i}{2\pi}
    \prod_{\substack{i<j\\ n_i=n_j}}
    \left(2\sin\frac{\alpha_i-\alpha_j}{2}\right)^2,
    \label{eq:ABJM-holonomy-measure-review-draft4}
\end{equation}
with an identical definition for
\([d\widetilde{\boldsymbol\alpha}]_{\widetilde{\boldsymbol n}}\).
This is the eigenvalue form of the normalized Haar measure of the unbroken
gauge group, as in the localization prescription of \cite{Kim:2009wb};
for background on Lie groups and their representations, see
\cite{Hall2013}.  Each flux vector in the sum is represented once up
to its own Weyl group.
The full finite-\(N\), finite-\(k\) result is \cite{Kim:2009wb}
\begin{equation}
\begin{aligned}
    \mathcal I^{\mathrm{ABJM}}_{N,k}(x,y_1,y_2,y_3)
    ={}&
    \sum_{\substack{
        \boldsymbol n,\widetilde{\boldsymbol n}\in\mathbb Z^N/S_N\\
        \sum_i n_i=\sum_i\widetilde n_i}}
    x^{E_0(\boldsymbol n,\widetilde{\boldsymbol n})}
    y_3^{\frac{k}{2}\sum_i n_i}
    \int[d\boldsymbol\alpha]_{\boldsymbol n}
         [d\widetilde{\boldsymbol\alpha}]_{\widetilde{\boldsymbol n}}
    \\
    &\times
    \exp\left[
        ik\sum_{i=1}^N
        \left(n_i\alpha_i-\widetilde n_i\widetilde\alpha_i\right)
    \right]
    \exp\left[
        \sum_{m=1}^{\infty}\frac{1}{m}\,
        \mathcal F_m
    \right],
    \label{eq:ABJM-localization-formula-review-draft4}
\end{aligned}
\end{equation}
where
\begin{equation}
\begin{aligned}
    \mathcal F_m
    =\sum_{i,j=1}^N\Big[{}
       &f^+_{ij}(x^m,y_1^m,y_2^m)
        e^{im(\widetilde\alpha_j-\alpha_i)}
       +f^-_{ij}(x^m,y_1^m,y_2^m)
        e^{im(\alpha_i-\widetilde\alpha_j)}
    \\
       &+f^{\mathrm{adj}}_{ij}(x^m)
        e^{-im(\alpha_i-\alpha_j)}
       +\widetilde f^{\mathrm{adj}}_{ij}(x^m)
        e^{-im(\widetilde\alpha_i-\widetilde\alpha_j)}
    \Big].
    \label{eq:ABJM-plethystic-kernel-review-draft4}
\end{aligned}
\end{equation}
Here \(m\) is the plethystic summation index, and the fluxes are held fixed
when the fugacities are raised to their \(m\)-th powers.  The holonomy
integrals impose the gauge-singlet constraints,
while the Chern--Simons phase supplies the electric charge that must be
screened by bifundamental letters.  The fixed-\(q\) quantity used in the main
text is obtained by restricting the sum in
\eqref{eq:ABJM-localization-formula-review-draft4} to
\eqref{eq:ABJM-total-flux-review-draft4}, before applying the fugacity map
\eqref{eq:Penrose-fugacity-map-draft4}.

\subsection{The BMN matrix model and its index}
\label{app:BMN-index-review}

The BMN matrix model is a mass deformation of matrix quantum mechanics
with sixteen dynamical supercharges and is proposed to describe M-theory on
the compact eleven-dimensional pp-wave \cite{Berenstein:2002jq}.
At \(q\) units of longitudinal momentum, \(p^+=q/R_-\), its gauge group is
\(U(q)\).  Its fields are nine Hermitian matrices \(X^I(x^+)\), a gauge
field \(A_+(x^+)\), and sixteen real adjoint fermions.  We split
\begin{equation}
    X^I=(X^i,X^a),
    \qquad
    i=1,2,3,
    \qquad
    a=4,\ldots,9,
    \label{eq:BMN-matrix-splitting-review-draft4}
\end{equation}
according to the manifest \(SO(3)\times SO(6)\) symmetry.  The triplet and
sextet have masses \(\mu/3\) and \(\mu/6\), respectively.  We use
gauge-theory normalization for the matrices: the bosonic kinetic term is
\((2g_{\mathrm B}^2)^{-1}\operatorname{Tr}(D_+X^I)^2\), and the potential
energy is \(g_{\mathrm B}^{-2}V_{\mathrm B}\), where
\begin{equation}
\begin{aligned}
    V_{\mathrm B}
    =\operatorname{Tr}\Bigg[{}
       &\frac12\left(
          \frac{\mu}{3}X^i
          +\frac{i}{2}\epsilon^{ijk}[X^j,X^k]
        \right)^2
        +\frac12\left(\frac{\mu}{6}\right)^2X^aX^a
    \\
       &-\frac12[X^i,X^a]^2
        -\frac14[X^a,X^b]^2
    \Bigg].
    \label{eq:BMN-bosonic-potential-review-draft4}
\end{aligned}
\end{equation}
Repeated indices are summed, \(\epsilon^{123}=1\), and
\(D_+X^I=\partial_{x^+}X^I-i[A_+,X^I]\).  For Hermitian matrices all
terms in the traced potential are nonnegative.  The fermion mass and Yukawa
couplings complete the action with sixteen supersymmetries, and variation
of \(A_+\) imposes Gauss' law.  These matrices have dimensions of inverse
length and should not be identified directly with the geometric transverse
coordinates in \eqref{eq:maximal-pp-wave-draft4}.

The physical light-cone Hamiltonian is \(p^-\).  As in
\eqref{eq:BMN-Hamiltonian-draft4}, we measure it in units of \(\mu/6\),
\begin{equation}
    H_{\mathrm{BMN}}=\frac{6}{\mu}p^-.
    \label{eq:BMN-dimensionless-Hamiltonian-review-draft4}
\end{equation}
Let \(M^{12}\) be the Cartan of \(SO(3)\), and let
\(M^{45},M^{67},M^{89}\) be a Cartan basis of \(SO(6)\), with orientations
fixed by \eqref{eq:BMN-Cartans-draft4}.  In the \(SU(2|4)\) superalgebra
we select \(\mathcal Q_{\mathrm B}=Q^4_-\) \cite{Chang:2024lkw}.  Its
charges under
\((H_{\mathrm{BMN}},M^{12},M^{45},M^{67},M^{89})\) are
\((\tfrac12,-\tfrac12,\tfrac12,\tfrac12,\tfrac12)\), and its
anticommutator with its conjugate is proportional to the positive operator
\begin{equation}
    \Delta_{\mathrm{BMN}}
    =H_{\mathrm{BMN}}
    -2M^{12}-M^{45}-M^{67}-M^{89}
    \geq0.
    \label{eq:BMN-positive-operator-review-draft4}
\end{equation}
The three independent commuting refinements are
\begin{equation}
    \ell_1=M^{12}+M^{45},
    \qquad
    \ell_2=M^{12}+M^{67},
    \qquad
    \ell_3=M^{12}+M^{89}.
    \label{eq:BMN-refinements-review-draft4}
\end{equation}

The supersymmetric vacua are obtained by setting each term in
\eqref{eq:BMN-bosonic-potential-review-draft4} to zero.  They satisfy
\begin{equation}
    X^a=0,
    \qquad
    X^i=\frac{\mu}{3}L^i_{\lambda},
    \qquad
    [L^i_{\lambda},L^j_{\lambda}]
    =i\epsilon^{ijk}L^k_{\lambda}.
    \label{eq:BMN-fuzzy-sphere-vacuum-review-draft4}
\end{equation}
Thus gauge-inequivalent classical vacua are classified by \(q\)-dimensional
unitary representations of \(SU(2)\), or equivalently by partitions
\(\lambda\vdash q\).  In
multiplicity notation,
\begin{equation}
    \lambda=(1^{r_1}2^{r_2}\cdots),
    \qquad
    r_s\in\mathbb Z_{\geq0},
    \qquad
    \sum_{s\geq1}sr_s=q,
    \label{eq:BMN-partition-multiplicities-review-draft4}
\end{equation}
and the defining representation decomposes as
\begin{equation}
    \mathbb C^q
    =\bigoplus_{s\geq1}
    \mathbb C^{r_s}\otimes\mathcal V_{\frac{s-1}{2}},
    \qquad
    L^i_{\lambda}
    =\bigoplus_{s\geq1}
    \boldsymbol 1_{r_s}\otimes L^i_{(s)}.
    \label{eq:BMN-vacuum-decomposition-review-draft4}
\end{equation}
Here \(\mathcal V_{(s-1)/2}\) is the \(s\)-dimensional irreducible
representation of \(SU(2)\).  The gauge group left unbroken by the vacuum is
the centralizer
\begin{equation}
    G_{\lambda}
    =\prod_{s:r_s>0}U(r_s).
    \label{eq:BMN-unbroken-gauge-group-review-draft4}
\end{equation}

At fixed \(q\) and nonzero \(\mu\), the refined index is invariant under
changes of the finite coupling.  It can therefore be evaluated as
\(g_{\mathrm B}^2/\mu^3\to0\), where the isolated classical vacua decouple
and the fluctuation problem becomes quadratic \cite{Chang:2024lkw}.
Let \(\mathcal H_{q,\lambda}\) denote the gauge-invariant fluctuation
Hilbert space around \(\lambda\) in this limit.  Its contribution is
\begin{equation}
\begin{aligned}
    \mathcal I^{\mathrm{BMN}}_{q,\lambda}
    (\Delta_1,\Delta_2,\Delta_3)
    \equiv
    \operatorname{Tr}_{\mathcal H_{q,\lambda}}\Big[{}
       &(-1)^{2M^{12}}
        e^{-\beta_{\mathrm B}\Delta_{\mathrm{BMN}}}
    \\
       &\times
        e^{-\Delta_1\ell_1-\Delta_2\ell_2-\Delta_3\ell_3}
    \Big].
    \label{eq:BMN-vacuum-index-review-draft4}
\end{aligned}
\end{equation}
On gauge-invariant states \((-1)^{2M^{12}}=(-1)^F\).  The index is
independent of \(\beta_{\mathrm B}\) and receives contributions only from
\(\Delta_{\mathrm{BMN}}=0\) states.  The vacuum labels organize this
weak-coupling evaluation; an exact decomposition of the interacting excited
Hilbert space into separate vacuum sectors is not needed.

The protected oscillator spectrum gives a finite-rank matrix integral for
\eqref{eq:BMN-vacuum-index-review-draft4} \cite{Chang:2024lkw}.  Introduce
\begin{equation}
    u_a=e^{-\Delta_a},
    \qquad
    \mathfrak u=u_1u_2u_3,
    \qquad
    p_m=(1-u_1^m)(1-u_2^m)(1-u_3^m).
    \label{eq:BMN-fugacity-abbreviations-review-draft4}
\end{equation}
The rectangular matrix harmonics between irreducible blocks of dimensions
\(s\) and \(t\) obey the tensor-product decomposition
\begin{equation}
    \operatorname{Hom}\left(
        \mathcal V_{\frac{t-1}{2}},\mathcal V_{\frac{s-1}{2}}
    \right)
    =
    \bigoplus_{j=\frac12|s-t|}^{\frac12(s+t)-1}\mathcal V_j.
    \label{eq:BMN-fuzzy-harmonic-range-review-draft4}
\end{equation}
The spin sum runs in integer steps.  Coupling these harmonics to the
bosonic and fermionic polarizations and retaining the BPS modes gives the
\(m\)-th plethystic kernel
\begin{equation}
    \iota_{st}^{(m)}
    =\delta_{st}
    +\sum_{j=\frac12|s-t|}^{\frac12(s+t)-1}
    (-1)^{2jm+1}\mathfrak u^{jm}p_m.
    \label{eq:BMN-block-kernel-review-draft4}
\end{equation}
There is no extra factor of \(2j+1\), since this expression counts protected
modes rather than the full harmonic multiplet.  The term \(\delta_{st}\)
accounts for an absent endpoint mode when \(s=t\); it cancels the constant
part of the \(j=0\) term.  Summing the finite progression gives
\begin{equation}
    \iota_{st}^{(m)}
    =\delta_{st}
    +(-1)^{m|s-t|+1}
     \mathfrak u^{\frac m2|s-t|}
     \frac{1-\mathfrak u^{m\min(s,t)}}{1-\mathfrak u^m}
     p_m.
    \label{eq:BMN-block-kernel-closed-review-draft4}
\end{equation}
The ratio is a finite polynomial in \(\mathfrak u^m\), so there is no pole
at \(\mathfrak u^m=1\).

The \(m\)-dependence of the sign in
\eqref{eq:BMN-block-kernel-review-draft4} is essential for blocks of
different parity.  On an \((s,t)\) oscillator the insertion
\((-1)^{2M^{12}}\) differs from fermion parity by
\((-1)^{s-t}\).  Combining this phase with bosonic and fermionic
plethystic statistics gives the factor \((-1)^{m|s-t|+1}\); it cannot be
replaced by its value at \(m=1\).  The extra block phases cancel after
gauge projection.

Let \(U_s\in U(r_s)\) be a holonomy of the unbroken group
\eqref{eq:BMN-unbroken-gauge-group-review-draft4}, and let \([dU_s]\) denote
normalized Haar measure.  Gauge projection gives
\begin{equation}
\begin{aligned}
    \mathcal I^{\mathrm{BMN}}_{q,\lambda}
    (\Delta_1,\Delta_2,\Delta_3)
    ={}&
    \int\prod_{s:r_s>0}[dU_s]
    \\
    &\times
    \exp\left[
        \sum_{m=1}^{\infty}\frac1m
        \sum_{s,t\geq1}
        \iota_{st}^{(m)}
        \operatorname{Tr}(U_s^{\dagger m})
        \operatorname{Tr}(U_t^m)
    \right].
    \label{eq:BMN-vacuum-matrix-integral-review-draft4}
\end{aligned}
\end{equation}
A trace is understood to vanish whenever the corresponding multiplicity is
zero.  Equivalently, the Haar-invariant change of variables
\(U_s\mapsto(-1)^sU_s\) removes the factor \((-1)^{m|s-t|}\) from the
nonconstant term in the kernel.  This relates
\eqref{eq:BMN-vacuum-matrix-integral-review-draft4} to the convention with
ordinary fermion parity on each oscillator.

Summing the vacuum contributions yields the rank-\(q\) index
\begin{equation}
    \mathcal I_q^{\mathrm{BMN}}
    (\Delta_1,\Delta_2,\Delta_3)
    =\sum_{\lambda\vdash q}
    \mathcal I^{\mathrm{BMN}}_{q,\lambda}
    (\Delta_1,\Delta_2,\Delta_3).
    \label{eq:BMN-all-vacuum-index-review-draft4}
\end{equation}
Each vacuum contribution has constant term one.  We retain
the full \(U(q)\) model, including the decoupled overall
\(U(1)\) trace multiplet describing its center-of-mass motion.
This multiplet is distinct from the neutral ABJM factor
appearing in \eqref{eq:intro-index-relation-draft4}.
Section~\ref{sec:double-scaling-index-draft4} identifies each
BMN partition \(\lambda\) with a matched nonzero GNO flux
sector and derives the corresponding matrix integral.

\section{Schur polynomials and character identities}
\label{app:character-identities-draft5}

We collect the character conventions and identities used in
section~\ref{sec:double-scaling-index-draft4}, following
\cite{bump2006averages}.  The notation distinguishes three
objects: \(s_\lambda\) is a Schur polynomial,
\(\chi_\alpha(U)\) is a \(U(r)\) character evaluated on a
matrix, and \(\chi^\tau(\eta)\) is an \(S_d\) character
evaluated on a permutation cycle type.  The rank \(r\)
specializes to \(N_0\) in the neutral sector and to \(r_s\)
in an occupied charged block.  All Haar measures are normalized
to unit volume.

We begin with the Schur-polynomial and unitary-character
conventions.
A partition \(\lambda\) is a nonincreasing sequence of
nonnegative integers with finitely many nonzero entries.
Its size is \(|\lambda|=\sum_i\lambda_i\), and its length
\(\ell(\lambda)\) is the number of nonzero entries.
For \(\ell(\lambda)\leq r\), pad the partition with zeros
to length \(r\).  The Schur polynomial is
\begin{equation}
    s_\lambda(z_1,\ldots,z_r)
    =\frac{\det\bigl(z_i^{\lambda_j+r-j}\bigr)_{i,j=1}^r}
           {\det\bigl(z_i^{r-j}\bigr)_{i,j=1}^r}.
    \label{eq:characters-Schur-definition-appB-draft5}
\end{equation}
The ratio is a symmetric polynomial and extends to coincident
eigenvalues.  We set \(s_\lambda=0\) when
\(\ell(\lambda)>r\), and \(s_\varnothing=1\).
For a matrix \(X\) with eigenvalues \(z_i\),
\(s_\lambda(X)\) means evaluation on those eigenvalues.
Restricted to \(U(r)\), it is the character of the polynomial
representation labelled by \(\lambda\).  In particular,
\(s_{(1)}(X)=\operatorname{Tr}X\) and
\(s_{(1^r)}(X)=\det X\).

The irreducible characters of \(U(r)\) are labelled more
generally by dominant integer weights
\(\alpha=(\alpha_1\geq\cdots\geq\alpha_r)\in\mathbb Z^r\).
For an invertible matrix \(X\), define
\begin{equation}
    \chi_\alpha(X)
    =(\det X)^{\alpha_r}
      s_{\alpha-\alpha_r\boldsymbol1_r}(X).
    \label{eq:characters-Laurent-definition-appB-draft5}
\end{equation}
Here \(\boldsymbol1_r=(1,\ldots,1)\), and the subscript of
\(s\) is a partition.  If \(\alpha_r\geq0\), then
\(\chi_\alpha=s_\alpha\); otherwise the character is a
symmetric Laurent polynomial.  The determinant shift is
\begin{equation}
    \chi_{\alpha+h\boldsymbol1_r}(X)
    =(\det X)^h\chi_\alpha(X),\qquad h\in\mathbb Z.
    \label{eq:characters-determinant-shift-appB-draft5}
\end{equation}
For example, at rank two,
\(\chi_{(0,-1)}(X)=(\det X)^{-1}\operatorname{Tr}X
=\operatorname{Tr}X^{-1}\).  Negative weights therefore
describe inverse-eigenvalue powers, rather than additional
partitions.

For \(U\in U(r)\),
\(\chi_\beta(U^{-1})=\overline{\chi_\beta(U)}\), and
normalized Haar orthogonality reads
\begin{equation}
    \int_{U(r)}[dU]\,
    \chi_\alpha(U)\chi_\beta(U^{-1})
    =\delta_{\alpha\beta}.
    \label{eq:characters-unitary-orthogonality-appB-draft5}
\end{equation}
For partition labels this reduces to the corresponding
identity for Schur polynomials.  The conjugation in this
statement acts on the character value of a unitary matrix;
formal fugacity coefficients in an insertion are held fixed.

We next relate these unitary characters to symmetric-group
characters and trace moments.
Let \(\eta=(1^{c_1}2^{c_2}\cdots)\vdash d\) label a
conjugacy class of \(S_d\): a permutation of this type has
\(c_m\) cycles of length \(m\).  Define
\begin{equation}
    d=\sum_{m\geq1}m c_m,
    \qquad z_\eta=\prod_{m\geq1}m^{c_m}c_m!.
    \label{eq:characters-cycle-normalization-appB-draft5}
\end{equation}
The conjugacy class has \(d!/z_\eta\) elements.
For \(\tau\vdash d\), \(\chi^\tau(\eta)\) is the trace
of a permutation of cycle type \(\eta\) in the irreducible
\(S_d\) representation labelled by \(\tau\).  These character
values are integers.

The Frobenius identity relates these values to the polynomial
characters of \(U(r)\):
\begin{equation}
    \prod_{m\geq1}(\operatorname{Tr}U^m)^{c_m}
    =\sum_{\substack{\tau\vdash d\\\ell(\tau)\leq r}}
      \chi^\tau(\eta)s_\tau(U).
    \label{eq:characters-Frobenius-appB-draft5}
\end{equation}
One way to obtain this identity is to take the trace of
\(U^{\otimes d}\) composed with a permutation of cycle
type \(\eta\).  Each cycle produces one trace on the left;
the decomposition under the commuting \(U(r)\) and \(S_d\)
actions gives the right-hand side.  In particular,
\(s_\tau(U)=\chi_\tau(U)\) is a unitary-group character,
whereas \(\chi^\tau(\eta)\) is a symmetric-group character.

The symmetric-group character table obeys
\begin{equation}
    \sum_{\tau\vdash d}
      \chi^\tau(\eta)\chi^\tau(\zeta)
    =z_\eta\,\delta_{\eta\zeta},
    \qquad \eta,\zeta\vdash d.
    \label{eq:characters-symmetric-orthogonality-appB-draft5}
\end{equation}
Combining \eqref{eq:characters-Frobenius-appB-draft5} with
\eqref{eq:characters-unitary-orthogonality-appB-draft5}
therefore gives, for
\(\zeta=(1^{e_1}2^{e_2}\cdots)\vdash d\),
\begin{equation}
    \int_{U(r)}[dU]\,
    \prod_{m\geq1}(\operatorname{Tr}U^m)^{c_m}
                  (\operatorname{Tr}U^{-m})^{e_m}
    =\sum_{\substack{\tau\vdash d\\\ell(\tau)\leq r}}
      \chi^\tau(\eta)\chi^\tau(\zeta).
    \label{eq:characters-finite-rank-moment-appB-draft5}
\end{equation}
When \(r\geq d\), every partition of \(d\) is allowed,
and \eqref{eq:characters-symmetric-orthogonality-appB-draft5}
reduces the answer to \(z_\eta\delta_{\eta\zeta}\).
If the two total trace degrees differ, the integral vanishes
by the central \(U(1)\) projection.  These statements give
\eqref{eq:stabilization-Haar-moment-draft4} with \(r=N_0\).

Finally, we record the Cauchy identity and the finite Laurent
expansions used in the charged-block projection.
For invertible rank-\(r\) matrices \(V,W\), with eigenvalues
\(v_i,w_j\), the Cauchy identity is
\begin{equation}
    \begin{aligned}
        \mathcal C(V,W)
        &\equiv\exp\left[\sum_{m\geq1}\frac1m
          \operatorname{Tr}V^{-m}\operatorname{Tr}W^m\right]
        \\
        &=\prod_{i,j=1}^r\left(1-\frac{w_j}{v_i}\right)^{-1}
         =\sum_{\lambda:\,\ell(\lambda)\leq r}
           s_\lambda(V^{-1})s_\lambda(W).
    \end{aligned}
    \label{eq:characters-Cauchy-appB-draft5}
\end{equation}
We use this as a formal expansion in the \(W\) eigenvalues,
with the \(V\) eigenvalues invertible.  Haar projection is
performed term by term in the character expansion.  This
is the convention for
\eqref{eq:projection-Cauchy-kernel-draft4}.

The determinant shift also explains the character bound in
\eqref{eq:projection-character-insertion-draft5}.
Let \(\Phi(V,W)\) be a finite Laurent polynomial invariant
under conjugation of \(W\), with each \(W\) eigenvalue
exponent in \([-L,L]\), where \(L\) is a nonnegative integer.
Then \((\det W)^{-L}\Phi(V,W)\) is an ordinary symmetric
polynomial in the eigenvalues of \(W^{-1}\), and hence
\begin{equation}
    \begin{aligned}
        (\det W)^{-L}\Phi(V,W)
        &=\sum_{\lambda:\,\ell(\lambda)\leq r}
          a_\lambda(V)s_\lambda(W^{-1}),
        \\
        \Phi(V,W)
        &=\sum_\beta c_\beta(V)\chi_\beta(W^{-1}),
        \qquad \beta=\lambda-L\boldsymbol1_r.
    \end{aligned}
    \label{eq:characters-Laurent-expansion-appB-draft5}
\end{equation}
The sums are finite.  Since \(\lambda_r\geq0\), every
nonzero character coefficient satisfies
\begin{equation}
    \beta_1\geq\cdots\geq\beta_r\geq-L,
    \qquad \beta_i\in\mathbb Z.
    \label{eq:characters-Laurent-support-appB-draft5}
\end{equation}
The equality \(\beta_r=-L\) is allowed.  For example,
\((\det W)^L=\chi_{(-L,\ldots,-L)}(W^{-1})\).
This lower bound follows from the upper bound on the original
\(W\) eigenvalue powers, because the character expansion is
in \(W^{-1}\).  With \(r=r_s\), it supplies the condition
used in the projection identity
\eqref{eq:projection-determinant-identity-draft4}.

\bibliographystyle{JHEP}
\bibliography{ref.bib}
\end{document}

%% file: preamble.tex
\usepackage{jheppub}
\usepackage{mathtools}
\usepackage{float}
\usepackage{tikz-cd}

\date{}

\newcommand{\s}{\psi}

 \usepackage[normalem]{ulem}
\usepackage[textsize=scriptsize,textwidth=2.5cm]{todonotes}

\let\exporig\exp
\usepackage{physics}
\usepackage{enumitem}
\let\exp\exporig

\usepackage{xspace}